\documentclass[usenatbib]{mn2e}
\usepackage{amsmath, amssymb}
\usepackage{amsfonts}
\usepackage{epsfig,rotating,subfigure}

\def\simeq{
\mathrel{\raise.3ex\hbox{$\sim$}\mkern-14mu\lower0.4ex\hbox{$-$}}
}

\def\ltsima{$\; \buildrel < \over \sim \;$}
\def\simlt{\lower.5ex\hbox{\ltsima}}
\def\gtsima{$\; \buildrel > \over \sim \;$}
\def\simgt{\lower.5ex\hbox{\gtsima}}

\def\msun{{\rm M_{\odot}}}

\def\be{\begin{equation}}
\def\ee{\end{equation}}

\def\del#1{{}}

\newcommand{\apj}{ApJ}
\newcommand{\apjs}{ApJS}
\newcommand{\mnras}{MNRAS}
\newcommand{\aap}{A\&A}
\newcommand{\araa}{ARA\&A}
\newcommand{\apjl}{ApJL}
\newcommand{\aj}{AJ}

\newcommand{\ssr}{Sp.Sci.Rev.}
\newcommand{\aapr}{Astron.Astrophys.Rev.}

\title{Collapse and fragmentation of molecular clouds under pressure}
\author[Kastytis~Zubovas, Kostas~Sabulis, Rokas~Naujalis]{Kastytis~Zubovas, Kostas~Sabulis and Rokas~Naujalis\\
Center for Physical Sciences and Technology, Savanori\c{u} 231, Vilnius LT-02300, Lithuania\\
{E-mail:~} {\rm kastytis.zubovas@ftmc.lt}}

\begin{document}

\maketitle

\begin{abstract}

Recent analytical and numerical models show that AGN outflows and jets
create ISM pressure in the host galaxy that is several orders of
magnitude larger than in quiescent systems. This pressure increase can
confine and compress molecular gas, thus accelerating star
formation. In this paper, we model the effects of increased ambient
ISM pressure on spherically symmetric turbulent molecular clouds. We
find that large external pressure confines the cloud and drives a
shockwave into it, which, together with instabilities behind the shock
front, significantly accelerates the fragmentation rate. The
compressed clouds therefore convert a larger fraction of their mass
into stars over the cloud lifetime, and produce clusters that are
initially more compact. Neither cloud rotation nor shear against the
ISM affect this result significantly, unless the shear velocity is
higher than the sound speed in the confining ISM. We conclude that
external pressure is an important element in the star formation
process, provided that it dominates over the internal pressure of the
cloud.

\end{abstract}


\section{Introduction}

Under normal circumstances, star formation is a slow process: on all
spatial scales larger than single pre-stellar cores, almost
independently of environment, only $1-3 \%$ of the molecular gas
available for star formation is converted into stars every dynamical
time \citep{McKee2007ARA&A, Krumholz2007ApJ,
  Krumholz2012ApJ}. Molecular clouds have lifetimes that depend on
their sizes, with the smallest clouds ($R_{\rm cl} \sim 1-3$~pc)
dispersing after only $1-3$~Myr due to stellar feedback
\citep{Allen2007conf, Hartmann2001ApJ}, while the largest clouds
survive up to $\sim 30$~Myr \citep{Williams1997ApJ, Kawamura2009ApJS,
  Dobbs2013MNRAS}. Despite these differences in absolute values, the
cloud lifetimes are typically equal to a few crossing times, defined
as the cloud size divided by the velocity dispersion in the cloud
\citep{Dobbs2013MNRAS}, and similar to the dynamical times for
turbulence-supported clouds. As a result, most clouds only convert $<
10\%$ of their gas mass into stars before dispersing
\citep{Williams1997ApJ, Hartmann2001ApJ, Kawamura2009ApJS,
  Dobbs2013MNRAS}.

Star formation must be more efficient than this in order to produce
clusters that remain bound after the parent cloud is dispersed; some
simulations show that $\sim 10-30\%$ of the cloud mass must be
converted into stars \citep[e.g.][]{Geyer2001MNRAS,
  Baumgardt2007MNRAS}. On the other hand, the hierarchical structure
of the ISM implies that stars forming in the densest parts of the
molecular clouds are more bound than the cloud as a whole, and so can
form bound star clusters even though the global star formation
efficiency stays low \citep{Kruijssen2012MNRAS,
  Kruijssen2013arXiv}. On a global scale, it has been suggested that
gravitational binding of massive clouds ($M \ga 7 \times 10^6 \msun$)
might be enough to withstand stellar feedback until enough gas is
converted into stars \citep{Kroupa2002MNRAS, Bressert2012ApJ}, leading
to formation of globular clusters, while less massive clouds probably
have to be compacted in some way in order to form stars more rapidly
\citep{Escala2008ApJ, Larsen2010RSPTA}. However, these suggestions are
unlikely to be correct, since they predict a cutoff of cluster
populations at low masses, which is not observed
\citep{Fall2012ApJ,Bastian2012A&A,Bastian2012MNRAS}.  Overall, a
picture emerges wherein regions of high density are important for the
formation of bound star clusters, but those regions do not necessarily
encompass whole clouds.

Several authors have suggested that the star formation rate is
ultimately governed by self-regulation, for example a balance between
pressure created by stellar feedback and self-gravity of the gas
\citep[e.g.,][]{Thompson2005ApJ}. If that is the case, then an
increase in the pressure of the ISM surrounding the cloud should
result in an increase of the star formation rate, as suggested
recently by, e.g., \citet{Zubovas2013MNRASb}. Numerous recent models
of the effect of AGN jets \citep{Silk2005MNRAS, Gaibler2012MNRAS} and
outflows \citep{Ciotti2007ApJ, Nayakshin2012MNRASb, Zubovas2013MNRASb,
  Ishibashi2012MNRAS} upon the host galaxy ISM suggest that these
processes can compress molecular gas to pressures several orders of
magnitude higher than typically found in the hot phase ISM of
quiescent galaxies. If this pressure translates into a linear increase
in star formation rate, gas-rich AGN hosts may experience starbursts
with star formation rates of several hundred $\msun$ yr$^{-1}$
\citep[e.g.][]{Drouart2014arXiv}.

Although the connection between higher ambient pressure and an
increased star formation rate seems robust, it rests on several
assumptions. First of all, it assumes that external pressure creates
more favourable conditions for star formation, i.e. that star
formation can be triggered by a pressure increase. The second
assumption is that there is enough material that can readily react to
an increase in pressure by forming stars, so that external pressure
accelerates ongoing star formation. Finally, the connection requires a
steady state to be established: star formation must not increase to
such rates that molecular gas is exhausted before feedback can
establish self-regulation. These assumptions cannot be tested in
large-scale models, because they require analysis of gas dynamics and
fragmentation on scales of molecular clouds, below the typical
resolution of galaxy-wide numerical simulations.

In this paper, we present results of numerical SPH simulations of
spherically symmetric turbulent clouds embedded in a hot ISM. We track
the collapse and fragmentation of the clouds, showing that under
pressure, fragmentation is caused by a combined action of a shockwave
driven into the cloud and instabilities behind it, leading to much
higher fragmentation rates than in uncompressed cloud. The net effect
of external compression is that confined clouds collapse and fragment
on a timescale shorter than the analytically-derived cloud-crushing
timescale $t_{\rm cr} \simeq t_{\rm dyn} \left(1 + P_{\rm ISM}/P_{\rm
  grav}\right)^{-1/2}$, where $P_{\rm grav}$ is the dynamical pressure
of the cloud material necessary to overcome gravitational collapse,
while $P_{\rm ISM}$ is the external pressure. Strongly compressed
clouds fragment and turn a significant fraction of their mass into
sink particles in $t < 1$~Myr, before the cloud can begin to disperse
due to stellar feedback.  Furthermore, external pressure may confine
even gravitationally unbound clouds, suggesting that compressed clouds
may survive for longer even in the presence of stellar feedback or
following the passage of an AGN shockwave. The resulting cluster of
sink particles forming in our compressed cloud simulations is more
massive and compact than the cluster born in uncompressed models. We
conclude that external pressure enhances star formation in the cold
ISM and produces clusters that are likely to survive for longer
periods of time.

The paper is organised as follows. We begin by describing in more
detail the physical basis of the connection between AGN activity and
enhanced star formation (Section \ref{sec:feedback}). Next, in Section
\ref{sec:model}, we present analytical estimates of the effect of
external pressure on the cloud. In Section \ref{sec:numsim}, we
describe the set up of numerical simulations, while their results are
shown in Section \ref{sec:results}. Discussion of our findings and
their implications is presented in Section \ref{sec:discuss}. Finally,
we summarize and conclude in Section \ref{sec:concl}.

\section{Positive AGN feedback on star formation} \label{sec:feedback}

The general picture of AGN effect upon the host galaxy is that of
negative feedback. AGN jets or wide-angle outflows heat the gas and
push it out of the host galaxy, quenching the star formation process
\citep{McNamara2007ARA&A, Feruglio2010A&A, Fabian2012ARA&A}. There is
growing evidence, however, that the real picture is more complex, and
that star formation can be enhanced by AGN activity as well.

One process through which AGN activity can enhance star formation is
the increase in hot phase ISM pressure. Simple photoionization and
Compton heating by the AGN radiation field can increase the
temperature of the diffuse gas to $10^7$~K or more. Using a typical
ISM density of $n_{\rm ISM} = 1 n_0$~cm$^{-3}$, we find that the
distance to which the gas is heated efficiently is
\citep{Sazonov2004MNRAS}
\begin{equation}
R_{\rm heat} \simeq \left(\frac{200 L_{\rm AGN}}{n_{\rm ISM}T_{\rm
    ISM}}\right)^{1/2} \simeq 140 L_{46}^{1/2} n_0^{-1/2} T_7^{-1/2}
\; {\rm pc},
\end{equation}
where $L_{\rm 46} \equiv L_{\rm AGN} / 10^{46}$~erg~s$^{-1}$ is the
AGN luminosity and $T_7 \equiv T/10^7$~K the ISM temperature. The only
clouds affected by direct AGN heating are those close to the centre of
the galaxy. Furthermore, these clouds are themselves exposed to the
dissociating and ionising AGN radiation. They can be heated,
maintaining pressure equilibrium with the surroundings and losing
molecular gas, leading to a lower star formation rate. Observations do
not show either enhancment or suppression of star formation in clouds
located in the centres of AGN hosts \citep{Davies2005ApJ}, so we
believe that any effect due to direct heating is small.

A more promising approach for increasing the ISM pressure is shock
heating. Shocks can be caused by a variety of processes, such as tidal
interactions with companion galaxies \citep{Ricker1998ApJ} and ram
pressure stripping in galaxy clusters \citep{Bekki2002MNRAS,
  Bekki2003ApJ}. AGN can also create shocks in the ISM by heating it
rapidly via jets \citep{Gaibler2012MNRAS} and/or outflows
\citep{Zubovas2012ApJ}. The physical model we use in this paper is
based on predictions of these positive AGN feedback models, and
considers three situations how AGN jets and/or outflows can interact
with cold dense gas.

A high-pressure outflow (created by either wind or jet) can overtake
dense clumps of gas and compress them. The outflow expands with a
velocity of order $10^3$~km/s \citep{Zubovas2012ApJ}, corresponding to
a shock temperature of order $10^7 - 10^8$~K. This causes the pressure
inside an energy-driven AGN outflow to be 2-3 orders of magnitude
higher than typical hot ISM pressure \citep{Nayakshin2012MNRASb,
  Zubovas2013MNRASb}. Observed jet-inflated cocoons in radio galaxies
have similarly high pressures \citep{Begelman1989ApJ,
  Gaibler2012MNRAS}. The interaction between a dense cloud and the
outflow is very similar to the interaction of a cloud with a passing
shockwave \citep[e.g.,][]{Klein1994ApJ}, except that the material
behind the shockwave compresses the cloud further.

An expanding outflow is generally thermally unstable, can cool and
form clouds \citep{Nayakshin2012MNRASb, Zubovas2014MNRASa}.  These
clouds form in pressure equilibrium with the surrounding flow and
hence are not necessarily bound by their own gravity. A similar
scenario was considered by \citet{Elmegreen1997ApJ}, who found an
increase in the star formation efficiency of clouds forming in
high-pressure environments.

Finally, the outflow expanding in the diffuse gas of the galactic
bulge and halo compresses the galactic disc. This creates a secondary
shockwave passing into the disc and compressing the clouds there
\citep{Zubovas2013MNRASb}. The shockwave can develop a complex
morphology due to the uneven density distribution of the disc ISM and
therefore the clouds experience a wide range of shockwave velocities
passing through them.

In the next section, we make analytic estimates of the effect of
external pressure in these three situations, starting with the
simplest, if somewhat unrealistic, scenario of negligible lateral
velocity of the shockwave.

\section{Pressure-enhanced cloud collapse} \label{sec:model}

\subsection{Cloud confinement by external pressure} \label{sec:confine_analyt}

Here we make rough estimates regarding the effect that external
pressure has on the internal dynamics of a molecular cloud in the
various configurations discussed above. We scale all results to a
cloud of mass $M = 10^5 \; \msun$; this is a typical, if somewhat
massive, example of a molecular cloud in the Milky Way
\citep{Roman-Duval2010ApJ}. Using the \citet{Larson1981MNRAS} and
\citet{Solomon1987ApJ} relations, we find the linear size of the cloud
to be $L \sim 17$~pc, so we choose the cloud radius as $R = 10$~pc
$\sim L/2$. This translates into a surface density $\Sigma_{\rm cl}
\simeq 318 \; \msun$~pc$^{-2}$. The velocity dispersion of a cloud of
this size should be $3.2$~km/s~$< \sigma_{\rm v} < 4.2$~km/s; we
choose $\sigma_{\rm turb} = 3.6$~km/s to have the cloud supported
against self-gravity (see below). For simplicity, we consider the
cloud to be spherical with uniform density.

The gravitational binding energy of the cloud is
\begin{equation}
e_{\rm b} \simeq \frac{3}{5}\frac{GM}{R} \simeq 2.5 \times 10^{11} M_5
R_{10}^{-1} \; {\rm erg} \; {\rm g}^{-1},
\end{equation}
where $M_5 = M/10^5 \; \msun$ and $R_{10} = R/(10 \; {\rm pc})$. The
virial temperature of the cloud is $T_{\rm vir} \simeq 3800 M_5
R_{10}^{-1}$~K, much higher than the typical gas temperature $T \sim
10$~K. Support against rapid gravitational collapse comes from
supersonic turbulence, with typical turbulent velocity dispersion
\begin{equation} \label{eq:sigmaturb}
\sigma_{\rm turb} \simeq \sqrt{\frac{3}{10}\frac{GM}{R}} \simeq 3.6
M_5^{1/2} R_{10}^{-1/2}\; {\rm km} \; {\rm s}^{-1}.
\end{equation}
The effective dynamical pressure of the cloud is
\begin{equation} \label{eq:intpres}
\frac{P_{\rm grav}}{k_{\rm b}} \simeq \frac{\rho \sigma_{\rm
    turb}^2}{k_{\rm b}} = n T_{\rm vir} \simeq 1.4 \times 10^6 \;
M_5^2 R_{10}^{-4} \; {\rm K}\; {\rm cm}^{-3}.
\end{equation}
Here, $\rho \simeq 1.6 \times 10^{-21} M_5 R_{10}^{-3}$~g~cm$^{-3}$
and $n \simeq 380 M_5 R_{10}^{-3}$~cm$^{-3}$ are the mass density and
particle density in the cloud, respectively.

From equation (\ref{eq:intpres}) we see that as long as the pressure
in the ISM surrounding the cloud is $P_{\rm ISM}/k_{\rm b} \ll 1.4
\times 10^6$~K cm$^{-3}$, the cloud evolution is unaffected by
external pressure. This is the case in most `normal' environments,
where $P_{\rm ISM}/k_{\rm b} < 10^5$~K cm$^{-3}$
\citep{Wolfire2003ApJ}. If the external pressure increases above this
value, the cloud is compressed. The exact situation depends on the
dynamics of the surrounding ISM.

\subsubsection{Negligible lateral velocity} \label{sec:zerovel}

In the simplest case, the external pressure around the molecular cloud
increases isotropically and homogeneously. This is an unlikely
scenario, since typically high pressure is caused by a shockwave
enveloping the cloud. There are, however, a few situations where the
velocity of the cloud with respect to its surroundings is low. First
of all, the vertical gas velocity dispersion in gas-rich starburst
galaxy discs can be as large as $\sim 50-100$~km/s
\citep{Quinn1993ApJ, Scoville1997ApJ, Bryant1999AJ}, so it is
conceivable for a given molecular cloud to be moving with a velocity
$> 100$~km/s vertically w.r.t. the rest frame of the galactic disc. A
low-density ($n_{\rm out} \sim 0.1-1$~cm$^{-3}$) outflow in the
galactic halo moving with velocity $v_{\rm out} = 1000$~km/s produces
a shockwave in the galactic disc \citep[which has a density $n_{\rm
    ISM} \sim 10-100$~cm$^{-3}$; c.f.][]{Thompson2005ApJ,
  Abramova2008ARep} with a velocity of order $v \sim v_{\rm out}
\left(n_{\rm out}/n_{\rm ISM}\right)^{1/2} \sim 100$~km/s. It is
therefore possible that in the reference frame moving with the cloud,
the shockwave is much slower than the sound speed of the shocked gas
behind it. The direct interaction between the molecular cloud and the
passing shockwave is mitigated by the atomic hydrogen envelope around
the cloud (see also Section \ref{sec:ic_improve}). In another case, a
cloud that forms due to cooling of gas inside the fast hot outflow
also experiences high external pressure without significant lateral
motion.

The effect of this high isotropic pressure is to compress the
cloud. In order to withstand this pressure, the cloud should have a
higher turbulent velocity dispersion. We can update equation
(\ref{eq:sigmaturb}) to include external confinement:
\begin{equation}
\sigma'_{\rm turb} \simeq \sqrt{\frac{3}{10}\frac{GM}{R} +
  \frac{P_{\rm ISM}}{\rho_{\rm cl}}};
\end{equation}
alternatively, one can express the updated velocity dispersion in
terms of the pressure ratio:
\begin{equation}\label{eq:sigmadash}
\sigma'_{\rm turb} = \sigma_{\rm turb} \sqrt{1 + \frac{n_{\rm ISM}
    T_{\rm ISM}}{n T_{\rm vir}}} \simeq 10.3 M_5^{-1/2} R_{10}^{3/2}
P_7^{1/2} \; {\rm km/s}.
\end{equation}
In this equation, $n_{\rm ISM}$ and $T_{\rm ISM}$ refer to the density
and temperature of the confining hot phase ISM, respectively, and $P_7
= P_{\rm ISM}/\left(10^7 k_{\rm b} \; {\rm K} \; {\rm
  cm}^{-3}\right)$. We further assume that the ISM pressure is purely
thermal. A corresponding timescale for the evolution of the compressed
cloud, which we term the effective dynamical timescale, is
\begin{equation} \label{eq:tdyndash}
t'_{\rm dyn} \sim t_{\rm dyn} \frac{\sigma_{\rm turb}}{\sigma'_{\rm
    turb}} = t_{\rm dyn} \left(1 + \frac{n_{\rm ISM} T_{\rm ISM}}{n
  T_{\rm vir}}\right)^{-1/2},
\end{equation}
where $t_{\rm dyn} \simeq 1.7 M_5^{-1/2} R_{10}^{3/2}$~Myr $\propto
\sigma_{\rm turb}^{-1}$ is the dynamical timescale of the cloud.

A shockwave is driven into the cloud approximately isotropically. It
has a predominantly radial velocity \citep[cf.,
  e.g.,][]{Jog1992ApJ,Spitzer1978book}
\begin{equation} \label{eq:vshock}
v_{\rm sh} \sim \left(\frac{P_{\rm ISM}}{\rho_{\rm cl}}\right)^{1/2}
\sim \left(\frac{\rho_{\rm ISM}}{\rho_{\rm cl}}\right)^{1/2}
c_{\rm s,ISM} \sim \sigma'_{\rm turb},
\end{equation}
where the last equality is valid if $P_{\rm ISM} \gg P_{\rm
  grav}$. The timescale for the shockwave to reach the centre of the
cloud, known as the cloud crushing timescale \citep{Klein1994ApJ}, is
the same as $t'_{\rm dyn}$ provided that $P_{\rm ISM} \gg P_{\rm
  grav}$. In the opposite case, the cloud crushing timescale becomes
longer than dynamical and crushing is essentially negligible, as
expected.

The passage of the shockwave heats the gas to temperatures
\begin{equation} \label{eq:tshock}
T_{\rm sh} \sim \frac{3 \mu_{\rm cl} m_{\rm p}}{16 k_{\rm B}} v_{\rm
  sh}^2 \sim \frac{3}{16} \frac{\rho_{\rm ISM}}{\rho_{\rm cl}} T_{\rm
  ISM}.
\end{equation}
For an ISM pressure $P_{\rm ISM}/k_{\rm B} = 10^7$~K cm$^{-3}$, this
translates into a postshock temperature $T_{\rm sh} \sim 2 \times
10^3$~K. Using the postshock gas density $n_{\rm sh} = 4 n_{\rm cl}$
(strong shock approximation; we further assume that gas remains mostly
molecular in the shock, since the temperature increase is not enough
to completely dissociate H$_2$ or CO) and the cooling function
approximation from \citet[Table 1]{Mckee1977ApJb}, we find the gas
cooling time $t_{\rm cool} \sim 400$~yr $\ll t'_{\rm dyn}$. Therefore
the postshock gas can be assumed to cool instantaneously and attain a
density \citep[cf.][]{Jog1992ApJ}
\begin{equation} \label{eq:nfinal}
n_{\rm final} \simeq \frac{16}{3} \frac{T_{\rm sh}}{T_{\rm vir}}
n_{\rm cl} \simeq \frac{P_{\rm ISM}}{P_{\rm grav}} n_{\rm cl} \simeq
2.7 \times 10^3 P_7 M_5^{-1} R_{10} \; {\rm cm}^{-3}.
\end{equation}
This final density increase by $\sim 7$ times over the mean cloud
density lowers the Jeans' length and mass of the post-shock gas by a
factor $\sim 2.6$, assuming that the temperature stays the same. This
means that smaller density perturbations become unstable and collapse
to form stars, leading to rapid star formation in the shell driven
into the cloud.

If the cloud forms under conditions of high external pressure, its
turbulent velocity should have a value as given by equation
(\ref{eq:sigmadash}). Such a cloud would not be bound by its own
gravity, but as long as the external pressure persists, it is able to
fragment and form stars. The timescale of the cloud evolution is
$R/\sigma'_{\rm turb} = t'_{\rm dyn}$. Only a weak shockwave is driven
into the cloud, so star formation starts in the central parts of the
cloud, where the local dynamical time is shortest.

\subsubsection{Large lateral velocity}

If the cloud is compressed by a passage of a shockwave, such that the
shear velocity $v_{\rm lat}$ of the postshock ISM gas with respect to
the cloud is significant compared with the sound speed in this gas,
the shear affects the cloud evolution. Direct interaction between a
strong shock and the GMC is unlikely to occur: GMCs are typically
surrounded by warm atomic hydrogen envelopes, which slow down and
weaken the shockwave. The shockwave driven into the cloud is no longer
approximately spherical. In the leading edge of the cloud, the
shockwave velocity is approximately
\begin{equation} \label{eq:vshocklat}
v_{\rm sh,lat} \sim \left(\frac{\rho_{\rm ISM}}{\rho_{\rm
    cl}}\right)^{1/2} \left(c_{\rm s,ISM}^2 + v_{\rm
  lat}^2\right)^{1/2}.
\end{equation}
In other directions, the velocity is lower than this, but never lower
than $v_{\rm sh}$ (eq. \ref{eq:vshock}). The cloud is destroyed by the
shockwave on a timescale
\begin{equation} \label{eq:tdestr}
t_{\rm destr} \sim A \frac{R}{v_{\rm lat}} \left(\frac{\rho_{\rm
    ISM}}{\rho_{\rm cl}}\right)^{-1/2},
\end{equation}
where $A$ is a factor of order a few \citep[finds $A \simeq
  1.6$]{Klein1994ApJ, Agertz2007MNRAS}. The sound speed of the hot ISM
does not enter into the expression for the cloud destruction timescale
because cloud compression happens isotropically and does not disperse
the cloud. We can use this expression together with the effective
dynamical time $R/v_{\rm sh,lat}$ to estimate the mass of stars that
form in a cloud thus affected:
\begin{equation} \label{eq:mstarlat}
\begin{split}
M_{\rm *,lat} & \sim \epsilon_{\rm *,ff}M \frac{t_{\rm destr} v_{\rm
    sh,lat}}{R} \\
& \sim A \epsilon_{\rm *,ff} M \frac{\left(c_{\rm
    s,ISM}^2 + v_{\rm lat}^2\right)^{1/2}}{v_{\rm lat}} \geq A
\epsilon_{\rm *,ff} M.
\end{split}
\end{equation}
The net result is that even though the cloud is destroyed by the
shear, the mass of stars formed from the cloud is larger than in a
free-floating cloud. In a particular case of a strong shock and
stationary cloud, where $c_{\rm s,ISM}^2/v_{\rm lat}^2 = 5$, this
ratio is $\sim 2.5A$ and the fraction of mass converted into stars can
exceed $10\%$ before the cloud disperses.

\subsection{Cluster survival in high-pressure systems} \label{sec:survival_analytical}

The low integrated (that is, calculated over the lifetime of the cloud
rather than its dynamical time) efficiency of mass converstion into
stars in a GMC suggests that the clouds are rapidly destroyed by
stellar feedback. It is not well understood which of the many feedback
processes are most important. It has been recently proposed that
massive star clusters can form in molecular clouds that have escape
velocities higher than the sound speed in ionized gas $c_{\rm HII}
\sim 10$~km/s \citep{Bressert2012ApJ}. For unconfined clouds, this
condition translates to a critical mass $M_{\rm crit} \sim 7 \times
10^6 \msun$; clouds above this mass retain even photoionized gas
\citep{Kroupa2002MNRAS,Krumholz2009ApJb}. If no other feedback
processes were relevant, this would lead to such massive clouds having
star formation efficiencies of several times $10\%$. On the other
hand, these massive clouds may be destroyed by radiation pressure
\citep{Krumholz2009ApJb, Fall2010ApJ, Murray2010ApJ} and thus maintain
a low integrated star formation efficiency. It is, however, not clear
how important radiation pressure feedback is \citep{Krumholz2012ApJb,
  Krumholz2013MNRAS}.

No matter which feedback process disrupts the cloud, it must
counteract the forces holding the cloud together. For a free-floating
cloud, the only such force is the self-gravity of the cloud. Within
our model, external pressure acts as an additional factor preventing
gas escape and cloud dispersal. In the case of photoionization, the
cloud is unable to expand and disperse provided that the total
confining pressure (produced by both cloud self-gravity and ambient
ISM) is higher than $P_{\rm crit} \simeq \rho_{\rm cl} c_{\rm HII}^2 =
1.6 \times 10^{-9} M_5 R_{10}^{-3}$~erg cm$^{-3}$. For our fiducial
cloud parameters, this translates into a required ISM pressure
\begin{equation} \label{eq:pext_min}
P_{\rm ISM} = \rho_{\rm cl} \left(c_{\rm HII}^2 - \sigma_{\rm
  turb}^2\right) \simeq 1.4 \times 10^{-9} \; {\rm erg} \; {\rm
  cm}^{-3},
\end{equation}
or, equivalently
\begin{equation}
\frac{P_{\rm ISM}}{k_{\rm b}} = 9.9 \times 10^6 \; {\rm K} \; {\rm
  cm}^{-3}.
\end{equation}
The balance between photoionization heating and stellar winds on one
side and external pressure on the other allows the cloud to survive
the photoionizing radiation of young stars. Similar estimates based on
pressure balance can be made for other forms of feedback, but these
are beyond the scope of this paper. We merely wish to point out that
as a result of external pressure, the cloud survives for longer
against feedback than if it were not compressed.  Therefore, the
integrated star formation efficiency $M_* / M_{\rm cl}$ is higher than
in unconfined clouds even if the star formation efficiency per
dynamical time ($\epsilon_{\rm ff, *}$) were the same. Even small
clouds can have large SFEs, giving rise to more strongly bound
clusters. A similar result was found by \citet{Elmegreen1997ApJ}, who
suggested that high external pressure reduces mass loss from nascent
globular clusters and so enhances their survival prospects.

\subsection{Summary}

The calculations above reveal three major effects that confining
external pressure has on a molecular cloud:

\begin{enumerate}

\item The cloud is compressed, reducing the effective dynamical
  timescale and thus increasing the rates of fragmentation and star
  formation. This should be a general effect of higher ambient
  pressure, independent of its source, the timescale over which the
  pressure increases or the shear velocity of the hot ISM w.r.t. the
  cloud.

\item A shockwave is driven into the cloud from the sides toward the
  centre; the density in the postshock region exceeds that of the
  undisturbed cloud medium by a factor $\sim7$, facilitating star
  formation there. As a result, stars form more rapidly in compressed
  clouds than in undisturbed ones, so that the cloud evolves on the
  effective dynamical timescale. The shockwave is approximately
  spherical if the lateral motion of the ISM past the cloud is
  slow. If this velocity is large, the cloud is destroyed by the
  shockwave in a few effective dynamical times. The presence of the
  shockwave is guaranteed only if the external pressure increases
  around the cloud on a timescale shorter than the cloud dynamical
  time; otherwise, the cloud has time to establish virial equilibrium
  with the higher surrounding pressure.

\item As long as the total (external plus gravitational) pressure
  confining the cloud exceeds the pressure created by stellar
  feedback, the cloud is not disrupted and can continue to form
  stars. For the case of photoionizing feedback, this pressure is
  $\sim 1.6 \times 10^{-9} M_5 R_{10}^{-3}$~erg cm$^{-3}$, easily
  reached in ISM heated by supernovae or AGN activity. The fraction of
  gas converted into stars is larger in confined clouds than in
  uncompressed ones, leading to formation of more tightly bound
  clusters.

\end{enumerate}

Although these conclusions seem robust based on analytical
calculations alone, we wish to investigate the evolution of compressed
clouds in more detail. Therefore, we turn to numerical simulations.

\section{Numerical simulations} \label{sec:numsim}

\begin{table*}
\begin{tabular}{c | c c c c | c c c c}
Model ID & $v_{\rm turb}$ (km/s) & $T_{\rm ISM}$ (K) & $\Omega_{\rm rot}$ (km/s/pc) & $v_{\rm lat}$ (km/s) & $t_{\rm sink}$ (Myr) & $t_{\rm frag}$ (Myr) & $r_{\rm h}$ (pc) & $\epsilon_{\rm ff, sink}$ \\
\hline
\hline
t4T5 & 4 & $10^5$ & 0 & 0 & $1.26$ & $1.78$ & $4.26$ & $0.076$ \\
t4T7 & 4 & $10^7$ & 0 & 0 & $0.37$ & $0.43$ & $1.37$ & $> 0.9*$ \\
t10T5 & 10 & $10^5$ & 0 & 0 & $1.51$ & $2.51$ & $8.29$ & $0.016$ \\
t10T7 & 10 & $10^7$ & 0 & 0 & $0.40$ & $0.56$ & $1.68$ & $> 0.9*$ \\
t2.8r4.2T5 & 2.8 & $10^5$ & 0.42 & 0 & $1.33$ & $1.98$ & $3.81$ & $0.036$ \\
t2.8r4.2T7 & 2.8 & $10^7$ & 0.42 & 0 & $0.34$ & $0.42$ & $1.04$ & $0.75$ \\
\hline
t4v10T5 & 4 & $10^5$ & 0 & 10 & $1.33$ & $1.66$ & $2.68$ & $0.28$ \\
t4v10T7 & 4 & $10^7$ & 0 & 10 & $0.36$ & $0.42$ & $1.30$ & $0.98$ \\
t4v30T5 & 4 & $10^5$ & 0 & 30 & $1.33$ & $1.66$ & $2.80$ & $0.27$ \\
t4v30T7 & 4 & $10^7$ & 0 & 30 & $0.36$ & $0.42$ & $1.30$ & $0.96$ \\ 
t4v100T5 & 4 & $10^5$ & 0 & 100 & $1.23$ & $1.59$ & $2.94$ & $0.43$ \\
t4v100T7 & 4 & $10^7$ & 0 & 100 & $0.29$ & $0.43$ & $2.25$ & $0.92$ \\
t4v300T5 & 4 & $10^5$ & 0 & 300 & $1.08$ & $1.39$ & $3.85$ & $0.52$ \\
t4v300T7 & 4 & $10^7$ & 0 & 300 & $0.60$ & $0.89$ & $1.48$ & $0.80$ \\ 
\hline
\hline

\end{tabular}
\caption{Parameters of the numerical models and most important
  results. The first column shows the model ID. The next four columns
  give the parameters: cloud turbulent velocity, confining ISM
  temperature, angular velocity of cloud rotation and linear velocity
  of shearing cloud motion, respectively. The final three columns are
  the primary results: time of formation of the first sink particles
  $t_{\rm sink}$, fragmentation timescale $t_{\rm frag}$, half mass
  radius $r_{\rm h}$ at fragmentation time and efficiency of gas
  conversion into sink particles in one dynamical time $\epsilon_{\rm
    ff,sink}$. Numbers with asterisks are extrapolated from earlier
  snapshots.}
\label{table:param}
\end{table*}

We run a number of simulations using the hybrid N-body/SPH code
GADGET-3 \citep[an updated version of the publicly available code
  from][]{Springel2005MNRAS}. We utilize the SPHS method
\citep{Read2012MNRAS}, which is specifically designed to remove
artificial conductivity errors in standard SPH and resolve mixing of
multiphase material \citep{Read2010MNRAS} and had been used
succesfully in modelling multiphase flows \citep{Hobbs2013MNRAS}. We
employ the fourth-order HOCT4 kernel with 442 neighbours, and use
adaptive smoothing and gravitational softening lengths.

Each model starts with a spherically symmetric cloud with $M_{\rm cl}
= 10^5 \msun$ and $R_{\rm cl} = 10$~pc, giving a mean particle density
of molecular hydrogen $n_{\rm cl} \simeq 380$~cm$^{-3}$. We assume the
cloud to have uniform density initially; we comment on this assumption
in the Discussion (Section \ref{sec:ic_improve}). The dynamical time of
the cloud is $t_{\rm dyn} \simeq 1.7$~Myr. The cloud is supported
against self-gravity by a large-scale turbulent velocity field with a
characteristic velocity $\sigma_{\rm turb}$.

We choose an implementation of turbulent velocities that produces a
purely solenoidal (divergence-free) turbulent velocity spectrum
\citep{Dubinski1995ApJ, Hobbs2011MNRAS}. This means that turbulence is
incompressible; another extreme would be a purely compressive
(curl-free) turbulence. Although supersonic turbulence is generally at
least partially compressive, a large fraction of the turbulent energy
is expected to be in solenoidal modes \citep{Federrath2010HiA,
  Hennebelle2012A&ARv}, so we are confident that our choice of the
velocity spectrum is not totally unrealistic. Furthermore, solenoidal
turbulence has a shallower power spectrum than compressive
one. Numerical simulations tend to steepen the spectrum as time goes
by, since turbulence decays artificially starting from the smallest
length scales (highest wavenumbers), therefore our choice of turbulent
power spectrum should produce more realistic results than the opposite
extreme. Finally, it is important to note that density perturbations
grow $\sim 10$ times slower for solenoidal turbulence than with purely
compressive turbulence \citep{Federrath2010HiA}, thus our results of
fragmentation rates are most likely underestimates.

From a technical point of view, turbulence is implemented as
follows. The velocity field has a Kolmogorov power spectrum
\begin{equation}
P_{\rm v}\left(k\right) \propto k^{-11/3},
\end{equation}
where $k$ is the wavenumber. The velocity can be described as a curl
of a vector potential $A$ (this menas that the velocity field is
homogeneous and incompressible) and so the power spectrum can be
expressed as
\begin{equation} \label{eq:ak}
\left<\left|A_{\rm k}\right|^2\right> = C \left(k^2 + k_{\rm min}^2\right)^{-17/6},
\end{equation}
where $k_{\rm min} \simeq R_{\rm cl}^{-1}$ is the minimum wavelength
of turbulence and $C$ is an arbitrary constant which is set later in
order to give the characteristic velocity $\sigma_{\rm turb}$. The
vector potential is sampled in Fourier space on a periodic cubic grid
of $256^3$ cells, calculating the value of $A_{\rm k}$ using
eq. (\ref{eq:ak}). The curl of $A_{\rm k}$ then gives the velocity
field in Fourier space, which is Fourier-transformed into real
space. We then use tricubic interpolation to calculate the velocity of
each SPH particle.

Once the turbulent velocities are set up, we scale them to give the
desired characteristic velocity (and hence turbulent energy). We
consider two values of $\sigma_{\rm turb}$. The lower value,
$\sigma_{\rm turb} = 4$~km/s, supports the cloud against self gravity
and creates a dynamical pressure inside the cloud $P_{\rm dyn}/k_{\rm
  b} \simeq 1.7 \times 10^6$~K~cm$^{-3}$. The higher value,
$\sigma_{\rm turb} = 10$~km/s, creates the same dynamical pressure as
a could filled with photoionised gas would have: $P_{\rm dyn}/k_{\rm
  b} \simeq 1.06 \times 10^7$~K~cm$^{-3}$. We choose to represent
photoionized gas with a higher turbulent velocity, rather than higher
gas temperature, because photoionization predominantly affects diffuse
gas \citep{Dale2011MNRAS} and does not necessarily stop the collapse
of already dense regions \citep{Dale2012MNRAS}; turbulence mimics this
behaviour better than a uniform increase in gas internal energy.
Alternatively, the large value of turbulence may represent a cloud
which forms within a high-pressure outflow \citep{Zubovas2014MNRASa}.

The cloud is surrounded by an ISM with particle density $n_{\rm ISM} =
1$~cm$^{-3}$ and temperature of either $10^5$~K or $10^7$~K. This
produces a pressure either much lower than the dynamical pressure of
the cloud ($P_{\rm ISM}/k_{\rm b} = 10^5$~K~cm$^{-3}$) or pressure
high enough to confine even the highly turbulent cloud ($P_{\rm
  ISM}/k_{\rm b} = 10^7$~K~cm$^{-3}$). Accordingly, the models are
called ``uncompressed'' and ``compressed'' respectively. The high
external pressure is also higher than the ISM pressure necessary to
prevent cloud dispersal by photoionization (see
eq. \ref{eq:pext_min}). The whole system is set up in a periodic box
of side length $80$~pc (models with shearing motion use a box of side
length $160$~pc).

All the models use the same number of particles, $N = 10^6$, to
represent the cloud, giving a mass resolution $m_{\rm res} = 442m_{\rm
  SPH} = 44.2 \; \msun$. This resolution is good enough to resolve
very massive stars and small stellar associations. We implement a
cooling function appropriate for dense gas at temperatures between
$10$ and $10^4$~K \citep{Inoue2008ApJ}, which we modify so that
cooling is turned off for gas at temperatures between $3 \times
10^4$~K and $T_{\rm ISM}$. With this prescription, the cloud gas is
modelled with reasonable accuracy, while the surrounding ISM stays
isothermal.

In order to speed up simulations and track the fragmentation within
the cloud, we introduce sink particles in regions where the density
exceeds $\rho_{\rm crit} = 10^{-17}$~g~cm$^{-3} \simeq 1.5 \times 10^5
\; \msun$~pc$^{-3}$. At temperature $T = 10$~K, this corresponds to a
Jeans' mass $M_{\rm J} \simeq 0.4 \; \msun = 4 m_{\rm SPH}$. This mass
is similar to that of pre-stellar cores, so our simulations should
not overproduce the number and total mass of fragments. The low Jeans'
mass allow us to track gas dynamics down to the resolution limit and
below (albeit with lower accuracy below $\sim 40 \; \msun$).

The models analysed are listed in Table \ref{table:param}. We first
consider models with zero lateral velocity - t4T5, t4T7, t10T5, t10T7,
t2.8T5r4.2 and t2.8T7r4.2, where each model is labelled by the value
of turbulence (``t'', in km/s), logarithm of surrounding ISM pressure
(``T'') and rotational velocity at the edge of the cloud (``r'', in
km/s). These simulations are designed to show the basic behaviour of
clouds compressed by the hot ISM. Next, we model the more realistic
cases of non-zero shear, with relative velocities of the ISM w.r.t the
cloud (``v'') of 10, 30, 100 and 300 km/s. The duration for which we
run each simulation is determined by numerical resources, but in all
cases, by the end of the simulation at least $70\%$ of the cloud gas
is converted into sink particles.

\section{Results} \label{sec:results}

\begin{figure*}
  \centering
    \includegraphics[trim = 6mm 23mm 6mm 0, clip, width=0.33 \textwidth]{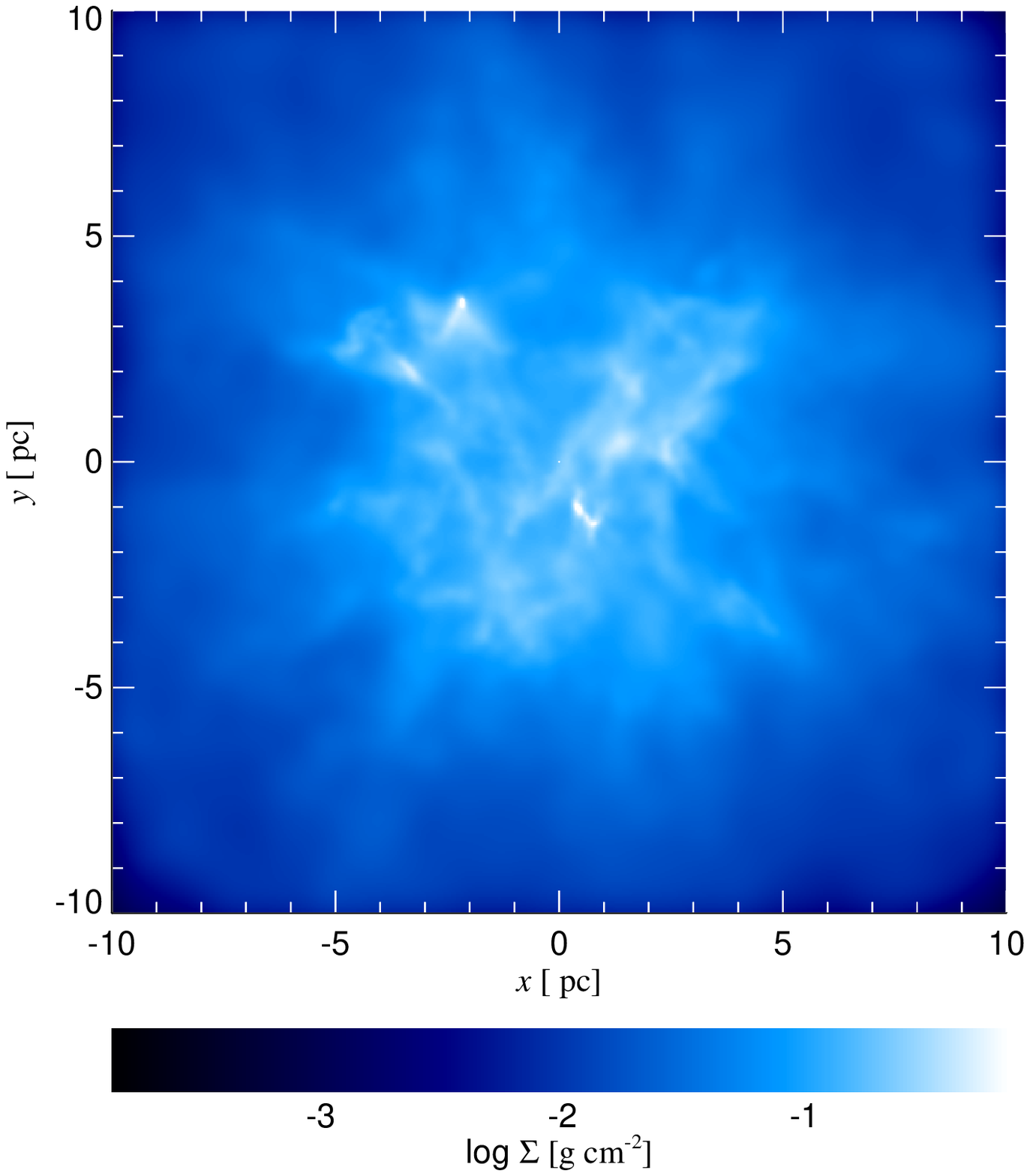}
    \includegraphics[trim = 6mm 23mm 6mm 0, clip, width=0.33 \textwidth]{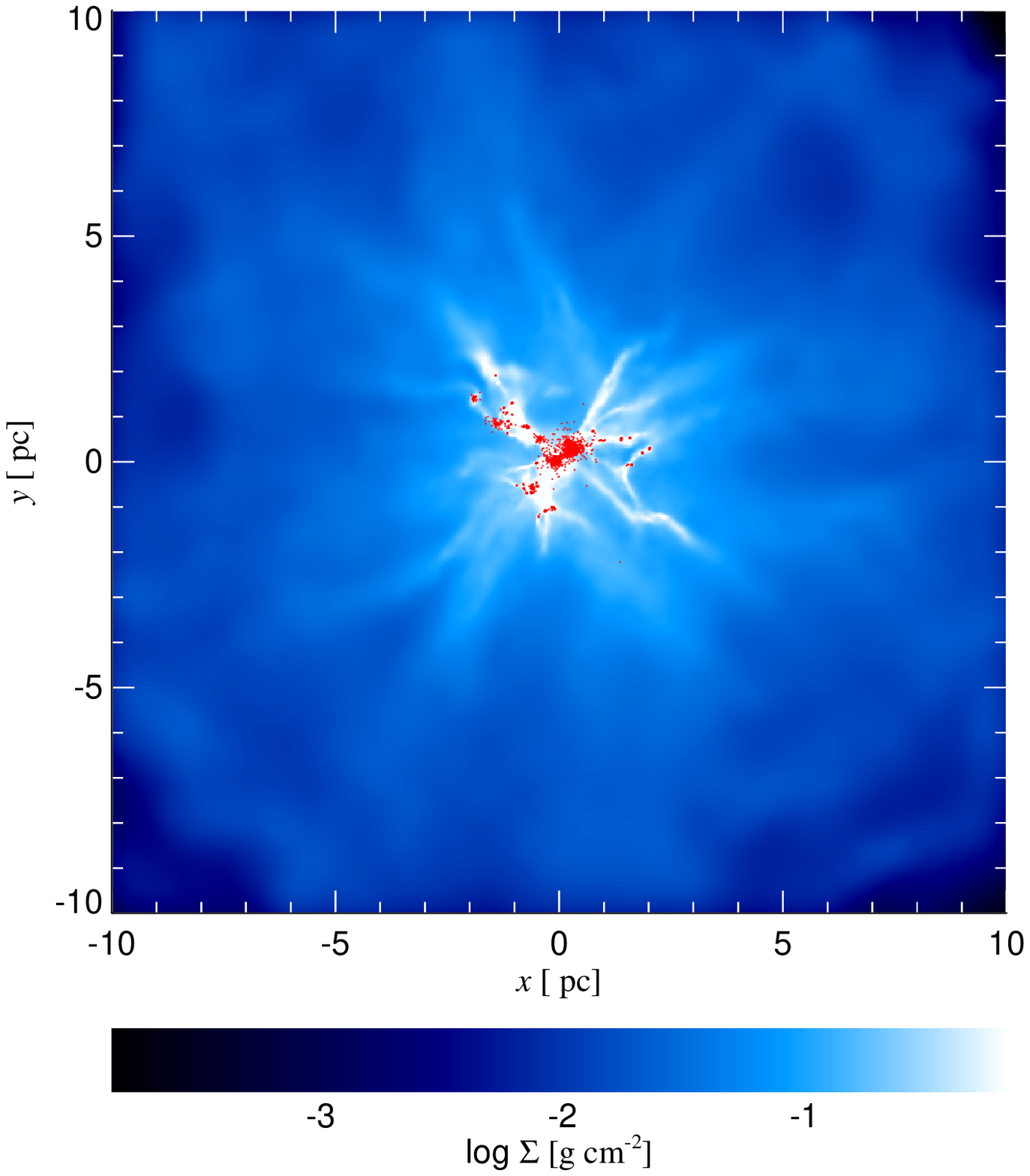}
    \includegraphics[trim = 6mm 23mm 6mm 0, clip, width=0.33 \textwidth]{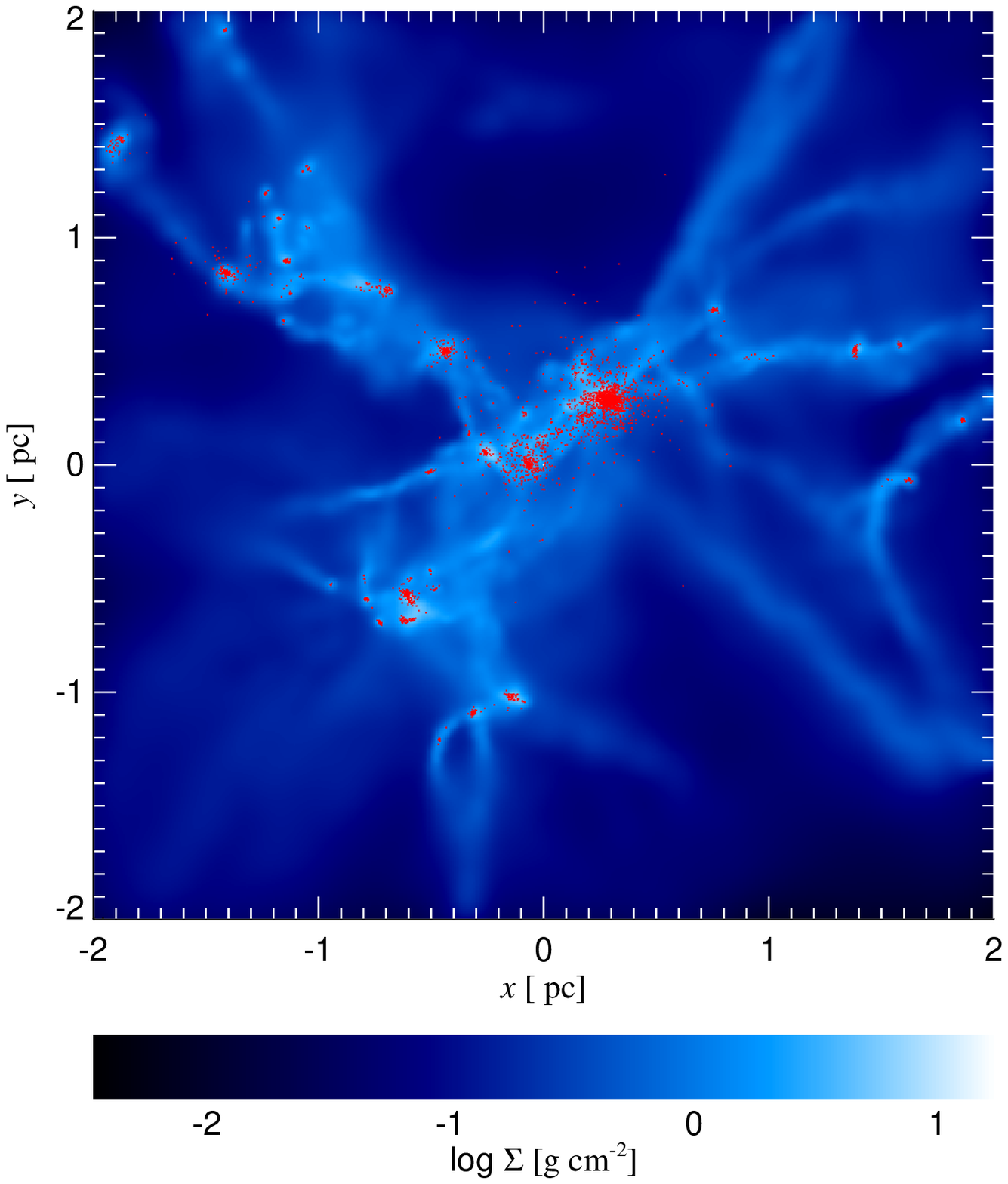}
  \caption{Evolution of the uncompressed cloud model, t4T5. {\bf Left
      panel:} $t = 1.26$~Myr, just before the first sink particles
    appear; the cloud initially expands slightly and develops an
    uneven density structure due to turbulence, with higher density in
    the centre and several high-density blobs and filaments. {\bf
      Middle panel:} $t = 1.78$~Myr, when $M_{\rm sink} =
    0.2\left(M_{\rm cl} + M_{\rm sink}\right)$; sink particles form
    predominantly in the centre of the cloud, where the dynamical time
    is shortest and densities are highest. {\bf Right panel:} zoom in
    to the centre of the cloud at $t = 1.78$~Myr; sink particles have
    formed in several clumps where turbulent motions created overdense
    regions. Note the density scale change in this panel compared with
    the previous two. In the last panel, only $10\%$ of sink particles
    are shown for clarity.}
  \label{fig:T5evol}
\end{figure*}

\begin{figure*}
  \centering
    \includegraphics[trim = 6mm 23mm 6mm 0, clip, width=0.33 \textwidth]{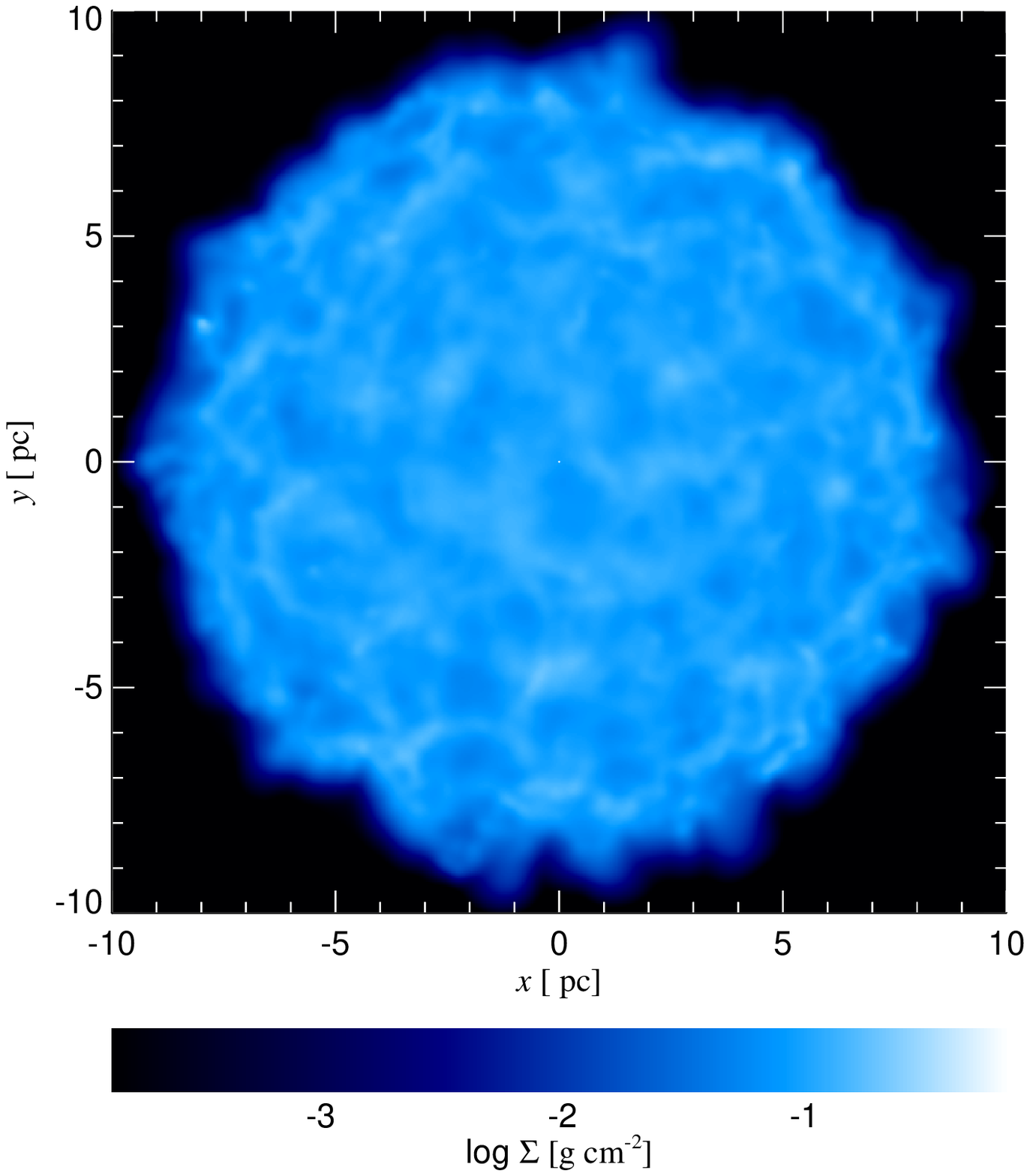}
    \includegraphics[trim = 6mm 23mm 6mm 0, clip, width=0.33 \textwidth]{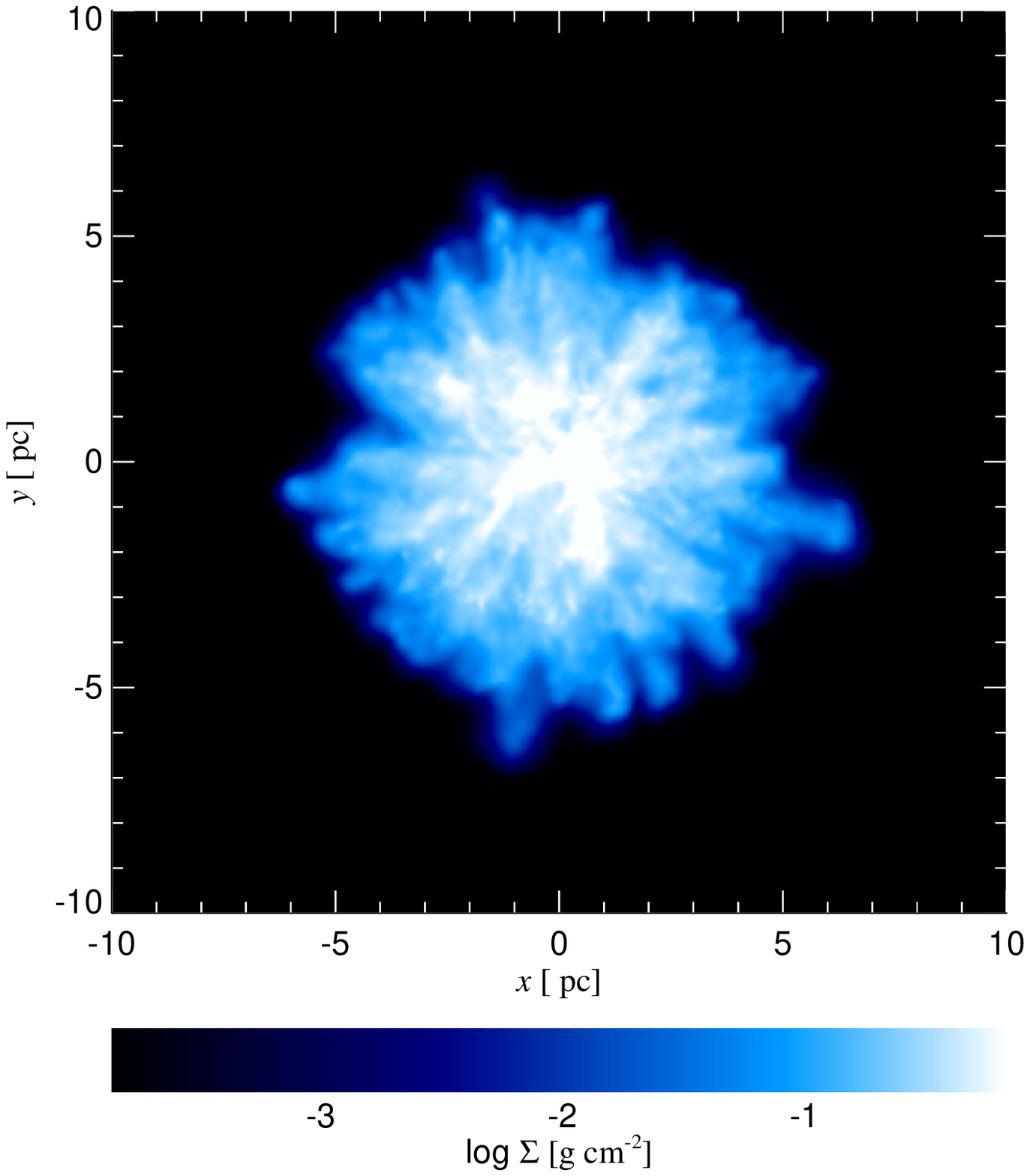}
    \includegraphics[trim = 6mm 23mm 6mm 0, clip, width=0.33 \textwidth]{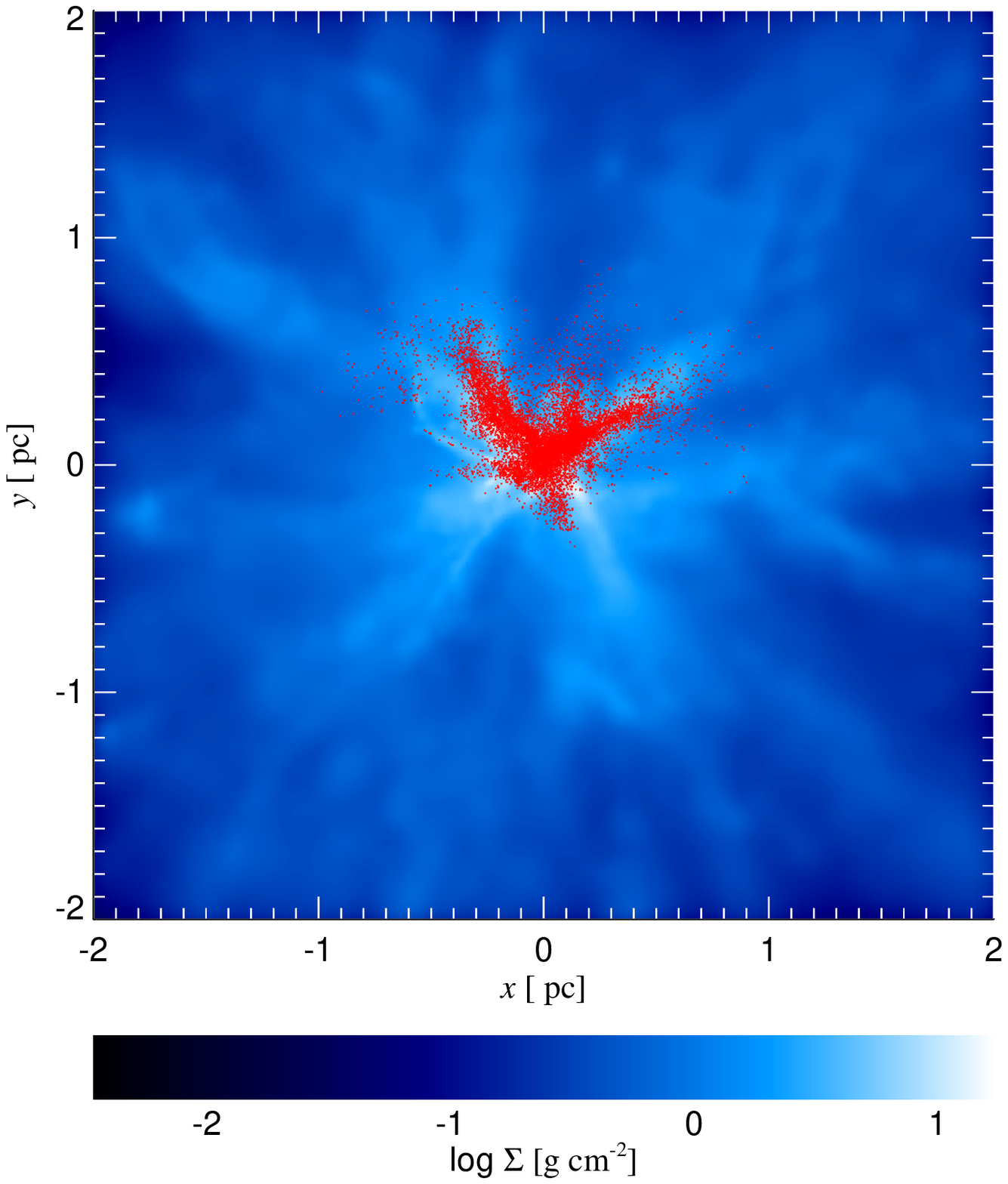}
  \caption{Evolution of the compressed cloud model, t4T7. {\bf Left
      panel:} $t = 0.14$~Myr; the cloud is confined and slowly
    compressed by the surrounding medium, developing an overdense
    shell at the interface. {\bf Middle panel:} $t = 0.37$~Myr, just
    before the first sink particles appear; the cloud radius is
    decreasing, but Richtmeyer-Meshkov instabilities destroy a
    coherent shockwave. {\bf Right panel:} $t = 0.43$~Myr, when
    $M_{\rm sink} = 0.2\left(M_{\rm cl} + M_{\rm sink}\right)$; sink
    particles are forming vigorously in a single clump in the centre
    of the cloud. Note the density and size scale change in this panel
    compared with the previous two. In the last panel, only $10\%$ of
    sink particles are shown for clarity.}
  \label{fig:T7evol}
\end{figure*}

\begin{figure*}
  \centering
    \includegraphics[trim = 3mm 0 5mm 0, clip, width=0.33 \textwidth]{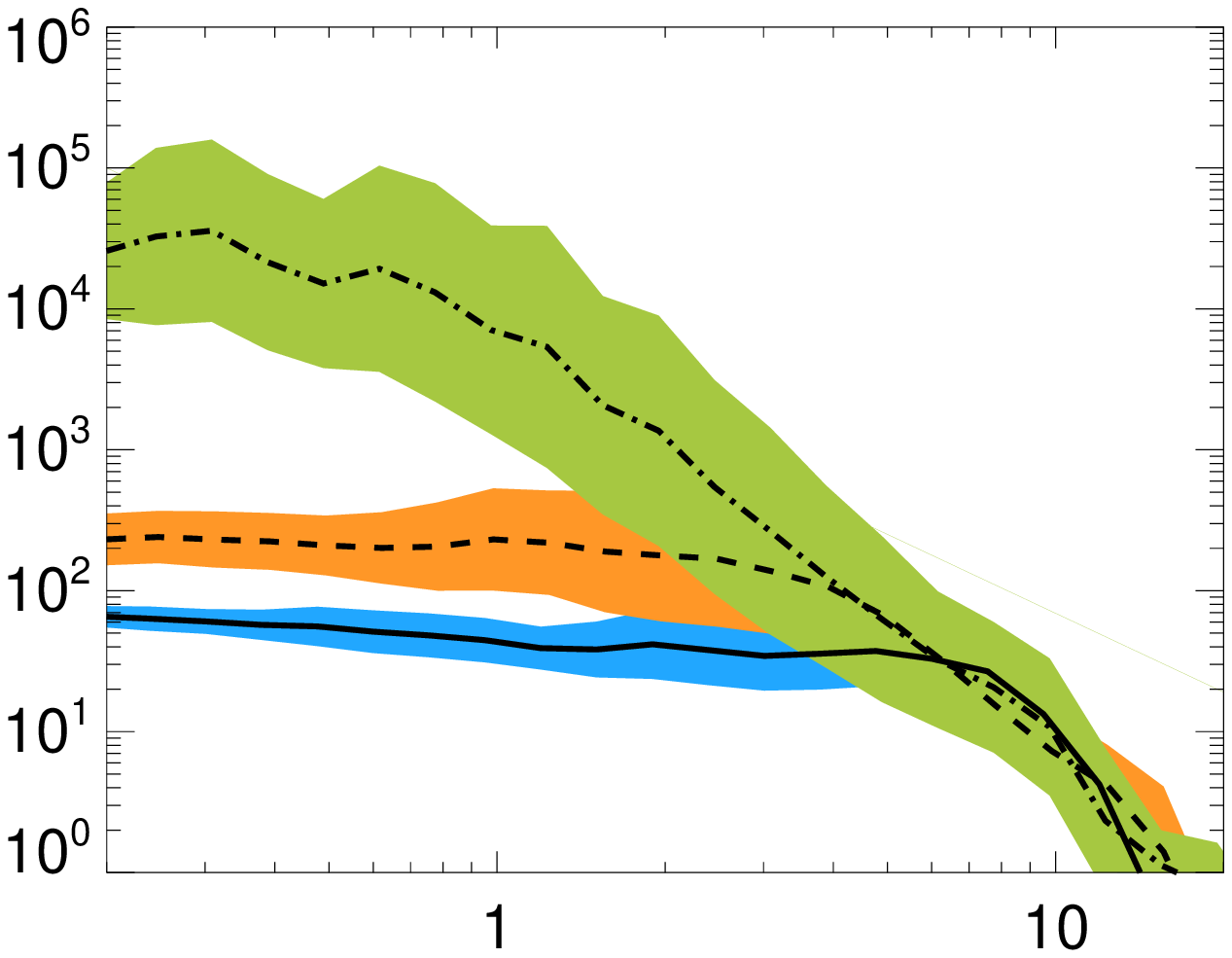}
    \includegraphics[trim = 3mm 0 5mm 0, clip, width=0.33 \textwidth]{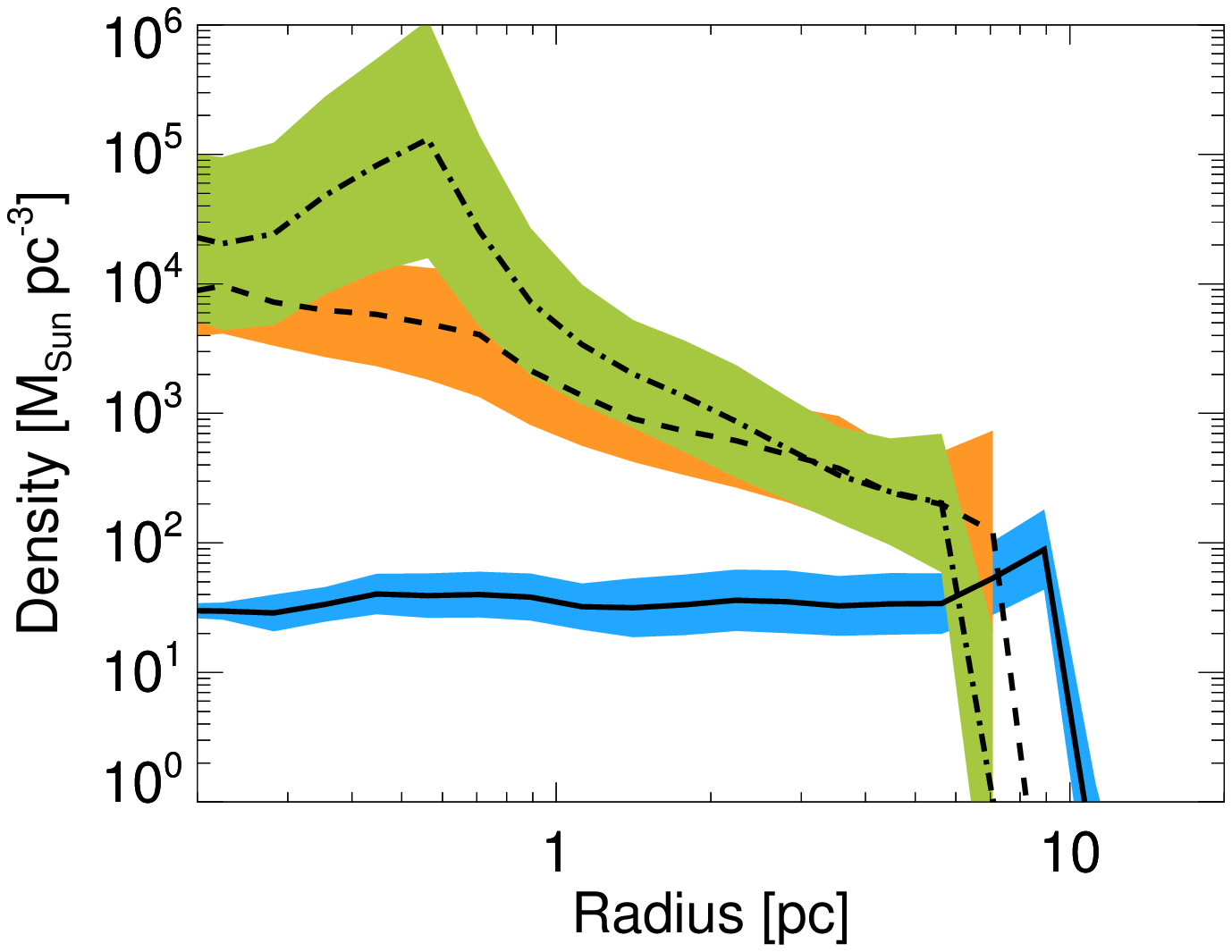}
    \includegraphics[trim = 3mm 0 5mm 0, clip, width=0.33 \textwidth]{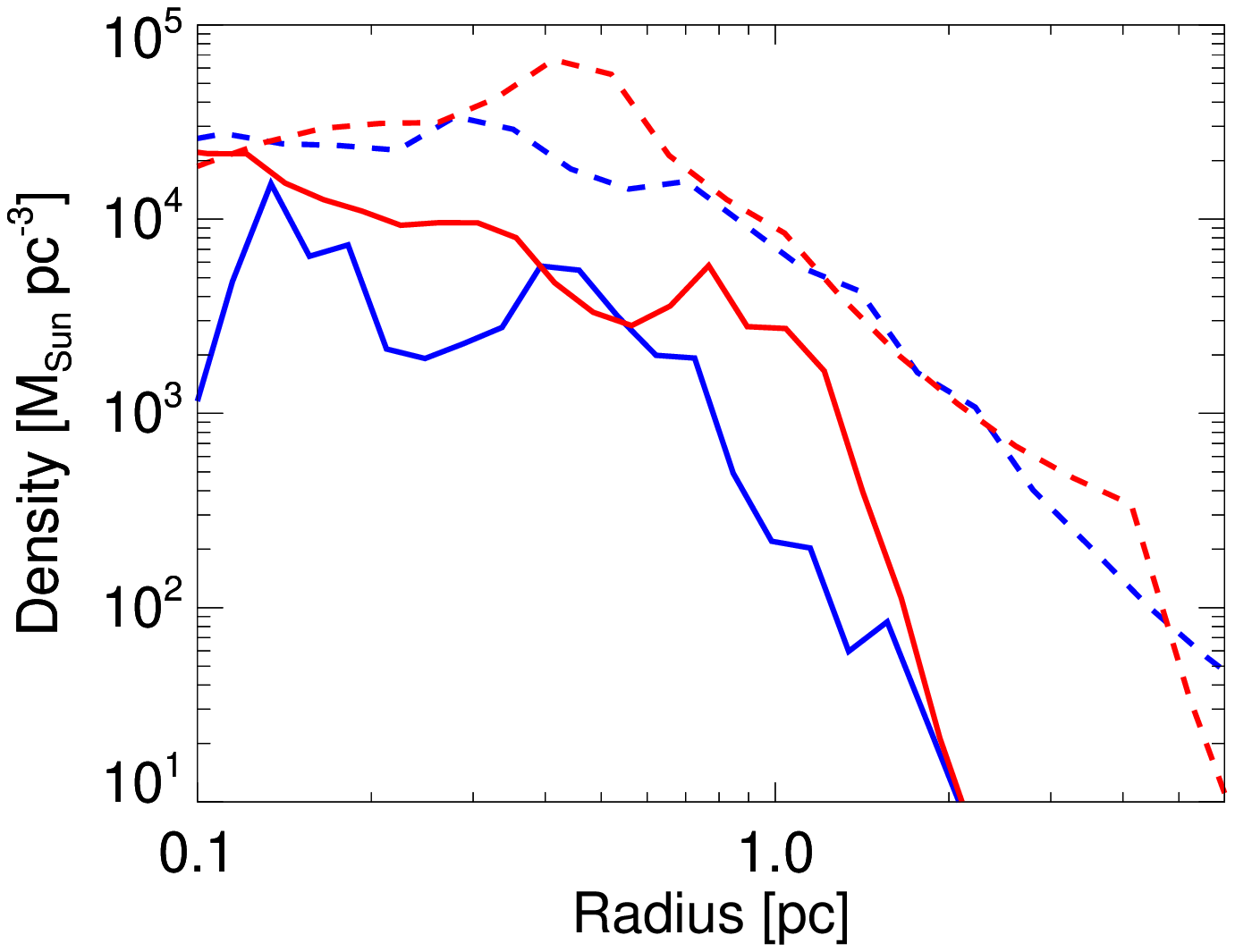}

  \caption{Radial density profiles. {\bf Left panel:} model t4T5 at
    $0.47$, $1.26$ and $1.78$~Myr (solid, dashed and dot-dashed lines,
    respectively; each line is created by averaging the values of
    three subsequent snapshots, in order to reduce numerical
    noise). Coloured regions indicate $\pm 1 \sigma$ deviation from
    the mean of log $\rho$. The cloud develops an approximately
    isothermal density profile, and its density exceeds the star
    formation threshold only in the very centre. {\bf Middle panel:}
    model t4T7 at $0.14$, $0.37$ and $0.43$~Myr. Line styles and
    colours as in previous panel. The cloud is compressed, with a weak
    shockwave (density ratio $\sim 2$) moving inward. The threshold
    density for star formation is still reached only in the
    centre. {\bf Right panel:} sink particle radial profiles (solid
    lines) and gas radial profiles (dashed lines) at $t = t_{\rm
      frag}$: blue lines indicate t4T5, red lines indicate t4T7. The
    gas density profiles are identical except for the presence of a
    shockwave in model t4T7, while the sink particles in t4T7 are
    distributed slightly more widely than in the uncompressed model.}
  \label{fig:radprof}
\end{figure*}

\begin{figure*}
  \centering
    \includegraphics[width=0.33 \textwidth]{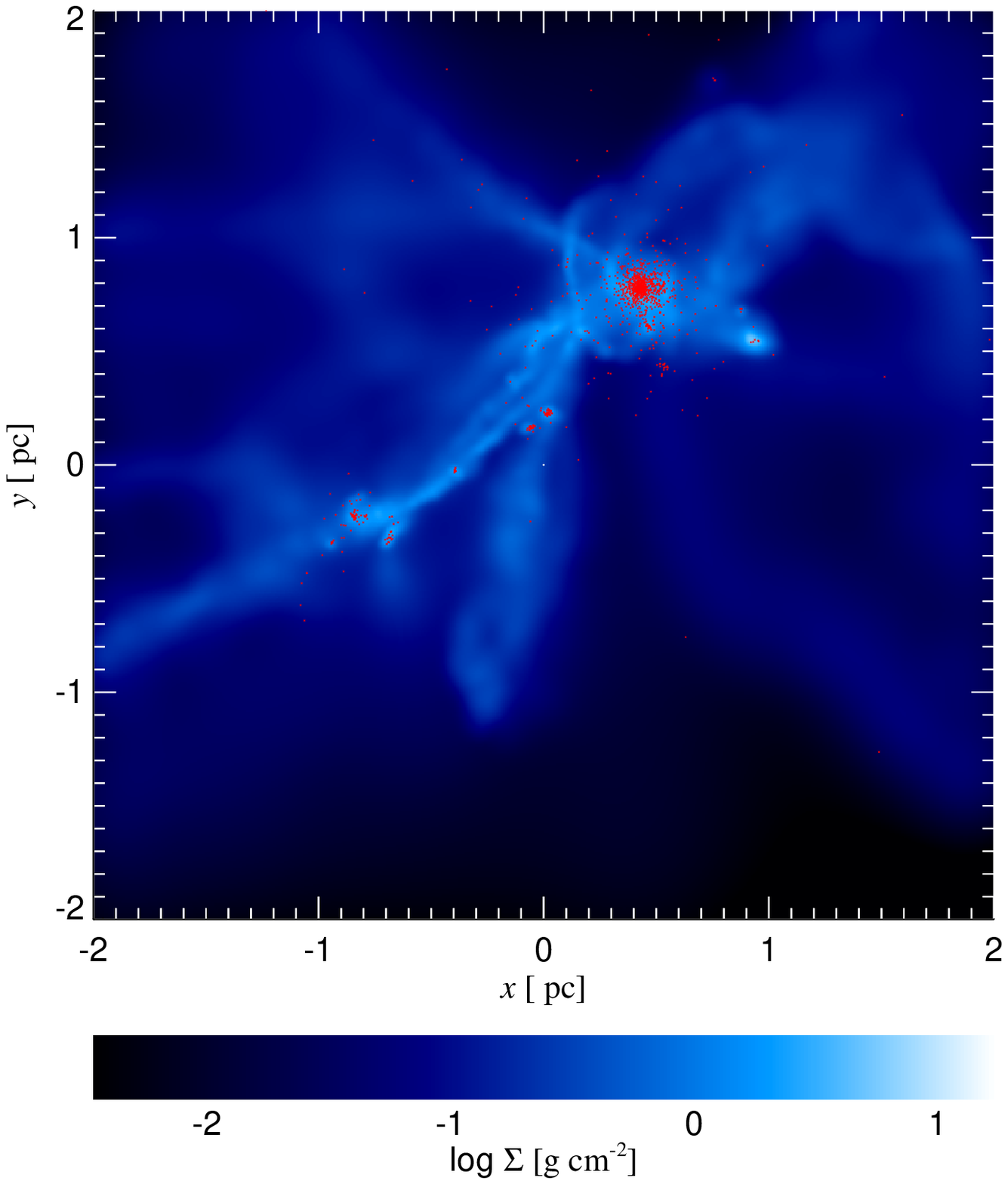}
    \includegraphics[width=0.33 \textwidth]{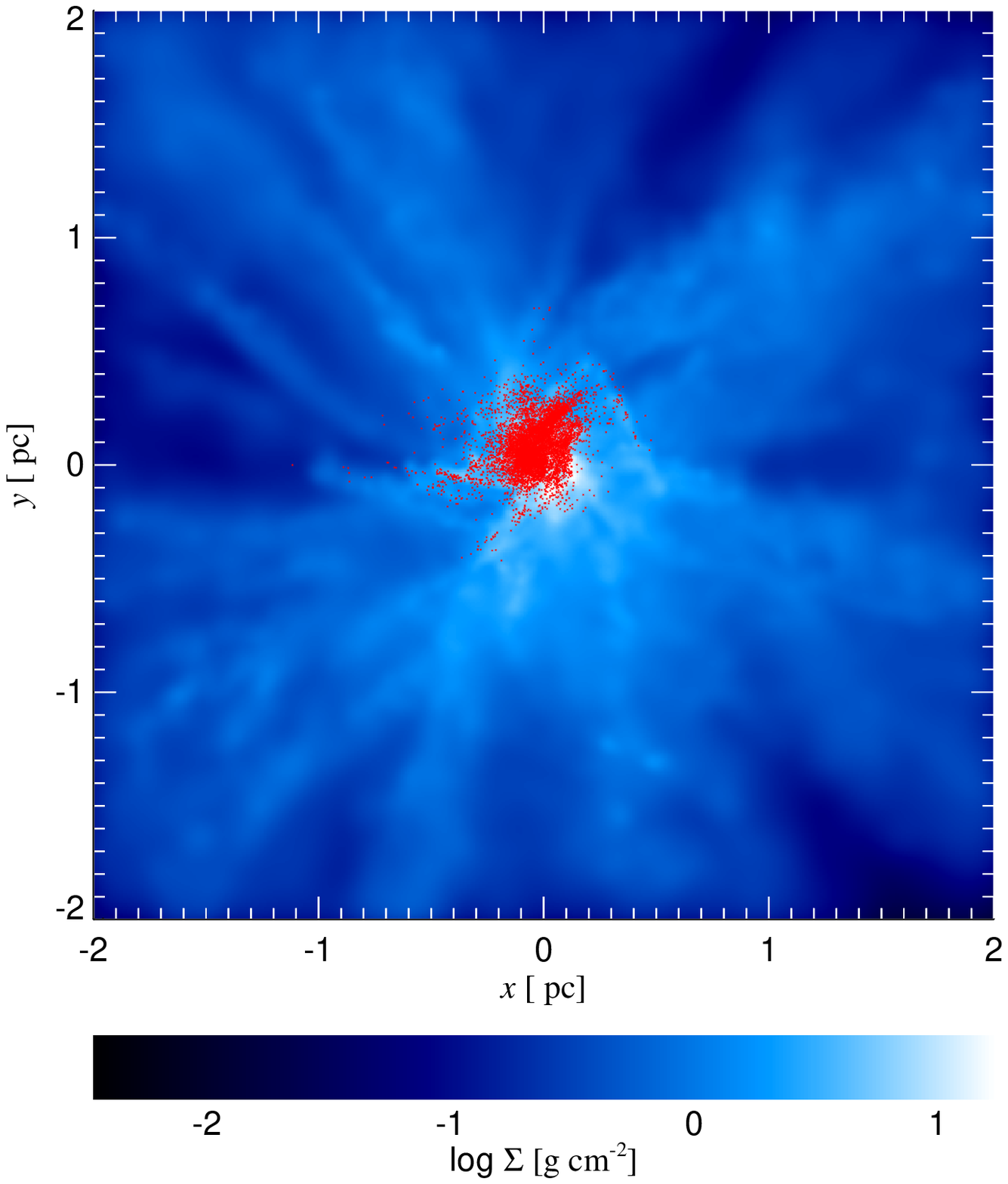}
  \caption{Morphology of the two high turbulence models, t10T5 (left
    panel) and t10T7 (right panel), when the sink particle mass
    fraction is $20\%$. This happens at $t = 2.51$~Myr for the
    uncompressed model and at $t = 0.56$~Myr for the compressed
    cloud. Even high turbulence is unable to prevent confinement and
    collapse of the cloud embedded in a high-pressure hot ISM. Only
    $10\%$ of sink particles are shown for clarity.}
  \label{fig:highturb}
\end{figure*}

We divide the result presentation into two parts. First we analyze the
effects of external pressure without shear, including cases of static
gravitationally bound clouds (models t4T5 and t4T7), static
gravitationally unbound clouds (models t10T5 and t10T7) and rotating
gravitationally bound clouds (t2.8T5r4.2 and t2.8T7r4.2). Next, we
consider the effects of progressively stronger shear upon
self-gravitating clouds (models t4vXT5 and t4vXT7).

For each model, we derive four parameters which allow for easy
quantitative comparison of their progress. The first parameter is the
time when the first sink particle forms, which we use as a proxy for
the onset of star formation. Secondly, we define the fragmentation
time, $t_{\rm frag}$, as the time when the sink particle mass fraction
reaches $20\%$; the choice of the particular mass fraction is
arbitrary, but choosing either $10\%$ or $30\%$ does not affect our
conclusions. The third parameter is the half-mass radius of the
system, $r_{\rm h}$, at $t = t_{\rm frag}$. Finally, we define the
efficiency of sink particle formation $\epsilon_{\rm ff,sink}$ as the
mass fraction of sink particles after one dynamical time of the cloud,
i.e. $1.7$~Myr. The numerical values of these parameters are given in
the last three columns of Table \ref{table:param}. The error due to
time resolution of the simulations is $\pm 0.02$~Myr, while fractional
errors on distances are $\pm 0.02$.

\subsection{Models with no shear} \label{sec:result_noshear}

\subsubsection{Triggering of fragmentation} \label{sec:fragtrig}

Figures \ref{fig:T5evol} and \ref{fig:T7evol} show the column density
plots which depict the evolution of the models t4T5 and t4T7,
respectively. The uncompressed model quickly develops an uneven
density structure and expands slightly, before starting to collapse as
the turbulence decays. The density increases mainly in the central
parts of the cloud, which develop an isothermal ($\rho \propto
R^{-2}$) density structure (left panel, also Figure \ref{fig:radprof},
left panel). Star formation begins in the central regions, where the
density is highest due to convergent turbulent flows. Sink particles
form along filaments and are gradually absorbed into a central
elliptical cluster (right panel). The fraction of mass converted into
sink particles in one dynamical time of the cloud is $\epsilon_{\rm
  ff, sink} \simeq 7\%$. Multiplying this value by the single-core
star formation efficiency of $25-75\%$
\citep{Matzner2000ApJ,Alves2007A&A} gives a star formation efficiency
$\epsilon_{\rm ff,*} \simeq 2-5\%$. The similarity of this result to
the observationally derived value \citep{McKee2007ARA&A} is partly
coincidental, and depends sensitively on the initial conditions,
especially the characteristic turbulent velocity of the cloud
\citep{Bate2003MNRAS}. Furthermore, the decay of turbulence in the
cloud is partly responsible for the high fragmentation rate (see
Section \ref{sec:turbdrive}).

The evolution of the compressed cloud is notably different in two
aspects. As expected, the hot ISM confines the cloud and compresses it
(Figure \ref{fig:T7evol}, left panel and Figure \ref{fig:radprof},
middle panel). However, there is no clearly visible shockwave moving
inward through the cloud. This happens because Richtmeier-Meshkov (RM)
instabilities begin growing along the interface between the cloud and
the ISM on a timescale comparable to the cloud crushing timescale
\citep{Klein1994ApJ}. These instabilities manifest as thick fingers
visible in the middle panel of Figure \ref{fig:T7evol}. The first sink
particles appear at $t \sim 0.37$~Myr. Much like in the uncompressed
model, they form in the centre of the cloud, where growing
instabilities increase mixing rate and promote the formation of
high-density clumps. Unlike the uncompressed model, however, the sink
particles are more likely to escape their parent clumps, so the mean
mass of sink particles is lower, and all sink particles form a single
large cluster, rather than keeping a complex substructure. All of the
cloud gas is converted into sink particles well before $1.7$~Myr,
giving a formal fragmentation efficiency $\epsilon_{\rm ff, sink} >
100\%$.

We compare the radial density profiles of gas (dashed lines) and sink
particles (solid lines) in the two models at $t = t_{\rm frag}$ in the
right panel of Figure \ref{fig:radprof}. The gas density profiles are
almost identical in the two models, except for a weak shockwave
(density ratio $\sim 2$) in model t4T7 (red line) at $0.3$~pc $< R <
0.7$~pc. Both gas density profiles are slightly steeper than
isothermal, $\rho \propto R^{-2.5}$, outside $R \simeq 0.7$~pc. They
should flatten over time. The sink particles are slightly more
centraly concentrated than the gas in both models; furthermore, the
uncompressed cloud has a smaller ``core'' of sink particles, going out
to $\sim 0.5$~pc, as opposed to the $\sim 1$~pc-wide core in the
compressed model. The outer slopes of sink particle density profiles
are very steep in both models, d ln$\rho /$ d ln$R \simeq -4$ --$-5$,
which should later expand and relax to a shallower distribution.

The morphology of high turbulence models, t10T5 and t10T7, is
presented in Figure \ref{fig:highturb}. The unconfined cloud expands
significantly, with some filaments reaching radii of
$>30$~pc. Meanwhile, the density in the central parts keeps increasing
and the first sink particles appear at $1.51$~Myr, not much later than
in the low turbulence model (t4T5, $1.26$ Myr). Later on, however, the
fragmentation rate stays much lower than in t4T5 (see Section
\ref{sec:dyntime}). As a result, the fragmentation efficiency is only
$\epsilon_{\rm ff,sink} \simeq 1.6\%$. This is much lower than the
$5-10\%$ found by \citet{Clark2005MNRAS}, presumably because those
authors considered a cloud with $E_{\rm turb}/E_{\rm grav} = 2$,
whereas in our model, the ratio is $7.7$. The sink particle cluster
that eventually forms in the central parts (Figure \ref{fig:highturb},
left panel) is not a realistic result due to the decay of turbulence
(see Section \ref{sec:turbdrive}) and lack of stellar feedback (see
Section \ref{sec:clusters}), which would presumably destroy the cloud.

The confined highly turbulent cloud (Figure \ref{fig:highturb}, right
panel) evolves in a very similar way to its low turbulence
analogue. The cloud expands very little at first, but is quickly
confined and compressed; even though the shockwave is weak, RM
instabilities form behind it and help increase gas density, which
leads to formation of the first sink particles at $t \simeq 0.40$~Myr,
a very similar time to model t4T7. Subsequently, both the
fragmentation rate and fragment mass fraction increase
exponentionally, with $20\%$ sink particle mass fraction achieved by
$t_{\rm frag} = 0.56$~Myr (compare with $t_{\rm frag} = 0.43$~Myr of
t4T7). The half-mass radius of the cloud is slightly larger than in
the model t4T7, but this difference does not affect the global
evolution: just as in t4T7, all of the cloud gas turns into sink
particles within one dynamical time.

These results show that the surrounding hot ISM pressure can have a
dominant effect on molecular cloud evolution, provided it is larger
than the dynamical pressure of the cloud. The affected cloud is
confined, preventing gas dispersal (and reducing its tidal radius; see
Section \ref{sec:clusters}); a shockwave is driven into the cloud,
followed by RM instabilities, which accelerate and trigger
fragmentation.

\subsubsection{Fragmentation of rotating clouds} \label{sec:rotation}

\begin{figure*}
  \centering
    \includegraphics[trim = 6mm 23mm 6mm 0, clip, width=0.23 \textwidth]{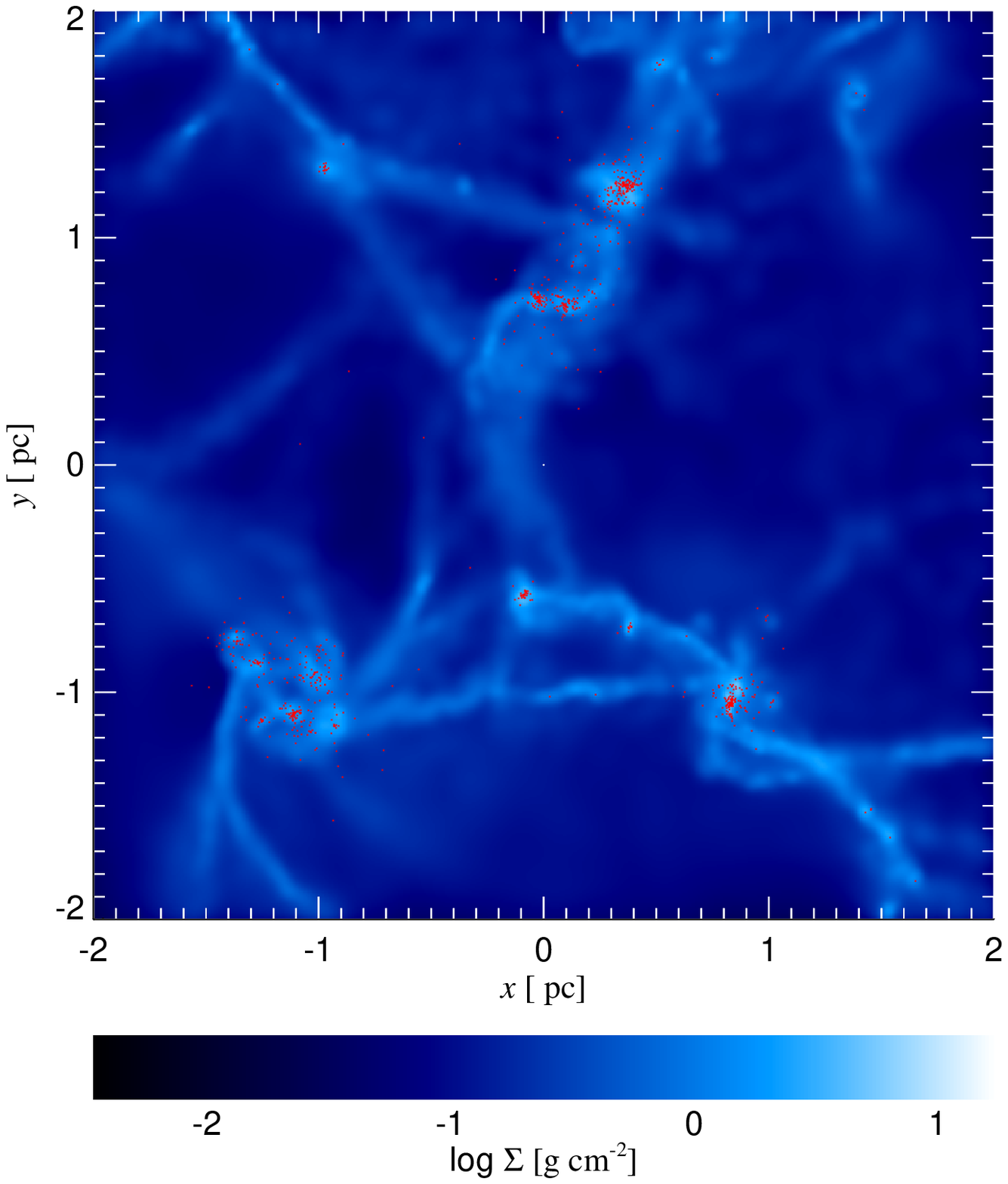}
    \includegraphics[trim = 6mm 23mm 6mm 0, clip, width=0.23 \textwidth]{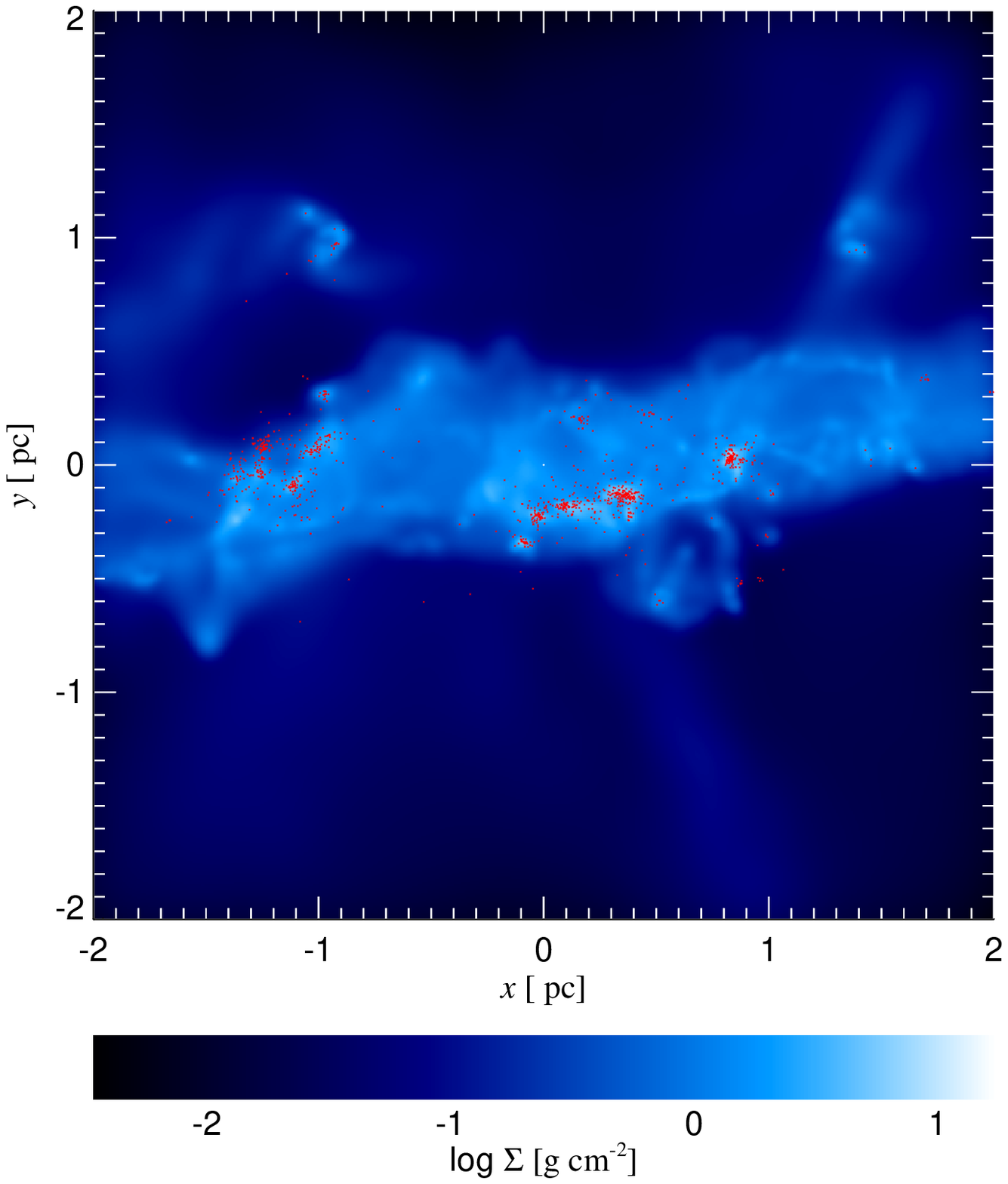}
    \includegraphics[trim = 6mm 23mm 6mm 0, clip, width=0.23 \textwidth]{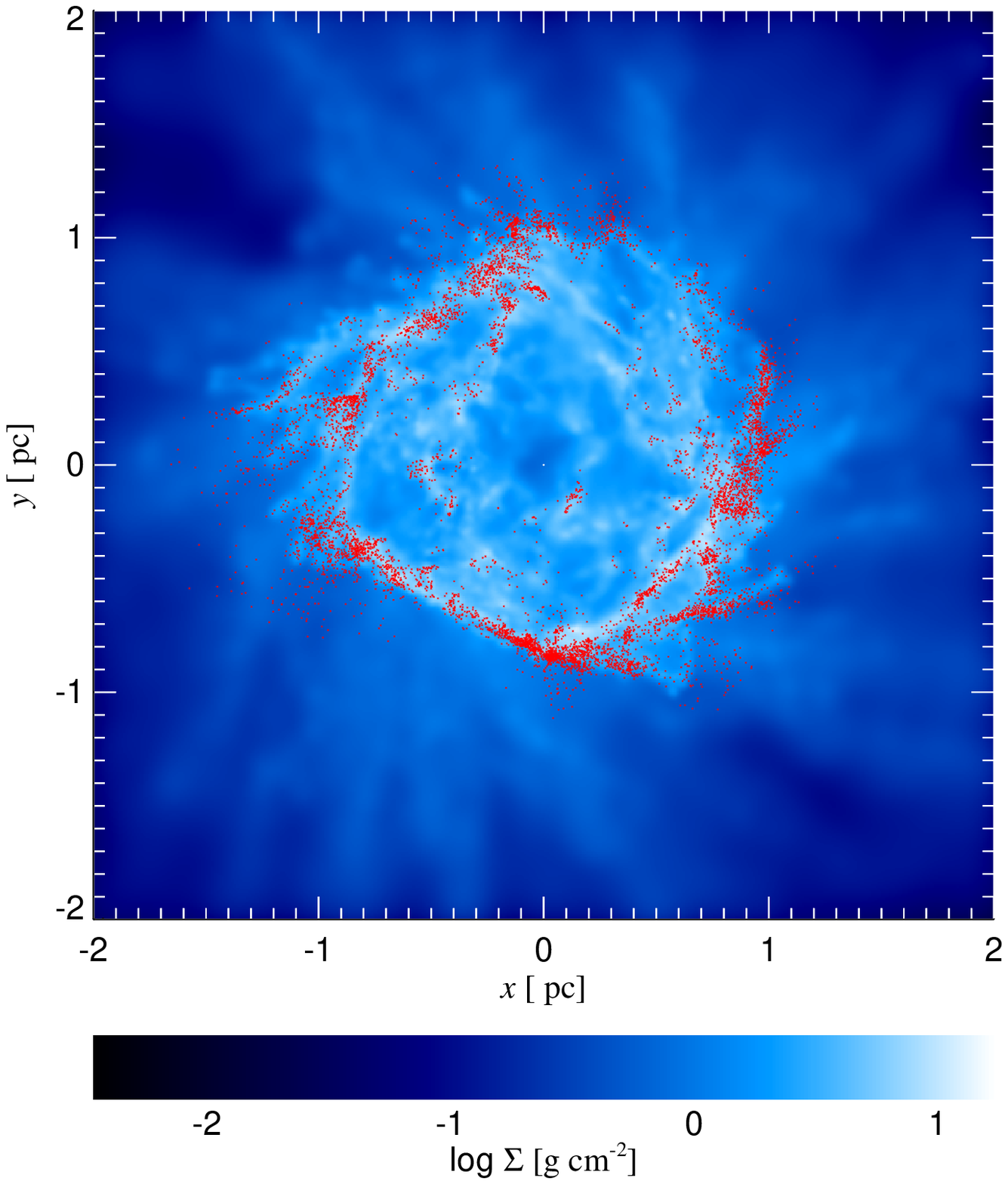}
    \includegraphics[trim = 6mm 23mm 6mm 0, clip, width=0.23 \textwidth]{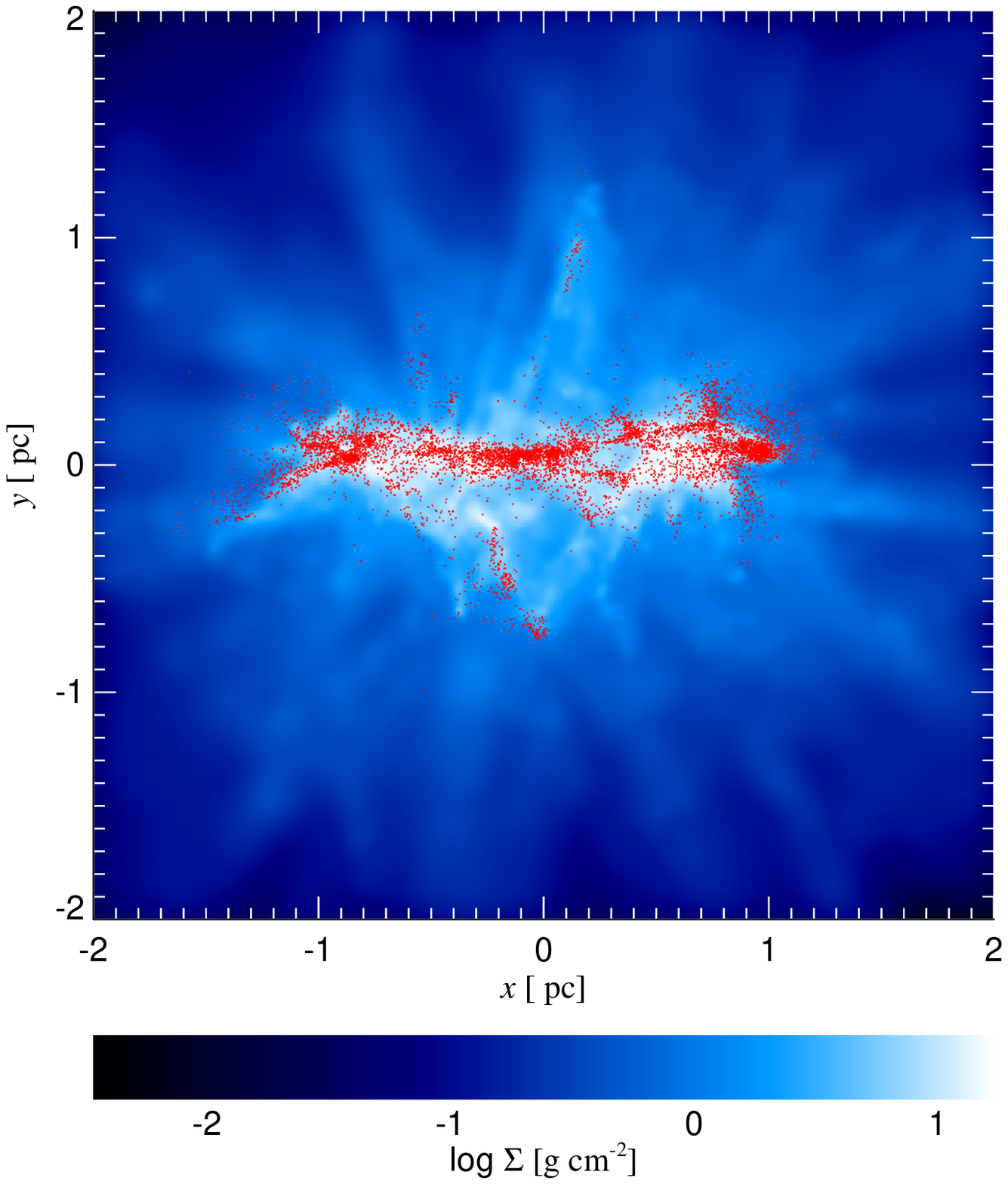}

  \caption{Morphology of rotating cloud models, t2.8r4.2T5 at $t =
    2.03$~Myr (two left panels) and t2.8r4.2T7 at $0.42$~Myr (two
    right panels). First and third panels show the top-down view of
    the XY plane, second and fourth panels show the side view of the
    XZ plane. Rotation has a noticeable effect on the morphology of
    both uncompressed and compressed clouds, distributing the gas in a
    larger volume and preventing instability growth. Only $10\%$ of
    sink particles are shown for clarity.}
  \label{fig:rotating}
\end{figure*}

\begin{figure*}
  \centering
    \includegraphics[width=0.33 \textwidth]{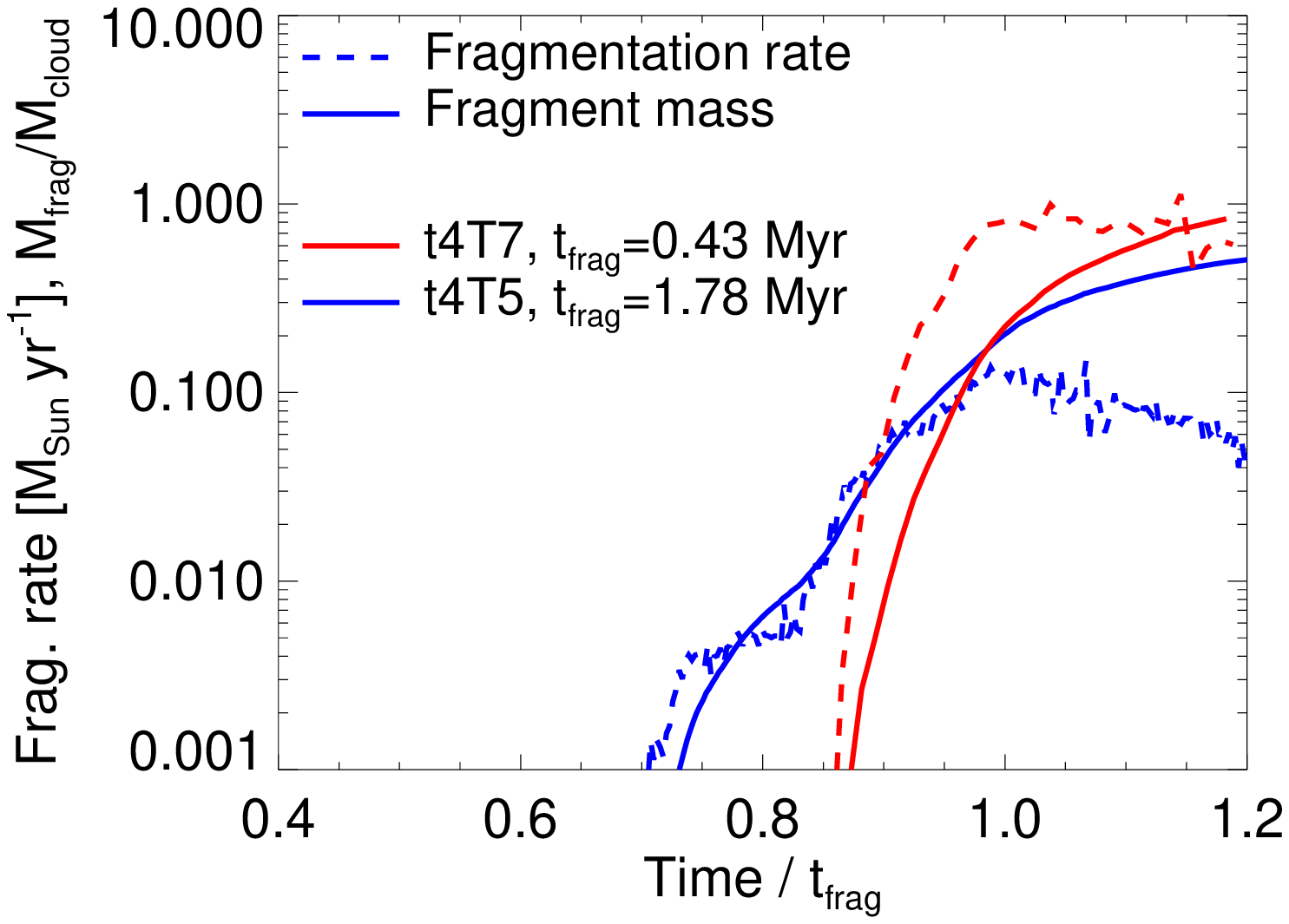}
    \includegraphics[width=0.33 \textwidth]{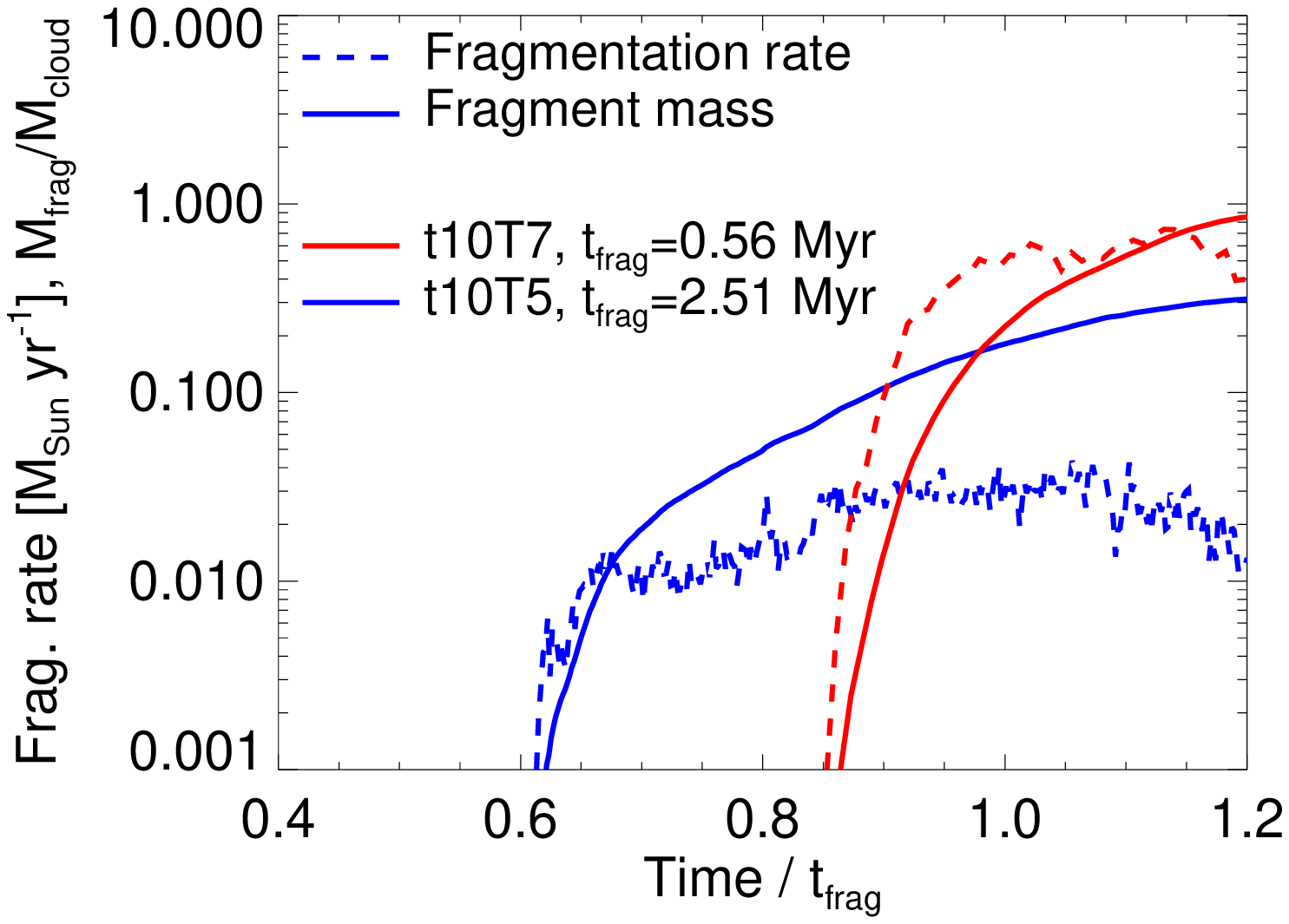}
    \includegraphics[width=0.33 \textwidth]{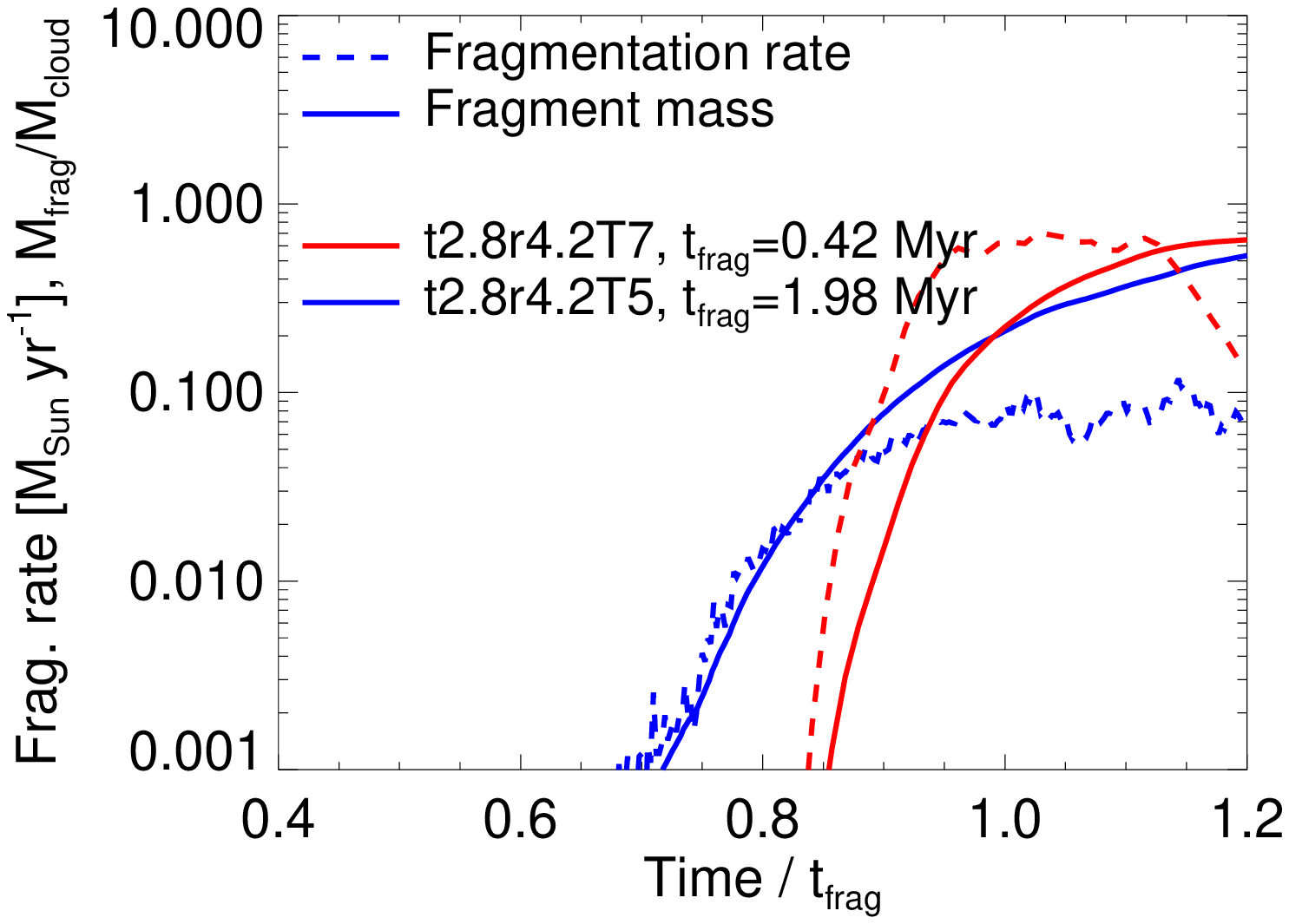}
  \caption{Fragmentation rates and growth of the sink particle
    population for the low turbulence (left panel), high turbulence
    (middle panel) and rotating (right panel) models. The time axis is
    scaled to the fragmentation timescale $t_{\rm frag}$ (see Table
    \ref{table:param}). In all three cases, compressed clouds show
    qualitatively faster fragmentation, with a much shorter time
    between first sink particles appearing and $t_{\rm frag}$ and a
    $10-20$ times larger fragmentation rate. The compressed models
    evolve similarly independently of characteristic turbulent
    velocity or rotation.}
  \label{fig:frag}
\end{figure*}

We run two simulations, t2.8r4.2T5 and t2.8r4.2T7 (see Table
\ref{table:param}), to investigate the effects of cloud rotation. The
cloud is set up to have solid-body rotation around the Z axis with an
angular velocity of $\omega_{\rm rot} =0.42$~km~s$^{-1}$~pc$^{-1}$.
The characteristic turbulent velocity is reduced to
$2.8$~km~s$^{-1}$. This ensures that the contributions to dynamical
pressure from rotation and turbulence are approximately the same in
the XY plane and that the total kinetic energy of the cloud is the
same as in t4 models.

Even though the cloud rotation is slow, with $\omega_{\rm rot} t_{\rm
  dyn} \simeq 0.7$, the uncompressed cloud evolves rather differently
from the non-rotating case. The first two panels of Figure
\ref{fig:rotating} show the cloud morphology at $t = t_{\rm frag} =
1.98$~Myr. The cloud collapses vertically, forming a disc partially
supported by rotation, which then fragments and starts producing sink
particles in small clusters throughout the disc. The mean density of
the gas disc is lower than of a spherically collapsing cloud, so the
sink particles begin appearing slightly later, after $\sim
1.33$~Myr. The fragmentation efficiency is accordingly lower,
$\epsilon_{\rm ff,sink} = 3.6\%$.

Since the compressed rotating model evolves much faster, rotation does
not have time to break the cloud into separate clumps. However,
rotation has another important effect; the RM fingers that
significantly affect the cloud evolution in the non-rotating models
are sheared away, and the central parts of the cloud are effectively
shielded from the shockwave. As a result, the highest densities are
achieved at the edge of the cloud; this is also where most sink
particles form (third panel of Figure \ref{fig:rotating}). In the
vertical direction, the cloud is strongly compressed (fourth panel of
Figure \ref{fig:rotating}), but the shockwave does not produce large
enough densities for rapid fragmentation. It is interesting that the
global parameters of cloud fragmentation are very similar to those of
the non-rotating compressed model. The first sink particles appear
after $\sim 0.34$~Myr and $t_{\rm frag} = 0.42$, leading to a large
fragmentation efficiency $\epsilon_{\rm ff,sink} = 75\%$; all these
numbers are very similar to those of model t4T7.

\subsubsection{Reduced effective dynamical time} \label{sec:dyntime}

Figure \ref{fig:frag} shows the mass fraction of sink particles (solid
lines) and fragmentation rate in Solar masses per year (dashed lines)
for the six models described above (left panel - t4, middle panel -
t10, right panel - t2.8r4.2 blue lines - T5, red lines - T7). The
horizontal axis is scaled to the fragmentation timescale $t_{\rm
  frag}$ (see Table \ref{table:param} and the beginning of this
Section).

We see immediately that the time evolution of compressed and
uncompressed models is qualitatively different. Model t4T7 (left
panel, red lines) forms stars $\sim 10$ times more rapidly than model
t4T5, so even when scaled to the fragmentation timescale $t_{\rm
  frag}$, the growth of sink mass fraction is much faster in the
compressed cloud. Similar differences appear in the high-turbulence
and rotating models.

It is interesting to compare the ratio of certain timescales for both
models with analytical predictions. The analytical prediction for the
effective dynamical time (Section \ref{sec:zerovel} and
eq. \ref{eq:tdyndash}) is $t_{\rm dyn} / t'_{\rm dyn} \simeq
\left(P_{\rm ISM} + P_{\rm grav} / P_{\rm grav}\right)^{1/2} \simeq
2.9$. The ratio of the times for the first sinks to form, $1.26/0.37
\simeq 3.4$, is similar, but somewhat larger. The discrepancy is even
greater when we consider the fragmentation timescales: $t_{\rm frag,
  t4T5} / t_{\rm frag, t4T7} = 1.78/0.43 \simeq 4.1$. We see that the
compressed cloud evolves progressively faster than the analytical
estimate predicts. This accelerated fragmentation cannot be the result
of increased mean cloud density, because both models have the same
density profiles at $t_{\rm frag}$ (Figure \ref{fig:radprof}, right
panel). A possible explanation is the action of RM instabilities,
which create denser and more diffuse regions while keeping the radial
density profile the same. This facilitates the fragmentation of gas
into sink particles. We tested this hypothesis by running these two
simulations with the standard SPH formalism, where artificial surface
tension suppresses instability growth, and found that the compressed
cloud there evolves consistently with the analytical prediction.

The difference in evolution between the two high turbulence models is
even larger. As expected, the uncompressed high-turbulence cloud,
model t10T5, fragments very slowly, reaching only $\dot{M}_{\rm frag}
\sim 0.04 \; \msun$~yr$^{-1}$ and maintaining this value for a long
time, as turbulence decays and material gradually accumulates back in
the centre of the cloud. The compressed cloud, however, evolves almost
identically to its low-turbulence counterpart, exhibiting similar peak
fragmentation rate and the ratio of $t_{\rm frag}$ to the time of the
formation of the first sinks.

Both rotating models evolve in a similar way to the t4 models.
Rotation slows down fragmentation in the uncompressed model, but
otherwise the time evolution is almost identical. This is striking
when considering how different the morphologies of the rotating and
non-rotating compressed models are. Such similarity suggests that RM
instabilities only increase the stratification of densities in the
cloud, rather than compacting the cloud as a whole. In the rotating
model, instabilities are confined to the outskirts of the cloud, but
their effects still manifest.

Overall, the time evolution reveals that external compression enhances
molecular cloud fragmentation in two ways. First of all, it pushes the
cloud together, reducing the effective dynamical time. In addition,
the shockwave generated by the high pressure ISM facilitates the
formation of instabilities, which stratify gas density in some
regions, accelerating fragmentation. Finally, if the cloud is confined
by external pressure, the characteristic turbulent velocity has very
little effect on the cloud evolution.

\subsection{Cloud with shearing motion} \label{sec:results_shear}

\begin{figure*}
  \centering
    \includegraphics[trim = 6mm 25mm 6mm 0, clip, width=0.32 \textwidth]{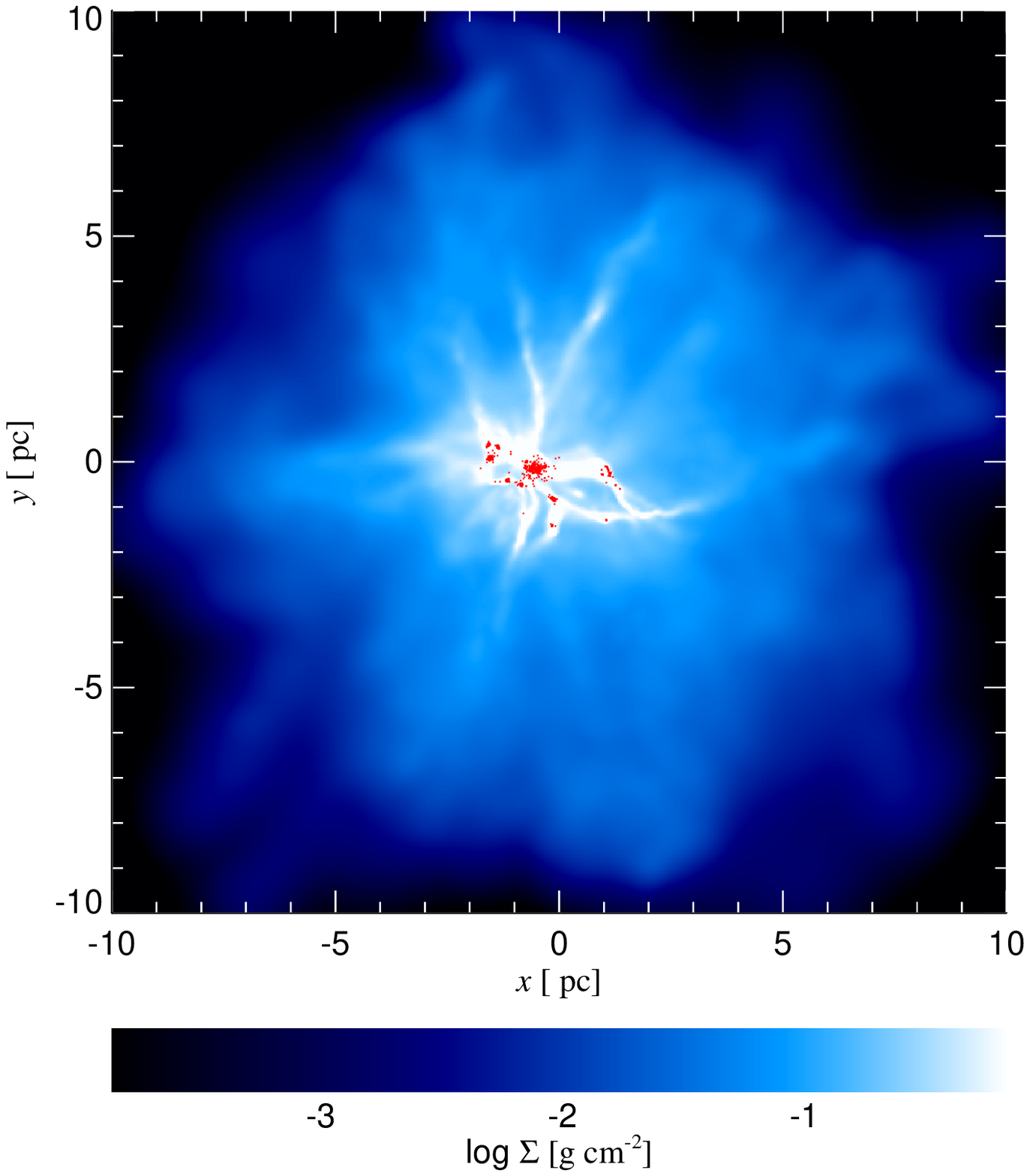}
    \includegraphics[trim = 6mm 25mm 6mm 0, clip, width=0.32 \textwidth]{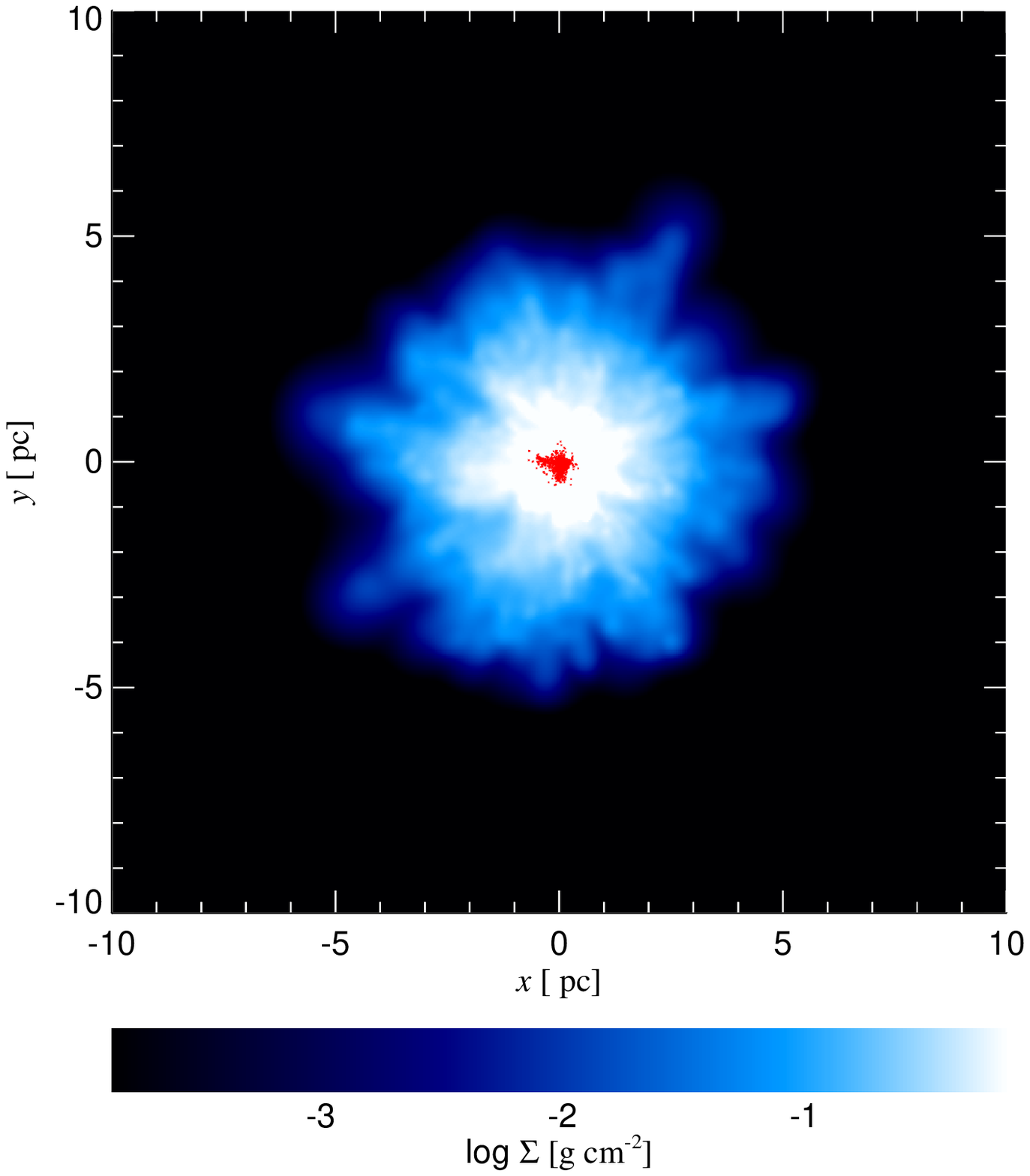}
    \includegraphics[width=0.34 \textwidth]{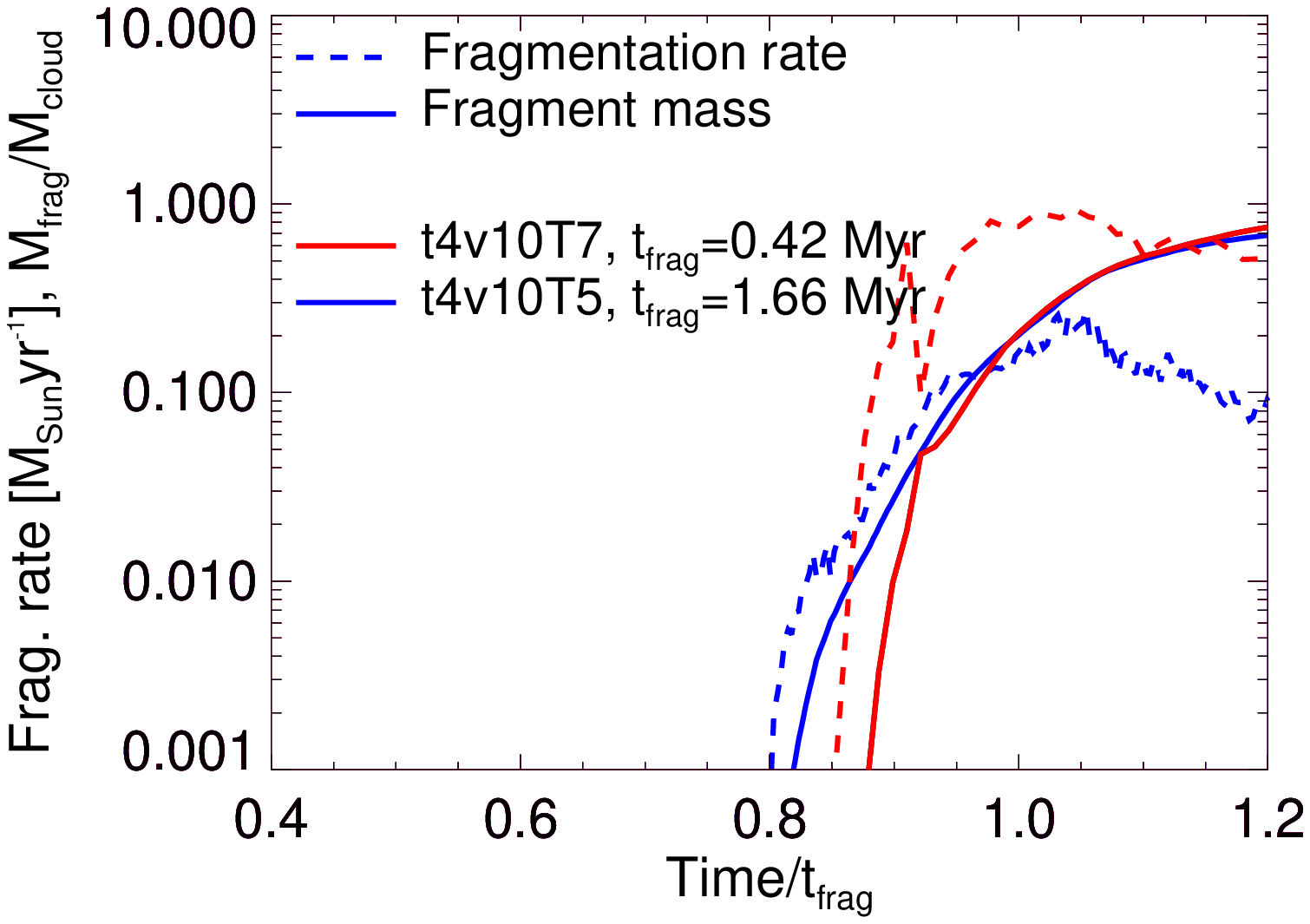}
    \includegraphics[trim = 6mm 25mm 6mm 0, clip, width=0.32 \textwidth]{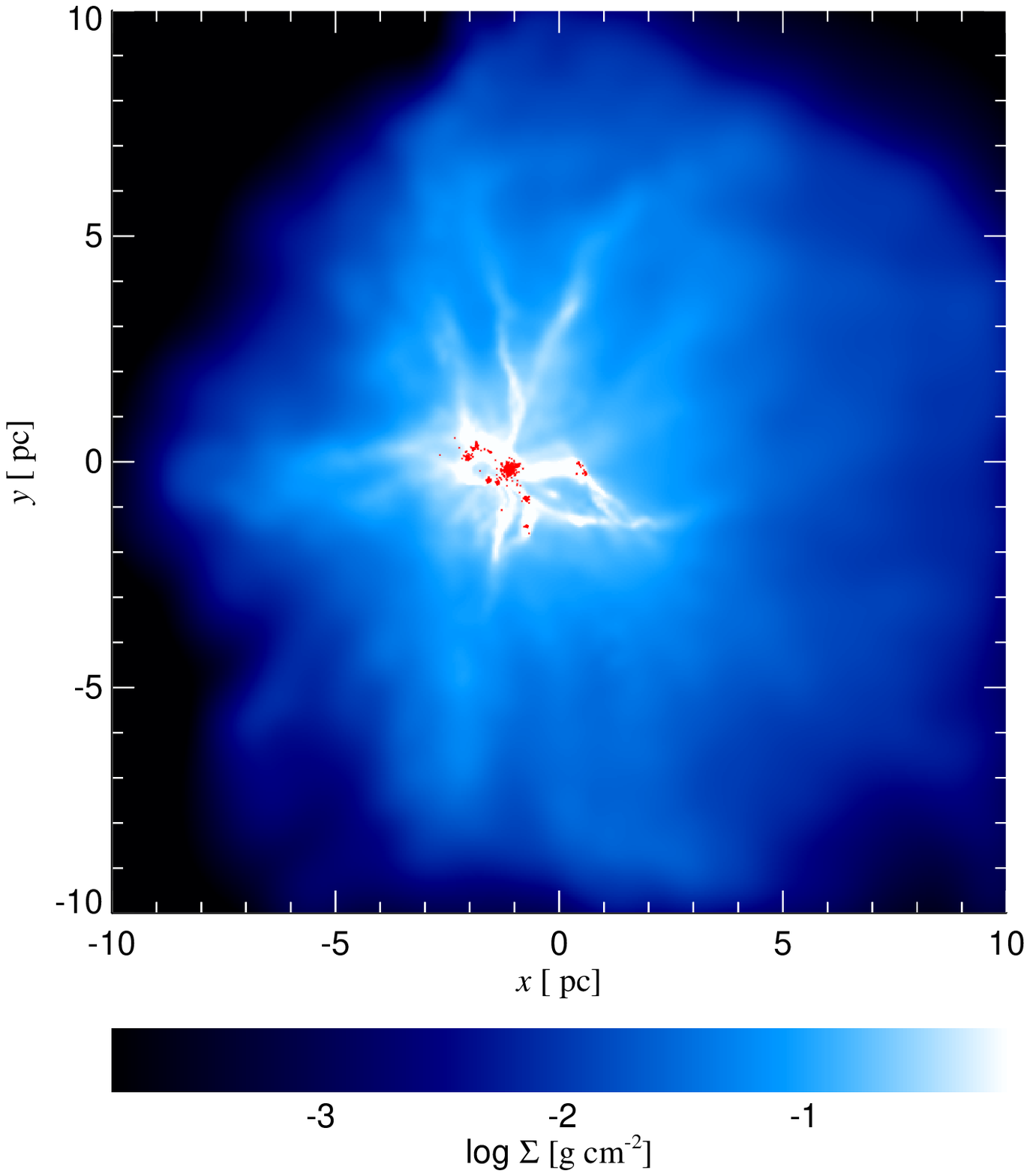}
    \includegraphics[trim = 6mm 25mm 6mm 0, clip, width=0.32 \textwidth]{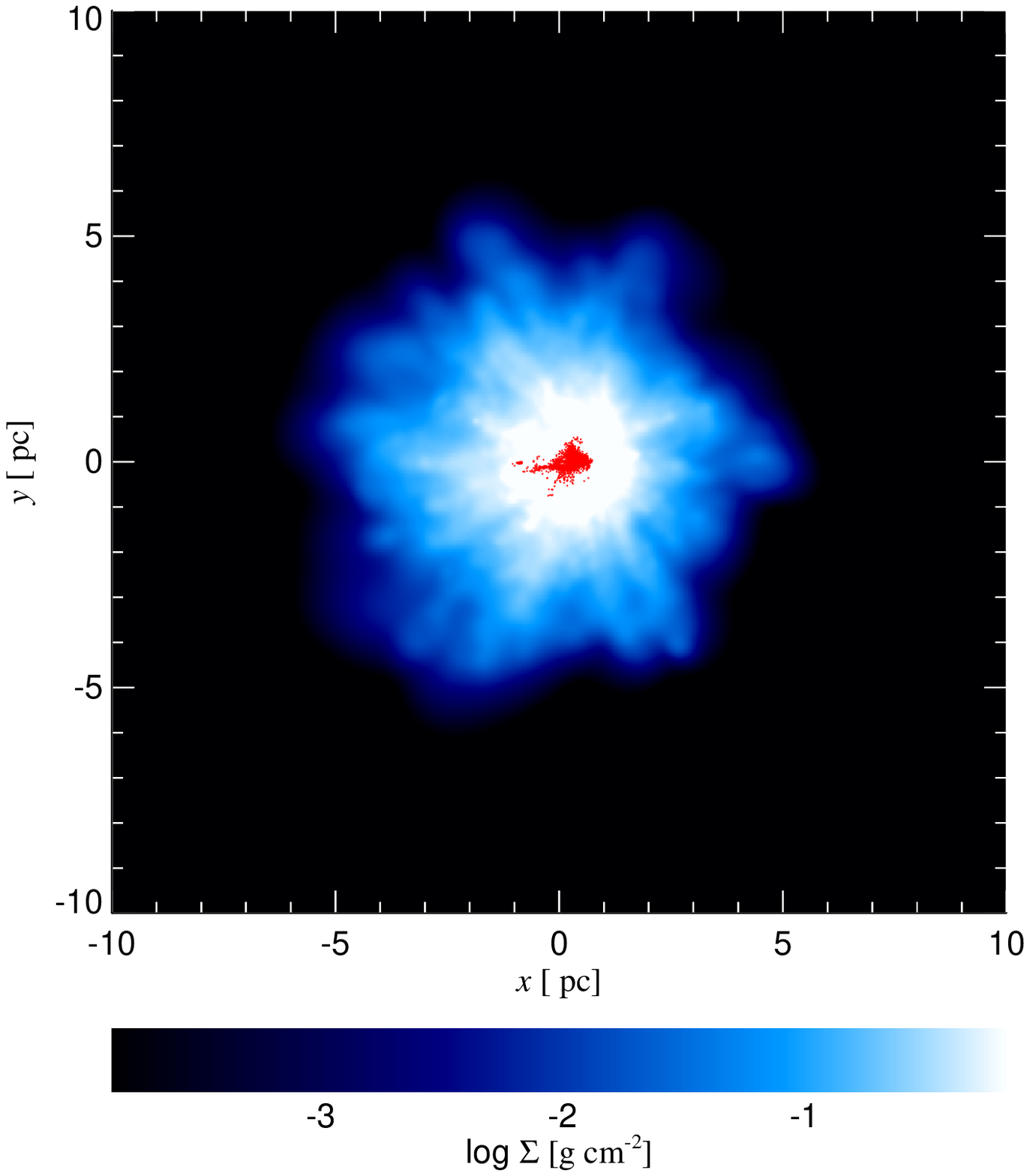}
    \includegraphics[width=0.34 \textwidth]{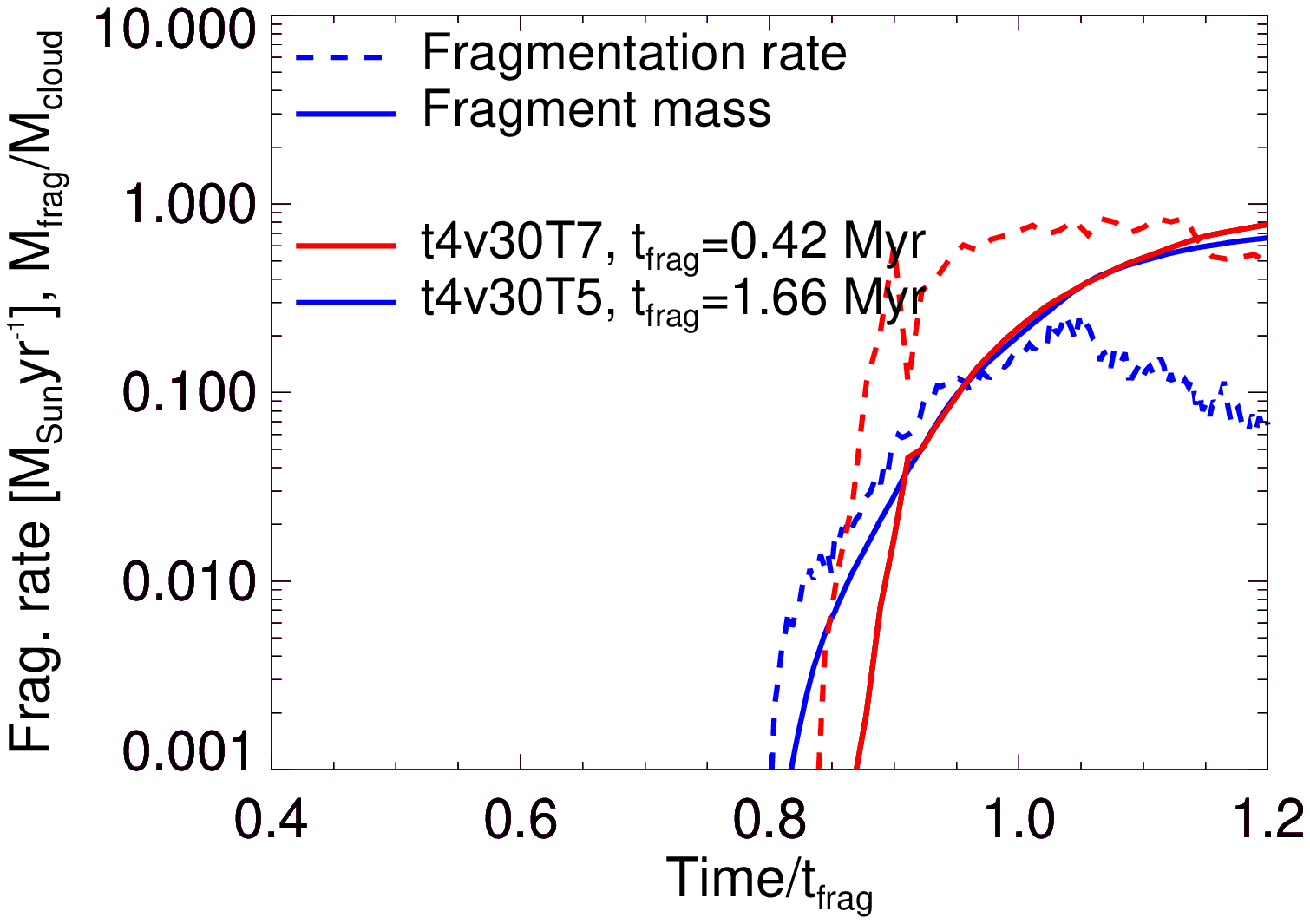}
  \caption{Morphology and time evolution of moving cloud models with
    lateral velocities of 10 km/s (top row) and 30 km/s (bottom
    row). Left column shows models with $T_{\rm ISM} = 10^5$~K,
    i.e. uncompressed clouds, middle column show clouds compressed
    with $T_{\rm ISM} = 10^7$~K (both at $t = t_{\rm frag}$, right
    column shows the time evolution scaled to the fragmentation
    time. The ISM moves to the right in all left- and middle-column
    plots.}
  \label{fig:lateral}
\end{figure*}

\begin{figure*}
  \centering
    \includegraphics[trim = 6mm 25mm 6mm 0, clip, width=0.32 \textwidth]{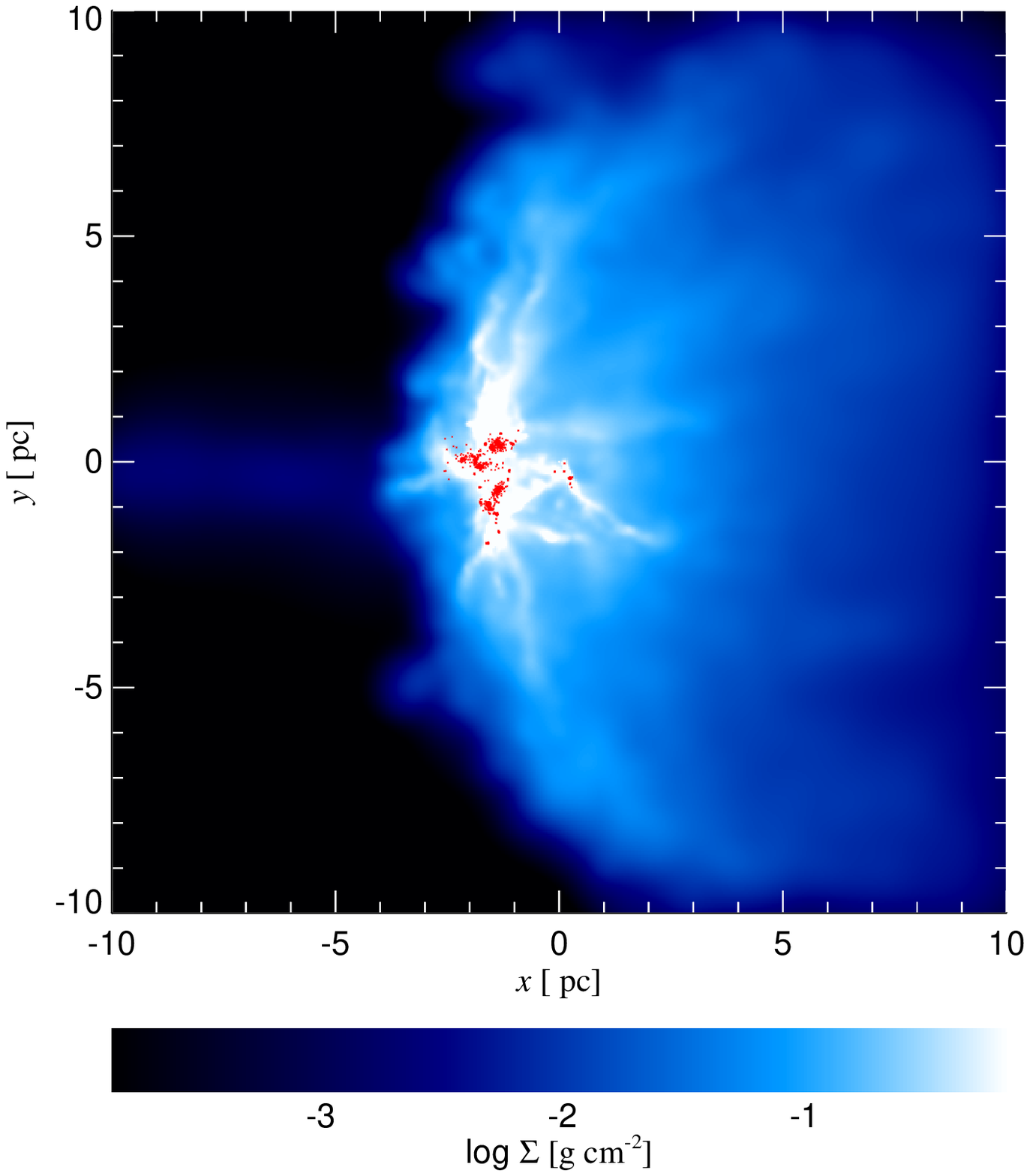}
    \includegraphics[trim = 6mm 25mm 6mm 0, clip, width=0.32 \textwidth]{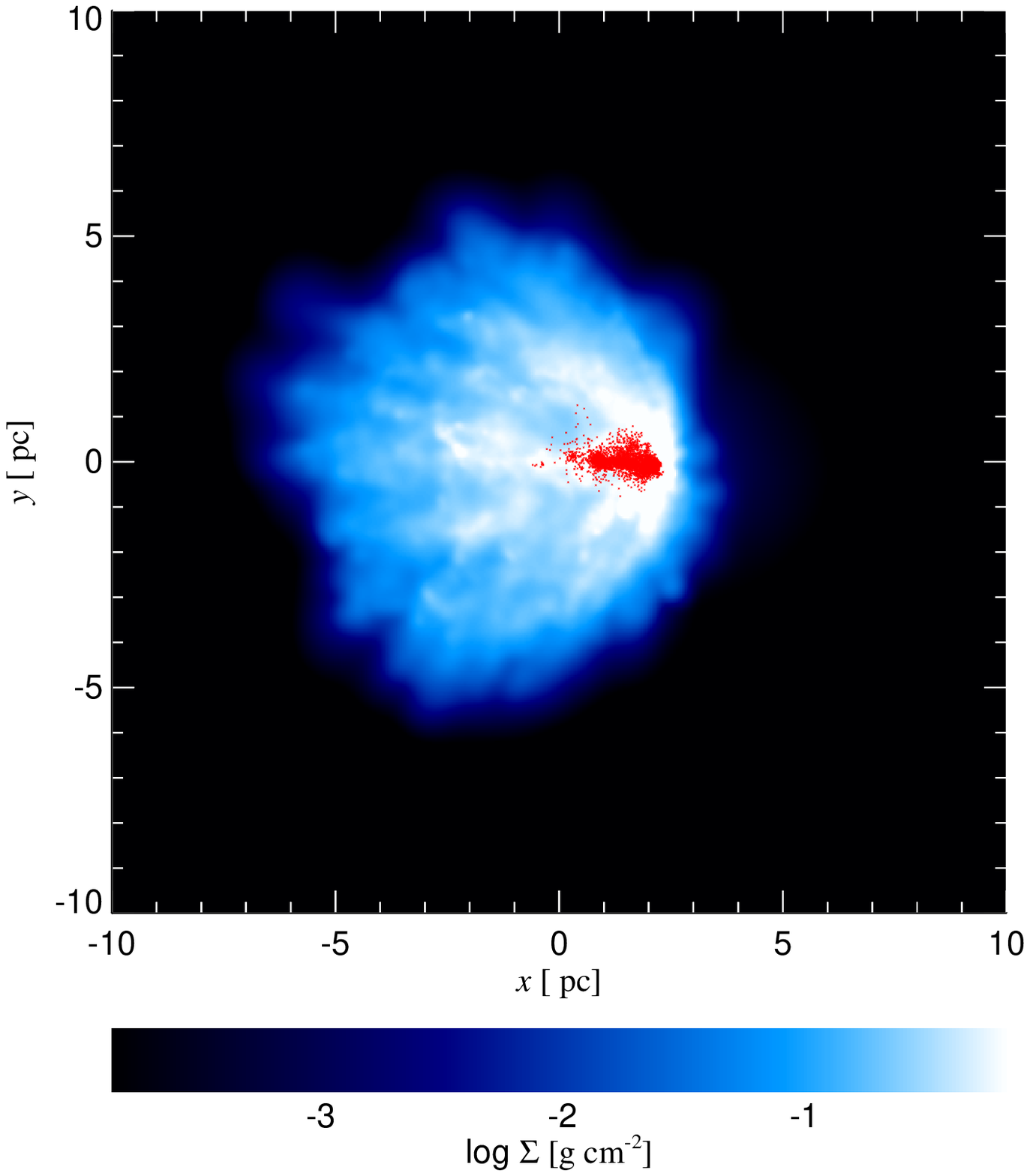}
    \includegraphics[width=0.34 \textwidth]{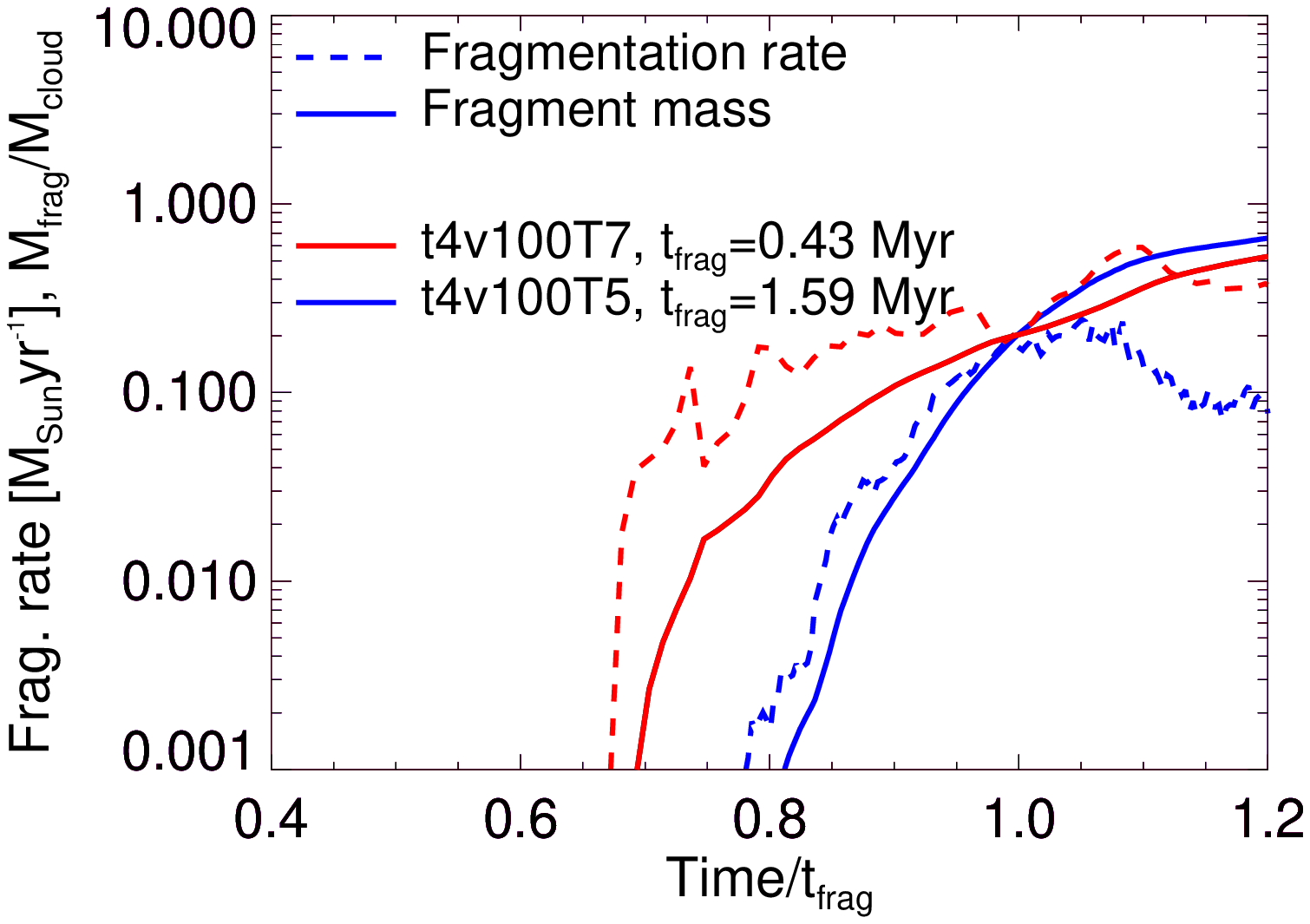}
    \includegraphics[trim = 6mm 25mm 6mm 0, clip, width=0.32 \textwidth]{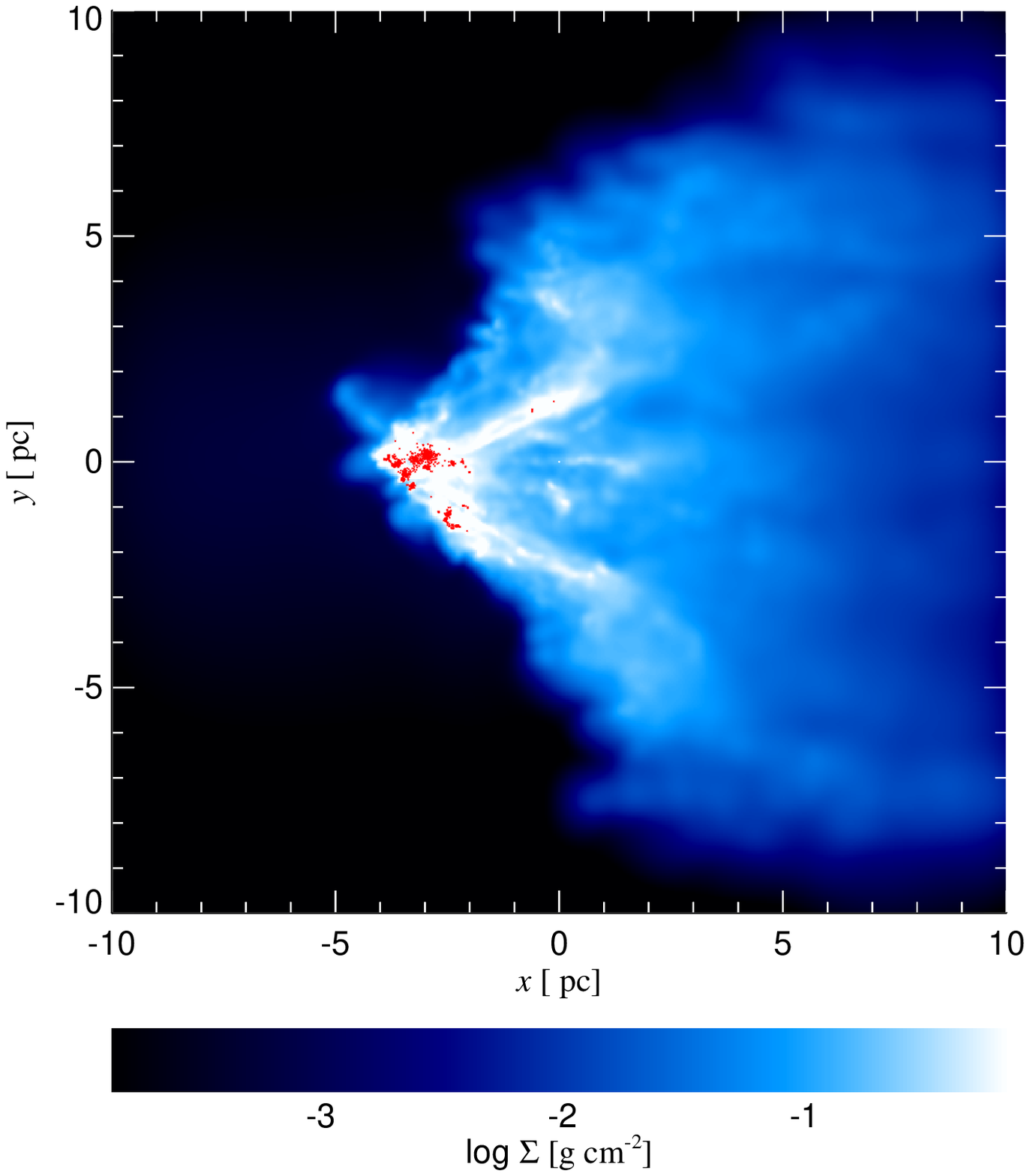}
    \includegraphics[trim = 6mm 25mm 6mm 0, clip, width=0.32 \textwidth]{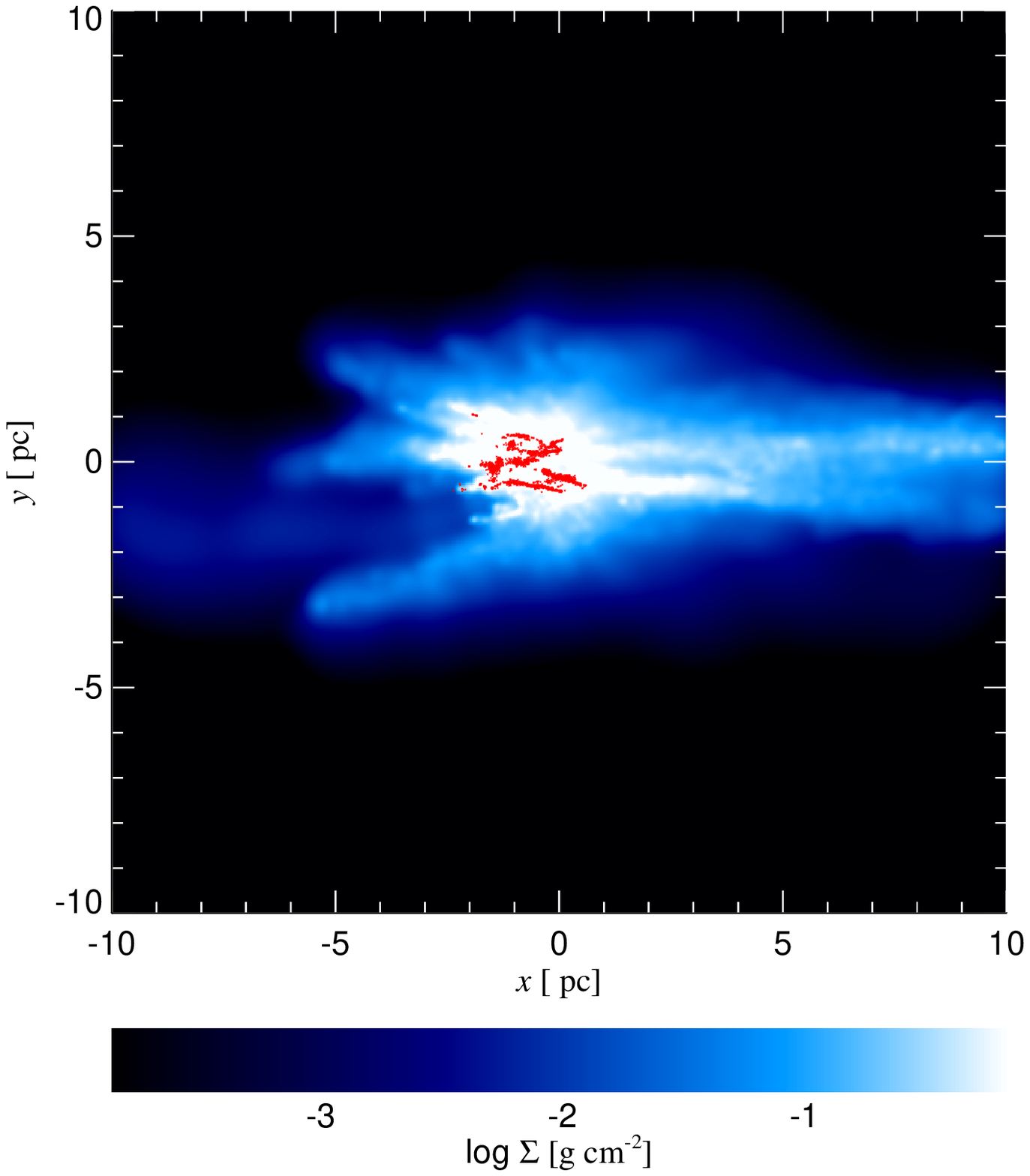}
    \includegraphics[width=0.34 \textwidth]{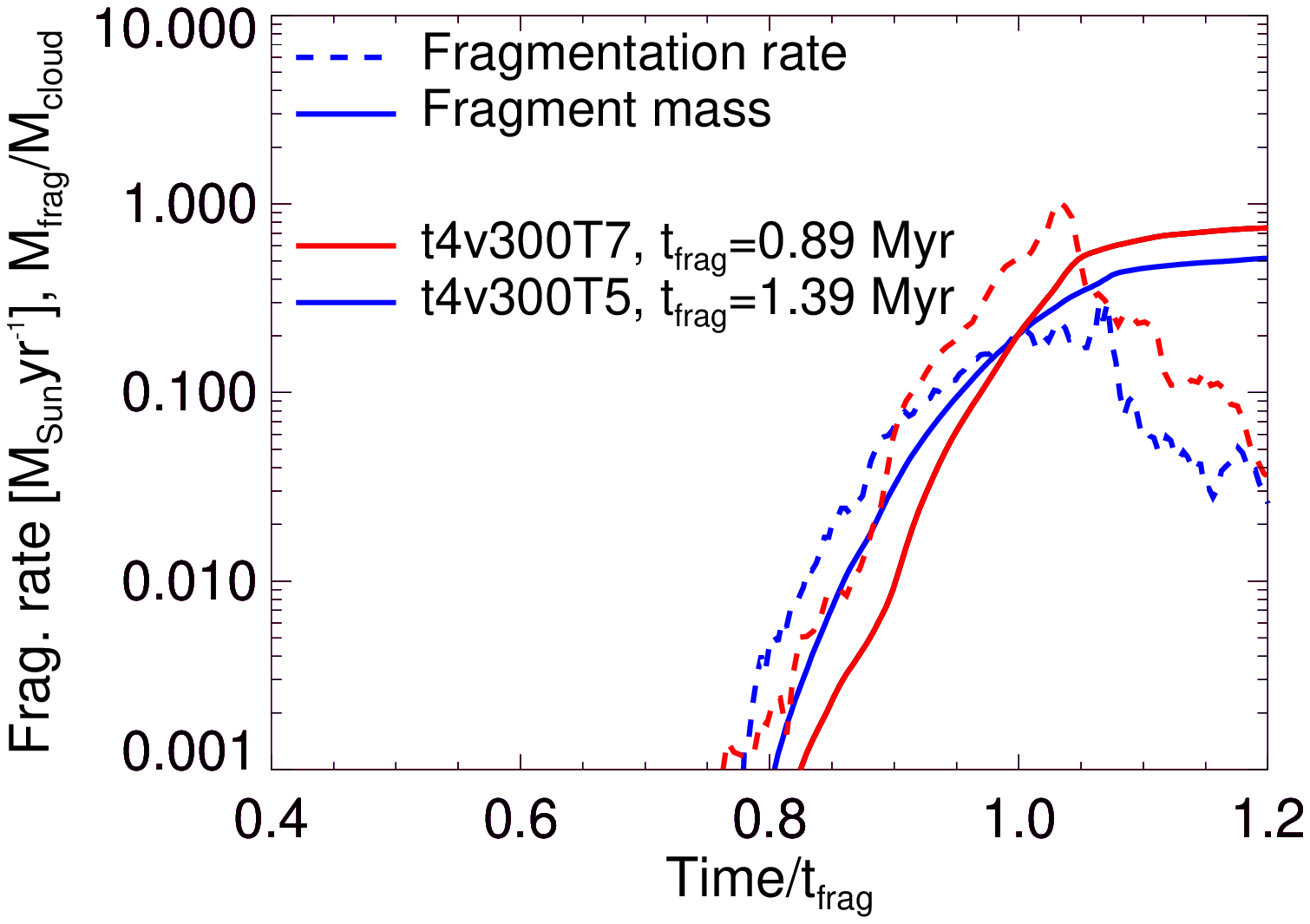}
  \caption{Same as Figure \ref{fig:lateral}, but for models moving
    with 100 km/s (top row) and 300 km/s (bottom row).}
  \label{fig:lateral2}
\end{figure*}

We perform eight simulations of clouds moving with respect to the
surrounding ISM (see Table \ref{table:param}) to evaluate the effects
of progressively stronger shearing motion. The low-pressure models in
this group are essentially a form of the standard ``blob test'' for
hydrodynamic codes \citep{Nakamura2006ApJS, Agertz2007MNRAS,
  Read2012MNRAS}, except that the cloud is turbulent. The timescale of
cloud destruction in such a system is \citep{Agertz2007MNRAS}
\begin{equation} \label{eq:clouddestr}
t_{\rm dest} \sim 1.6 \times \frac{2 R_{\rm cl}}{v_{\rm lat}}
\left(\frac{\rho_{\rm cl}}{\rho_{\rm ISM}}\right)^{1/2} \sim 12.5
M_5^{1/2} R_{10}^{-1/2} v_{100}^{-1} {\rm Myr},
\end{equation}
where $v_{100} = v_{\rm lat} / 100$~km/s. We see that the clouds
should not be destroyed by shear on the timescales relevant for our
models, especially when the post-shock gas further compresses the
cloud.

For numerical reasons, we set up the cloud as static, with the ISM
moving past it at a uniform velocity $v_{\rm lat}$ in the positive X
direction. We consider four ISM velocities: $v_{\rm lat} = 10, 30,
100$ and $300$~km s$^{-1}$. Direct acceleration of the cloud is
important only in the fastest case, where the cloud is accelerated to
a velocity of order $\sigma_{\rm turb}$ in $\sim 0.8$~Myr
\citep{McKee1978ApJ}; in the other cases, $t_{\rm accel} \gg t_{\rm
  dyn}$.

The results of these models are presented in Figures \ref{fig:lateral}
(cases with $v = 10$ and $30$~km/s) and \ref{fig:lateral2} ($v = 100$
and $300$~km/s). The first two columns in each figure show the
morphology of these models at $t = t_{\rm frag}$. In all the plots,
the ISM is moving to the right. Each row represents a different
lateral velocity, while the left and middle columns represent
uncompressed and compressed models, respectively.

There is an immediately obvious qualitative difference between the T5
and T7 models, namely that T5 models fragment closer to the leading
(left) edge of the cloud, while the T7 models fragment closer to the
trailing (right) edge. This is easy to understand once we consider how
shockwaves propagate through the clouds in various cases. In models
t4v10T5 and t4v30T5, there is no discernible shockwave, since the
lateral velocity is $v_{\rm lat} \ll \sigma_{\rm turb} \left(\rho_{\rm
  cl} / \rho_{\rm ISM}\right)^{1/2} \simeq 130$~km/s. The morphology
of the fragmenting central regions is identical to those of the static
cloud, and the parameters of fragmentation are almost identical as
well (see Table \ref{table:param}): first sink particles appear at $t
= 1.33$~Myr in both models; the fragmentation timescale is also
identical, $t_{\rm frag} = 1.66$~Myr, slightly smaller than in the
static model ($1.78$~Myr). This discrepancy arises because some of the
cloud material is removed and joins the ISM, so the mass of the cloud
drops and a lower total sink particle mass is required to bring the
mass fraction up to $20\%$. Cloud ablation and slight compression
along the leading edge also leads to the sink particle cluster
appearing off-centre and to the half-mass radius of the cloud being
much smaller ($2.68$~pc versus $4.26$~pc in model t4T5.

In the higher velocity models, t4v100T5 and t4v300T5 (Figure
\ref{fig:lateral2}, left column), the shockwave produced by the
lateral motion of the ISM affects the cloud significantly. The
shockwave moving through the cloud has a velocity comparable to or
larger than the characteristic turbulent velocity, and so the cloud
begins to break apart \citep[as in the standard ``blob
  test''][]{Agertz2007MNRAS}. Fragmentation is accelerated by the
shockwave, with first sink particles appearing at $t = 1.23$~Myr and
$t = 1.08$~Myr in the v100 and v300 models, respectively, and occurs
along the leading edge of the cloud. In particular, in the v300 model
the shockwave enhances gas compression by $\sim 20\%$, so that the
sink particles reach $20 \%$ by mass at only $t_{\rm frag} = 1.39$~Myr
(in model v100, $t_{\rm frag} = 1.59$~Myr). The increasing importance
of cloud ablation is also evident when considering the half-mass
radii, which also increase with increasing lateral velocity. In both
models, the faint line of material visible on the left is the extended
tail of the cloud arriving through the periodic boundary. The density
of this tail is low enough to be insignificant for the evolution of
the cloud.

The compressed models show different behaviour, because there is
always a shockwave moving in through the cloud. Once again, the low
velocity models evolve similarly to their static analogues, with both
the time of appearance of the first sinks ($0.36$~Myr) and
fragmentation time ($t_{\rm frag} = 0.42$~Myr) the same as in model
t4T7. Faster motion (model t4v100T7) produces noticeable change in
that the shockwave is anisotropic, moving in faster from the leading
(left-hand) side, resulting in maximum compression in the trailing
side of the cloud and more fragments forming there. The fragments also
start forming earlier, at $t = 0.29$~Myr, but $t_{\rm frag} =
0.43$~Myr is the same as in previous models. The fastest cloud, v300,
is disrupted by the shockwave and so fragmentation is actually
delayed, with first sink particles forming at $t = 0.6$~Myr and
$t_{\rm frag} = 0.89$.

The right-hand columns of Figures \ref{fig:lateral} and
\ref{fig:lateral2} show the time evolution of fragmentation in the
eight models. All four uncompressed models (blue lines) evolve
similarly, with first sinks appearing at around $0.8 t_{\rm frag}$,
the fragmentation rate increasing up to $\sim 0.2 \msun$~yr$^{-1}$ and
then dropping. Only in the fastest-moving model the fragmentation rate
drops more significantly due to rapid removal of gas from the
cloud. The two slow-moving compressed models evolve very similarly to
the static compressed one (t4T7, compare the left panel in Figure
\ref{fig:frag}), with a maximum fragmentation rate reaching $\sim 1
\msun$~yr$^{-1}$ at $t \simeq t_{\rm frag}$ before dropping
slightly. In the v100 model (Figure \ref{fig:lateral2}, top right
panel, red lines), fragmentation starts earlier, at $t \simeq 0.65
t_{\rm frag}$ and proceeds more slowly, with fragmentation rate
staying at $\sim 0.3 \msun$~yr$^{-1}$. This happens because the
anisotropic shockwave creates conditions for fragmentation as it moves
through the cloud; this is evident from the elongated shape of the
sink particle cluster (top middle panel). Finally, the compressed v300
model (Figure \ref{fig:lateral2}, bottom right panel, red lines)
evolves superficially similarly to the uncompressed model, because the
evolution is governed more by the shear than the isotropic
compression.

\section{Discussion} \label{sec:discuss}

Our simulations show that, under idealised conditions, high external
pressure has a dominant effect on the collapse and fragmentation of a
molecular cloud. Cloud fragmentation into pre-stellar cores is
significantly accelerated in a compressed cloud, more than a simple
pressure balance argument would suggest. Furthermore, even a highly
turbulent cloud, which would be unbound without external confinement,
is crushed and fragments in a very similar fashion to the
low-turbulence analogue. In a similar fashion, cloud rotation is also
unable to counteract external compression, although fragments form in
qualitatively different locations in rotating clouds.  Accelerated
fragmentation leads to a larger fraction of the initial cloud mass
converted into pre-stellar cores. Models of moving clouds show that
the influence of low shear velocities ($v_{\rm lat} \ll 100$~km/s)
upon cloud evolution is minimal, but ram pressure caused by large
velocity dominates over isotropic pressure and disrupts the cloud as
it fragments.

We consider the general implications of these results below, in
Section \ref{sec:confine}. We discuss the effect of external pressure
upon cloud dispersal and formation of compact star clusters in
Sections \ref{sec:starform} and \ref{sec:clusters}. On larger scales,
our results support the possibility of positive AGN feedback (Section
\ref{sec:agnfb}) and are consistent with other models of star
formation enhancement or triggering by pressure (Section
\ref{sec:other}). Finally, we briefly review the validity of
assumptions made in this work and discuss the possible improvements to
the models in Section \ref{sec:improv}.

\subsection{Cloud confinement} \label{sec:confine}

In all models of compressed clouds, the time evolution is almost
identical. The integrated fragmentation parameters - the onset of
fragmentation, the fragmentation rate and the timescale for a given
fraction of the cloud to be transformed into sink particles - are
hardly affected by differences in initial conditions, environment and
cloud morphology. The only noticeable difference is that the rotating
cloud models have lower mean sink particle mass, but the lack of
detailed star formation physics in our simulations prevents us from
drawing significant conclusions regarding this property.

This similarity suggests that as long as external compression
dominates over other sources of confinement, such as gravity or shock
due to shearing motion, cloud fragmentation is governed almost
exclusively by this compression. Furthermore, the acceleration of
fragmentation is greater than the analytical estimate based on
pressure balance predicts; this enhancement is due to instabilities
enhancing the density contrast within the cloud.

Empirically, the increase of total (external + gravitational) pressure
by a factor of $11.4/1.4 \simeq 8$ leads to an increase in the
fragmentation rate by a factor $\sim 10$, suggesting an almost linear
relationship. However, we cannot constrain it further without a wider
range of simulations, which are beyond the scope of this paper.

The results of even these idealised simulations reveal that external
pressure can have an important, even dominating effect upon molecular
cloud evolution. This effect should be accounted for in subgrid
prescriptions of large-scale simulations. Typically, the timescale
(and, equivalently, rate) of star formation in these prescriptions is
governed by gas density only \citep{Springel2003MNRAS,
  Fujita2003ApJ}. Instead, they should take the surrounding hot gas
pressure into account. In grid codes and some SPH codes
\citep[e.g.,][]{Springel2003MNRAS}, cold and hot gas can be present in
the same computational element, in which case the hot phase pressure
change can be used when calculating the star formation rate in the
cold component. In models where the cold phase is resolved with
separate particles, the pressure of hot gas surrounding a particular
cold gas clump should be used instead.

\subsection{Formation of stars and parent cloud evolution} \label{sec:starform}

In most compressed cloud models, cloud fragmentation is enhanced by
the shockwave moving in through the cloud together with the RM
instabilities further increasing the density contrast. The shockwave
only appears because we assume that the ISM pressure increases on a
timescale shorter than the dynamical time of the cloud, so that the
cloud does not have time to contract as a whole. Nevertheless, the
confinement of cold gas would happen even in the case of gradual ISM
density increase; \citet{Elmegreen1997ApJ} investigated such a
situation, finding a higher star formation rate in clouds that
virialize under high external pressure. Our simulation t10T7, with a
highly turbulent cloud in approximate pressure equilibrium with its
surroundings, where the shockwave would be weak, also shows
significantly enhanced fragmentation. Therefore, the acceleration of
star formation should be present independently of the existence of a
shockwave. In the rotating compressed cloud model, sink particles
appear in the shocked gas confined to the outskirts of the cloud. The
similarity of this fragmentation rate to that of the static cloud
depends on the initially uniform cloud density. In a more realistic
cloud with a centrally-peaked density distribution, we would expect
lower fragmentation rate. Since most molecular clouds should have
nonzero angular momentum, the fragmentation rates we find should be
taken as upper limits.

The overall sink particle formation rate in our compressed models is
$\sim 10$ times larger than in uncompressed clouds. Since the fraction
of pre-stellar core mass that ends up in a star is independent of core
mass \citep{Matzner2000ApJ}, the star formation rate should follow the
same trend. This increase is larger than the scatter in the KS
relation \citep[$\sigma_{\rm KS} \simeq
  0.2-0.3$~dex,][]{Bigiel2008AJ}, so it should be detectable. There is
some evidence that starburst galaxies have $\sim 3-4$ times higher SFR
surface densities at the same gas surface densities, and
correspondingly shorter depletion timescales of the cold gas
\citep{Garcia2012A&A}. This finding is consistent with the suggestion
that starbursts as a whole can be triggered by increased external
pressure \citep{Zubovas2013MNRASb}. A connection between higher star
formation rate and external pressure was also found in resolved
star-forming regions of M82 \citep{Keto2005ApJ}.

Stellar feedback, not included in our simulations, affects the
properties of the forming stellar population. Young massive stars heat
their surroundings, increasing the Jeans' mass and reducing the
fragmentation rate \citep{Bate2009MNRAS, Offner2009ApJ}, leading to a
top-heavy mass function \citep{Krumholz2011ApJ, Bate2012MNRAS}. Other
forms of stellar feedback, such as prestellar outflows
\citep{Krumholz2009ApJb} and radiation pressure \citep{Fall2010ApJ},
also tend to increase the mean stellar mass. In our compressed cloud
simulations, fragmentation happens much more rapidly, therefore more
sink particles may form before radiative heating is able to shut off
further fragmentation. On the other hand, external pressure confines
the gas for longer, so the fragments can grow to much larger masses
than in unconfined clouds. We cannot say which of the two processes is
more important and how different the mass functions of compressed and
uncompressed clouds would be without further simulations.

The global cloud dynamics is also affected by stellar feedback.
Outflows driven by photoionization can potentially remove a
significant fraction - more than $10\%$ - of the cloud mass
\citep{Wang2010ApJ,Dale2012MNRAS}, leaving it and the nascent star
cluster more prone to destruction \citep{Dale2012MNRAS,
  Pfalzner2011A&A}. The effects of prestellar outflows
\citep{Krumholz2012ApJc}, stellar winds \citep{Dale2013MNRAS} and
radiation pressure \citep{Krumholz2009ApJb} are also significant. In
particular, radiation pressure is probably the primary mode of
disruption of massive molecular clouds \citep{Murray2010ApJ} and can
drive significant turbulence \citep{Krumholz2012ApJc}.  Supernova
explosions begin $\sim 3$~Myr after the formation of first stars and
disperse most of the cloud gas within $\sim 6$~Myr
\citep{Rogers2013MNRAS}. All of these effects can also trigger
subsequent star formation by compressing the gas in other parts of the
cloud \citep[also see Section \ref{sec:rdi}
  below]{Dale2007MNRAS,Koenig2012ApJ}.  Ultimately, the star formation
rate might be determined by self-regulation, whereby turbulent and
stellar feedback pressure counteract gravity \citep{Thompson2005ApJ}.
In our model, external pressure aids gravity and must be compensated
as well (see also Section \ref{sec:agnfb}). We plan to investigate the
effects of stellar feedback in a future publication.

In addition to internal stellar feedback, molecular clouds may be
destroyed by external effects. Galactic-scale numerical simulations
show cloud lifetimes of a few to $\sim 20$~Myr
\citep{Dobbs2013MNRAS}. It is not clear how this dynamical evolution
would be affected by external pressure, if at all. On timescales
comparable to, or shorter than, the dynamical lifetime, the cloud can
evaporate due to heating by the ISM \citep{Cowie1977ApJ}. Equation
(22) of that paper, when rescaled to the parameters of our model
clouds, reads
\begin{equation}
t_{\rm evap} \sim 40 M_5 R_{10}^{-1} T_7^{-5/2} {\rm Myr}.
\end{equation}
So we may expect the compressed cloud to lose only a small fraction of
its material to evaporation during the time relevant in our
simulations. On the other hand, if the cloud was compressed by a much
hotter ISM, evaporation might take over ($t_{\rm evap} \simeq 1.3
\times 10^5$~yr at $T_{\rm ISM} = 10^8$~K) and destroy the cloud
before any significant star formation takes place. Shear also destroys
the cloud on a timescale $t_{\rm dest} \sim 12.5 M_5^{1/2}
R_{10}^{-1/2} v_{100}^{-1} $~Myr (eq. \ref{eq:clouddestr}). This
timescale is longer than the fragmentation timescales of all our
models so long as $v_{\rm lat} < 600$~km/s. For less massive and/or
more diffuse clouds, however, the destruction timescale is shorter and
disruption due to shear might become the dominant mechanism,
preventing rapid star formation. In addition, there are other
processes that affect the cloud destruction timescale: a smoother
density gradient increases this timescale \citep{Nakamura2006ApJS},
while turbulent motions in the post-shock ISM decrease it
\citep{Pittard2009MNRAS}.

Another result of cloud-shockwave interaction, as in blob tests, is
that the cloud expands laterally behind the shockwave and thus becomes
more susceptible to other forms of quasar feedback
\citep{Hopkins2010MNRAS}, enhancing the quenching effect that AGN
activity can have upon star formation. We do not find this behaviour
in our models for three reasons. First of all, we do not model the
initial shockwave interaction with the cloud; however, the evolution
of the fastest shearing cloud model, t4v300, is significantly affected
by the shear-induced shockwave, so a similar effect might be
expected. Secondly, the timescale of expansion is similar to the cloud
crushing timescale, and our clouds evolve faster than this. Finally,
the presence of isotropic pressure behind the shock helps confine the
cloud, and further compresses it in the T7 model, preventing expansion
and dispersal.

\subsection{Evolution of star clusters} \label{sec:clusters}

\begin{figure}
  \centering
    \includegraphics[width=0.49 \textwidth]{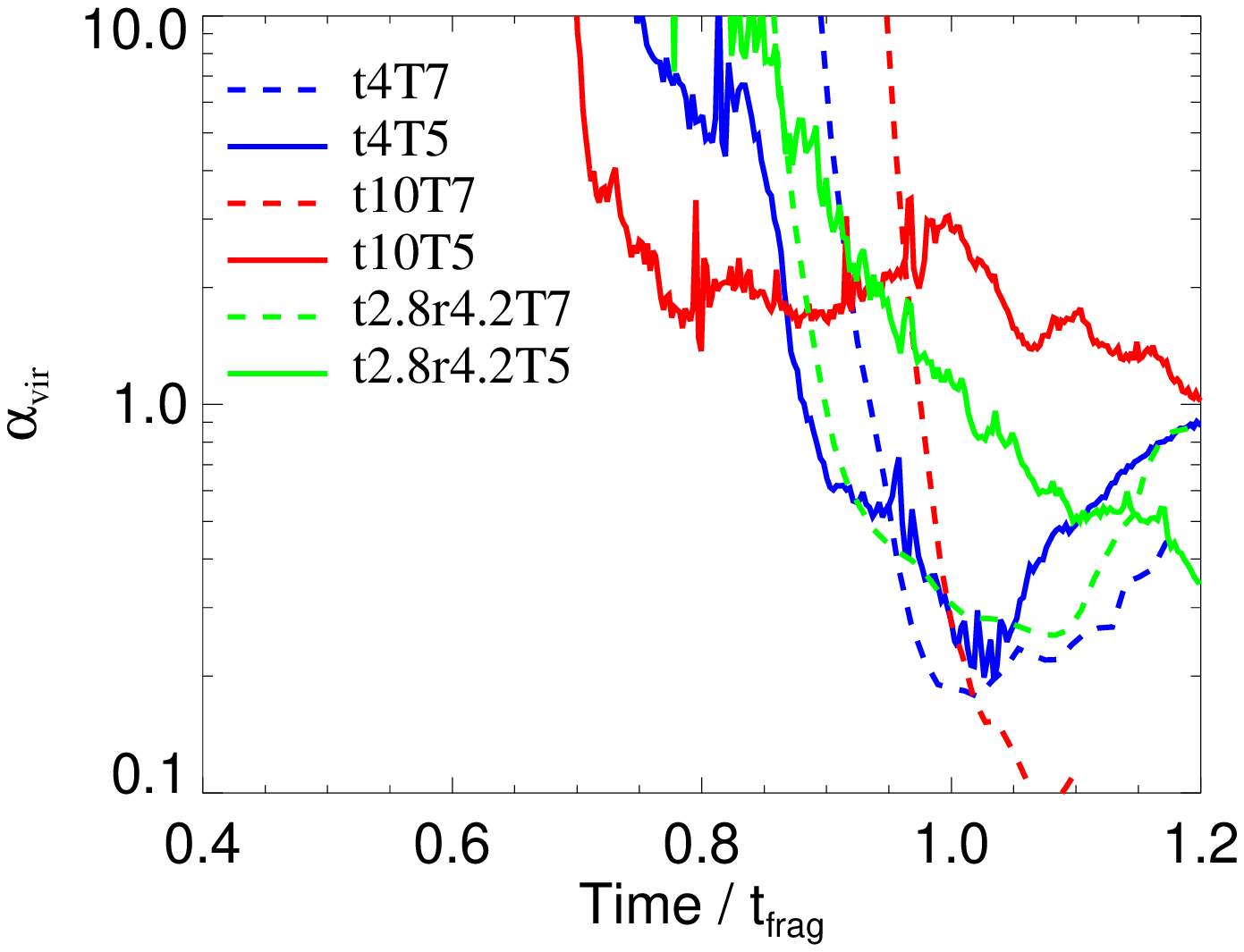}
    \includegraphics[width=0.49 \textwidth]{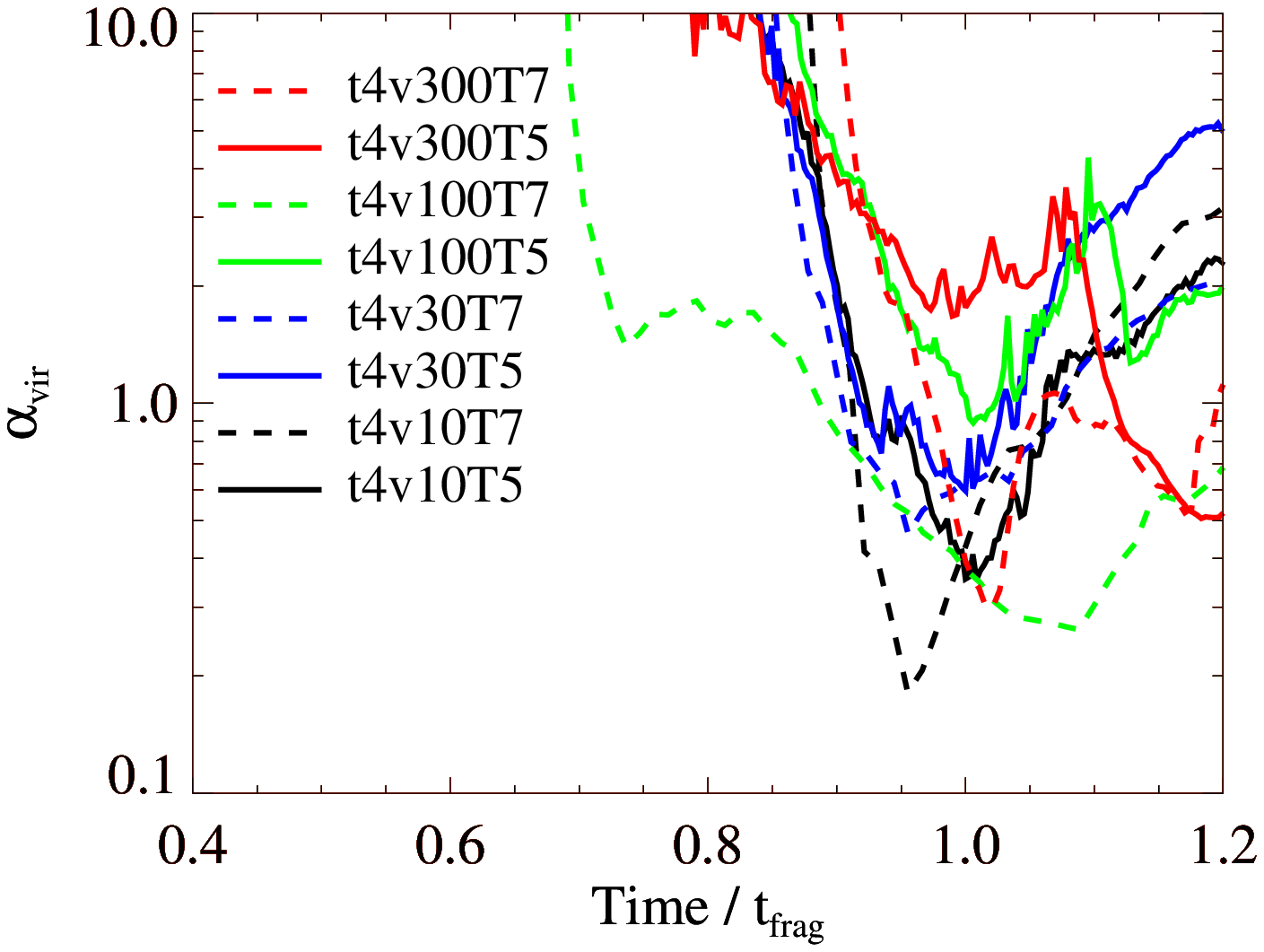}
  \caption{Virial parameter $\alpha$ of the sink particles in the six
    models without shear (top) and eight models with shear
    (bottom). Solid lines correspond to uncompressed models, dashed
    lines represent compressed clouds. In the top panel, blue lines
    are low-turbulence models, red lines are high-turbulence while
    green lines are rotating models. In the bottom panel, the blue,
    red, green and black lines correspond to v10, v30, v100 and v300
    models, respectively.}
  \label{fig:alpha}
\end{figure}

The sink particles tend to form in a single clump in all non-rotating
models except v300. The clump rapidly relaxes and the sink particles
appear to stay together in a cluster. In order to understand the
subsequent evolution of this cluster, we plot, in Figure
\ref{fig:alpha}, the time evolution of the virial parameter
$\alpha_{\rm vir} = E_{\rm kin}/\left|E_{\rm g}\right|$ of the sink
particles for the non-shearing (top panel) and shearing models (bottom
panel). In all pairs of models, the dashed line shows compressed
clouds and solid line shows uncompressed ones. In all models, the sink
particles are at first strongly unbound ($\alpha_{\rm vir} > 1$), but
the virial parameter rapidly decreases, falling below $1$ around $t =
t_{\rm frag}$. The only models which retain a formally unbound sink
particle population throughout are t10T5, as one might naively expect,
and t4v100T5, where the cloud disrupts as is it forming the
clusters. After $t_{\rm frag}$, the virial parameter increases in most
models; this represents the fragmentation of a single sink particle
cluster into subclusters, which disperse, since sink particles are not
affected by surrounding gas pressure. Some models, such as t4 and
t2.8r4.2 pairs, keep globally bound sink particle clusters throughout
the simulation, but this is not a common occurence, especially when
shear is considered. Comparing the compressed and uncompressed models
in each pair, we see that in most cases, compressed clouds tend to
have somewhat lower virial parameters. This suggests that compressed
cloud are somewhat, but not significantly, more likely to form bound
star clusters than uncompressed ones. Such clusters are then less
likely to disperse due to internal motions soon after the parent cloud
disperses \citep[so-called ``infant mortality'';][Section
  5.2]{Lada2003ARA&A}.

The sink particle clusters formed in the static cloud simulations have
visually similar density profiles (Figure \ref{fig:radprof}, right
panel), but their half-mass radii differ, with model t4T5 having
$r_{\rm h,sink} \simeq 0.79$~pc and model t4T7 having $r_{\rm h,sink}
\simeq 0.26$~pc. This difference is caused by the high-density core
present only in the compressed model and a low-density envelope
present only in the uncompressed one. The difference is not large, so
both clusters would appear very similar while young. However, their
long-term evolution may be significantly different, especially because
the compressed cloud converts a larger fraction of its mass into sink
particles (and, hence, stars). Both more massive
\citep{Kruijssen2009ApJ} and more compact
\citep{Spitzer1987book,Gieles2008MNRAS} clusters can withstand the
tidal field of the host galaxy more easily and lose a smaller fraction
of their mass during evolution, so we may expect compressively-formed
clusters to survive for longer. Since these clusters do not
necessarily have very large masses, they may appear as compact star
clusters \citep[CSCs;][]{Holtzman1992AJ}. These clusters are expected
to form from molecular clouds that have been compacted by some process
\citep{Escala2008ApJ,Larsen2010RSPTA}, consistent with the picture
presented in this paper. The higher star formation rate in the
progenitor cloud coinciding with a more long-lived cluster is also
consistent with observations that a larger fraction of stars stay in
clusters in regions of galactic discs with higher star formation rate
per unit area \citep{Larsen2000A&A}. Similar arguments apply to our
other models, both with and without shearing motion (see Table
\ref{table:param}).

\subsection{Implications for positive AGN feedback} \label{sec:agnfb}

The main motivation of the simulations presented above is to validate
the assumption that increased pressure leads to increased star
formation rates, which is central to several models of positive AGN
feedback \citep{Zubovas2013MNRASb, Silk2005MNRAS, Ciotti2007ApJ}. Such
a connection requires, first of all, that external pressure should
trigger and/or enhance molecular cloud fragmentation on
timescales much shorter than the flow timescale of the AGN
outflow. Otherwise, the decrease in outflow pressure as the outflow
expands would preclude any significant enhancement of the galactic
star formation rate. The flow timescale is of order several tens of
Myr \citep{King2011MNRAS, Zubovas2012ApJ}; our results show that
compressed molecular clouds evolve on much shorter timescales, so this
requirement is satisfied.

In addition, the high fragmentation rate under compression must
translate into a sustained high star formation rate. This requires
that self-regulation of star formation produces a higher SFR at higher
pressure. We may use the analytical model of \citet{Thompson2005ApJ}
to show that this is expected. In that model, the star formation rate
is set by the balance between pressure force from feedback processes
(stellar radiation, winds and supernova explosions) and self-gravity
of the gas. In our model, external pressure provides an extra force
working in tandem with gravity:
\begin{equation}
F_{\rm grav} + F_{\rm ISM} = F_{\rm fb},
\end{equation}
where the right-hand-side term is force from feedback. Assuming that
all forces act isotropically, we can rewrite this equation in terms of
pressures:
\begin{equation}
G \Sigma_{\rm tot}^2 + P_{\rm ISM} = \epsilon \dot{\Sigma}_* c,
\end{equation}
where $\Sigma_{\rm tot} = \Sigma_{\rm g} + \Sigma_*$ is the column
density of cloud gas and stars. It follows from this relation that the
star formation rate density increases when external confinement is
present. Our results corroborate this finding, by showing that
external pressure can easily confine clouds which have very high
dynamical pressures (such as would be present in photoionized gas; see
Section \ref{sec:confine}).

The shockwave of an AGN outflow can have a more direct impact upon
molecular clouds. \citet{Gray2011ApJ} found that AGN outflows can
trigger vigorous star formation in primordial minihaloes, leading to
formation of compact star clusters in systems that would otherwise not
form stars at all. Although we do not model the passage of a shockwave
around the cloud (see Section \ref{sec:rdi} for comments on other work
addressing such issues), the increased fragmentation rate in our
models is consistent with these results. On the other hand, star
formation triggering by passage of supernova remnants requires a
particular range of densities and SNR radii to work
\citep{Melioli2006MNRAS}; clouds that are too small are disrupted,
while overly large clouds dissipate the shockwave too efficiently. A
similar separation of various regimes may occur in the case of AGN
feedback, but we cannot constrain these parameters without a larger
study.

Therefore, our results support the models of positive AGN feedback due
to outflows that can overpressurize the ambient gas and induce or
increase star formation rates. In the future, we intend to run
larger-scale simulations that will self-consistently treat the
propagation of an AGN outflow and its effect on the dense clouds in a
galaxy environment.

\subsection{Other work} \label{sec:other}

Several authors have considered the connection between star formation
rates (both on GMC and on galactic scales) and the properties of the
ambient medium, including pressure. We briefly summarize those results
below and comment on their connection with our results.

\subsubsection{External confinement of star forming regions}

Recently, the has been mounting observational evidence that the
immediate surroundings of the molecular clouds are an important
element in their evolution. Observations of star-forming regions in
M82 \citep{Keto2005ApJ} show molecular clouds forming stars only when
compressed by hot interstellar medium (ISM), but not on their own
volition; this is the case even for parts of the same cloud which are
otherwise indistinguishable observationally. At high redshift, many
gas-rich galaxies show evidence of high ISM pressure
\citep{Swinbank2011ApJ}, which correlates with the presence of
starbursts. Observations of merger-triggered ULIRGs hint at star
formation being triggered by molecular cloud infall into a
high-pressure medium \citep{Solomon1997ApJ}. More recently, it was
proposed that ISM pressure differences among M51, M33 and LMC are
responsible for the properties of molecular clouds in those galaxies
\citep{Hughes2013ApJ}.

From a theoretical standpoint, \citet{Jog1992ApJ} have investigated
how GMCs react to an increase in HI cloud pressure due to galaxy
collisions. Using analytical calculations, they predicted cloud
crushing and enhanced star formation (see Section
\ref{sec:confine_analyt}). 

\citet{Elmegreen1997ApJ} proposed that all star clusters form by
the same mechanism, with the primary difference in initial conditions
being the external pressure affecting nascent molecular clouds. They
suggest that globular clusters form in high-pressure environments,
where the star formation efficiency is higher due to confinement. This
result is similar to our findings, however we find that a sudden
increase in external pressure directly increases the fragmentation
rate of the cloud, instead of simply preventing mass loss due to
stellar feeback.

On galactic scales, \citet{Krumholz2009ApJ} suggested an
explanation for the break in the star formation law (i.e. the
proportionality between $\Sigma_{\rm SFR}$ and $\Sigma_{\rm gas}$) in
galaxies based on external confinement of star-forming regions in
dense environments. In this model, the high pressure of ambient
galactic disc medium at large surface densities enhances star
formation and leads to a steeper slope of the KS law than in
lower-density systems. Although our simulations are not detailed
enough to make predictions regarding the KS law, the results are
consistent with this picture, except that we do not necessarily
require high surface densities, merely the presence of an external
pressure compressing the cloud.

\subsubsection{Radiatively-driven implosion} \label{sec:rdi}

Several aspects of our model are similar to radiation-driven implosion
\citep[RDI;][]{Bertoldi1989ApJ}. This process has been investigated in
great detail \citep{Klein1980SSRv, Kessel-Deynet2003MNRAS,
  Dale2007MNRASb, Bisbas2011ApJ}. Gas can be strongly compressed by
either ionizing radiation from massive stars, passage of supernova
shells, or both; molecular cloud fragmentation and star formation are
enhanced as a result. Although we model a different kind of external
pressure (almost isotropic and constant in time instead of directed
and rapidly changing), the main result of increased pressure resulting
in increased fragmentation rate is the same in both cases.

It is important to note that the observational evidence of RDI is
inconclusive. Some molecular cloud surveys suggest that clouds with an
increased star formation rate (compared with the background level) are
affected by either passages of supernova shells
\citep{Preibisch1999AJ} or ionizing radiation from nearby stellar
clusters \citep{Sugitani1989ApJ, Sugitani1991ApJS,
  Sugitani1994ApJS}. The locations of intermediate- and low-mass stars
in OB associations are consistent with their formation being triggered
by the radiation of massive stars \citep{Lee2007ApJ}. On the other
hand, recent large surveys of young stellar objects located around
infrared bubbles suggest that star formation on the edges of those
bubbles is triggered by the collect-and-collapse model instead of RDI
\citep{Kendrew2012ApJ, Thompson2012MNRAS}. Despite these issues, it
seems clear that RDI can enhance star formation in some environments
and is a physically sound mechanism similar to our model.

\subsection{Possible improvements to the model} \label{sec:improv}

In this paper, we are only interested in the basic dynamics, collapse
and fragmentation of a turbulent cloud. Therefore, we purposefully
neglected many physical processes that are relevant for the details of
cloud evolution. We mentioned some of them, particularly stellar
feedback, above; here we describe the other improvements we plan to
make in the future.

\subsubsection{Gas equation of state}

The models presented in this paper use a heating-cooling function
which includes most relevant processes. However, heating from
protostars, both ionizing and non-ionizing, is not accounted for. Such
heating can significantly change the mass function of stars
\citep{Bate2009MNRAS} and shut off further accretion and fragmentation
after several Myr. We hope to expand our simulations with inclusion of
these feedback mechanisms (radiative heating, photoionization,
radiation pressure and stellar winds) in order to more properly
simulate the properties of the star clusters forming in pressurized
clouds.

\subsubsection{Initial conditions and model scale} \label{sec:ic_improve}

We make several simplifying assumptions both regarding the cloud and
the ISM. First of all, we assume a uniform initial density of the
cloud. Although turbulent velocities create an uneven density
distribution very rapidly, more fundamental differences remain
unexplored. In particular, a cloud with a $\rho \propto R^{-2}$
density profile is more likely to form stars in the central
regions. The confining effect of external pressure would remain, but
triggering and acceleration of fragmentation would be mitigated. On
the other hand, clouds with smoother boundaries are more stable
against disruption due to shear \citep{Nakamura2006ApJS} and would
presumably be able to fragment for longer without dispersing.

Real molecular clouds have more complex density profiles. Ideally, the
simulation should follow the molecular cloud assembly as well, because
clouds exchange material with their surroundings throughout their
$\sim 5-20$~Myr lifetimes \citep[e.g.][]{Dobbs2013MNRAS,
  Bournaud2013arXiv}. In order to follow this process, the simulation
volume should encompass a region of linear size $l \sim 100$~pc, so
that the assembly and evolution of a whole cloud complex can be
followed \citep{Wilson2003ApJ}.

The hot ISM should also be implemented in a more realistic way, with
turbulent motions and uneven evolving density structure. More
importantly, the passage of the ISM shock around the cloud and its HI
envelope can have different effects from those of the uniform ISM,
even though the time for the ISM to envelope the cloud is much shorter
than the cloud crushing timescale.

Eventually, we plan to run a simulation tracking a significant part of
the whole galaxy. Such a simulation would allow for self-consistent
formation of molecular clouds and driving of the AGN outflow (or any
other process which generates increased external pressure).
Unfortunately, at this scale, a feasible simulation would have a mass
resolution of order $10^3 \; \msun$ or worse, preventing one from
investigating the details of cloud evolution. A balanced approach,
combining insights from detailed small-scale simulations with the
galactic context gleaned from large scale models, is needed.

\subsubsection{Driving of turbulence} \label{sec:turbdrive}

Currently, turbulent velocities are only implemented in the initial
conditions of the models. This leads to a decay of turbulent power on
the dynamical timescale of the cloud; this is a common feature of
similar models \citep[see, e.g.,][]{Bate2003MNRAS}. Therefore, our
results become unrealistic after at most one dynamical time
independently of any other numerical inaccuracies. The decaying
turbulence leads to an increase in fragmentation rate, especially in
the uncompressed high-turbulence model (t10T5). Both star formation
feedback \citep{Krumholz2006ApJ} and accretion of external material
\citep{Klessen2010A&A} maintain turbulent motions within the cloud,
and both processes probably have similar relative importance
\citep{Goldbaum2011ApJ}, so these processes should be included in more
long-term simulations.

\subsubsection{Numerical accuracy}

We used sink particles in our simulations partly for convenience of
analysis and partly in order to speed up the calculations. To test the
importance of this approximation, we ran a simulation identical to
t4T5, but without sink particles. We find that the global evolution of
the cloud is identical between the two runs, but the model without
sink particles retains more structure in the dense clumps than the
sink particle cluster in t4T5. We conclude that the presence of sink
particles does not affect our overall conclusions.

We also ran a simulation at higher resolution, using $N = 4 \times
10^6$ particles instead of $10^6$. As expected, we find more
small-scale clumps and filaments in the higher resolution model. The
first sink particles appear slightly later, at $t = 1.32$~Myr instead
of $1.26$~Myr. Overall, there is very little difference in the
large-scale cloud evolution. Therefore, we conclude that our
simulations are numerically converged.

The choice of hot-phase ISM density of $1$~cm$^{-3}$ is motivated by
numerical considerations: higher density leads to more particles in
the ISM and higher computational costs, while lower density reduces
resolution and can lead to unwanted low-particle-number
effects. Higher ISM temperature may lead to cloud evaporation on
timescales comparable to fragmentation (see Section
\ref{sec:clusters}), but otherwise there should be no physical
difference between pressure caused by a high-density ISM and a
proportionately more diffuse, but hotter, ISM. Nevertheless, we intend
to check for possible numerical differences in the future.

\section{Conclusion} \label{sec:concl}

In this paper, we presented results of numerical simulations following
the collapse and fragmentation of spherical turbulent molecular clouds
under different external ISM pressures. Our idealized initial
conditions contain spherical clouds with mass $10^5 \; \msun$, radius
$10$~pc and uniform density. Each cloud gas has an initial turbulent
velocity spectrum with characteristic velocity of either $4$ or
$10$~km/s to mimic self-gravitating and gravitationally unbound
clouds, respectively. The ambient ISM pressure is either $P_{\rm ISM}
= 10^5$ or $10^7$~K~cm$^{-3}$; the lower value is similar to typical
ISM pressures and does not affect cloud evolution, while the higher
one significantly compresses the cloud. We consider the effects of
pressure upon static and rotating clouds, as well as clouds moving
with various lateral velocities w.r.t. the surrounding ISM.

The following are the main results:

\begin{enumerate}

\item The compressed clouds collapse and fragment much more rapidly
  and efficiently than the uncompressed ones. The ratio between the
  times of comparable evolutionary state, i.e. the effective dynamical
  times of the uncompressed and compressed systems, is larger than the
  analytically predicted $t'_{\rm dyn}/t_{\rm dyn} = \left(1 + P_{\rm
    ISM} / P_{\rm cl}\right)^{-1/2}$. The difference arises due to
  instabilities which follow behind the shockwave into the compressed
  clouds, enhance the density contrasts there and thus promote faster
  fragmentation. This result shows that external pressure can both
  accelerate star formation by compacting the cloud, and trigger star
  formation by promoting density contrasts.

\item The fragmentation rate in the cloud rapidly attains a constant
  value. This value, and presumably the corresponding star formation
  rate, is approximately linearly proportional to the total (virial
  plus external) pressure affecting the cloud. This proportionality
  can be used in subgrid models in larger simulations, thus accounting
  for the effects of increased ambient pressure around star forming
  regions.

\item Even high turbulence is unable to withstand the external
  pressure. The compressed clouds evolve in an almost identical
  fashion despite the difference in characteristic turbulent
  velocity. This confirms that external pressure can trigger star
  formation in regions that would otherwise have dispersed.
  Furthermore, it shows that external compression can confine clouds
  that would be destroyed by stellar feedback and allow for
  star formation to continue even after the first massive stars heat
  up the cloud.

\item Cloud rotation changes the morphology of the forming sink
  particle cluster, but not the time evolution of the fragmentation
  rate or fragment mass.

\item The shearing motion of the cloud w.r.t. the surrounding ISM has
  a number of small effects. The cloud is ablated and progressively
  destroyed, reducing the final mass of the star cluster. Shear
  affects the shockwave driven into the cloud in compressed models,
  and can create a shockwave in uncompressed ones, provided that the
  shear velocity is large enough. The location of most vigorous
  fragmentation is affected by shear differently for different models:
  uncompressed clouds experience more fragmentation in the leading
  side, while compressed clouds fragment in the trailing part.

\item The sink particle clusters forming in compressed clouds are
  slightly more bound, more compact and more massive than those
  forming in the uncompressed clouds. This suggests that clusters
  formed in compressed clouds are more likely to survive as bound
  objects.

\end{enumerate}

All together, these results indicate that external pressure has a
strong effect on the evolution of cold-phase ISM and star
formation. Not only passing shock fronts (such as supernova shells),
but also large-scale isotropic pressure is important and should be
included in models of galaxy evolution. More quantitative predictions
require larger and more detailed simulations, which we plan to perform
in the future.

\section*{Acknowledgments}

We thank the anonymous referee for invaluable and extensive comments
on the manuscript, which led to significant improvement of the
paper. We thank Vladas Vansevi\v{c}ius, Sergei Nayakshin, Donatas
Narbutis and Walter Dehnen for helpful discussions. KZ acknowledges
the UK STFC for support in the form of a postdoctoral research
position at the University of Leicester. This work was funded by the
Research Council of Lithuania grant no. PRO-15/2012.

Numerical simulations presented in this work were carried out on two
computing clusters. Computations were performed on resources at the
High Performance Computing Center „HPC Sauletekis“ in Vilnius
University Faculty of Physics. This research also used the ALICE High
Performance Computing Facility at the University of Leicester. Some
resources on ALICE form part of the DiRAC Facility jointly funded by
STFC and the Large Facilities Capital Fund of BIS.

\end{document}